\documentclass{article}
\usepackage[main, final]{neurips_2026}
\usepackage[utf8]{inputenc}
\usepackage[T1]{fontenc}
\usepackage[hidelinks]{hyperref}
\usepackage{url}
\usepackage{booktabs}
\usepackage{array}   
\usepackage{nicefrac}
\usepackage[dvipsnames]{xcolor}
\usepackage{amsmath, amssymb, amsfonts, mathtools}
\usepackage{amsthm}
\usepackage{accents}
\usepackage{algorithm}
\usepackage{algpseudocode}
\usepackage{graphicx}
\usepackage{enumitem}
\usepackage{adjustbox}
\usepackage{tikz}
\usetikzlibrary{arrows.meta, positioning, shapes.geometric, fit}
\usepackage{tcolorbox}
\newcommand{\ubar}[1]{\underaccent{\bar}{#1}}
\newcommand{\ee}[1]{\mathbb{E}{#1}}
\newcommand{\E}{\mathrm{E}}

\newtheorem*{theorem*}{Theorem}    
\newtheorem{theoremapp}{Theorem}[section]  

\newcommand{\Var}{\operatorname{Var}}
\newcommand{\Cov}{\operatorname{Cov}}
\newcommand{\one}{\mathbf{1}}

\newcommand{\rev}[1]{#1}                                              
\newenvironment{revblock}{\par}{\par}                                 
\newcommand{\revcolor}{\relax}                                         
\newcommand{\lang}[1]{#1}

\newtheorem{theorem}{Theorem}
\newtheorem{lemma}{Lemma}
\newtheorem{proposition}{Proposition}

\title{Characterizing Underrepresentation when\\ Generalizing Causal Survival Estimates}

\author{%
  Bolun Liu\\Johns Hopkins University \And Sean McGrath \\Washington University in St. Louis \And Yiren Hou\\Yale University \And Elizabeth A. Stuart\\Johns Hopkins University \And Harsh Parikh\\Yale University  
}

\begin{document}

\maketitle

\begin{abstract}
Randomized trial findings are routinely generalized to broader target populations to inform decisions in
medicine and public policy. When the trial sample is misaligned with the target population, generalized effect
estimates can yield suboptimal decisions. We characterize the target subgroups that are
underrepresented in the trial cohort in time-to-event settings. Here, underrepresentation arises in two distinct
ways: baseline underrepresentation, when a subgroup's covariate profile is rare in the trial relative to the
target, and time-varying underrepresentation, when differential censoring erodes a subgroup over follow-up.
Existing diagnostics either conflate these mechanisms or \lang{apply} only \lang{to} the baseline case. We show that the asymptotic variance of the efficient generalizability estimator
admits a decomposition in which only two terms can diverge as representation weakens: a sampling-weight
term capturing baseline misalignment and a censoring-weight term capturing differential loss to follow-up.
Underrepresented subgroups are therefore \lang{those with imprecise generalized estimates}, and the
\lang{larger of the two terms gives} the responsible \lang{channel}. We \lang{use this decomposition} in \textit{TRACE} (\textit{T}emporal \textit{R}ashomon \textit{A}nalysis of \textit{CE}nsored data), which combines a self-normalized variance objective with a tree-based search to localize underrepresented subgroups, attribute each to baseline mismatch or censoring, and \lang{find} the follow-up intervals where censoring drives the variance inflation\lang{. We solve} the joint problem \lang{with} a parametric dynamic program nested inside a Rashomon ensemble of near-optimal trees. \lang{In synthetic benchmarks, TRACE recovers the planted underrepresented
regions, attributing each to its mechanism and locating the follow-up windows that drive time-varying
underrepresentation. In a case study generalizing the National Job Training
Partnership Act (JTPA) Study to U.S. JTPA-eligible adults, restricting to reliably generalizable subgroups narrows the 95\% confidence interval for the refined estimand by 60\%.}
\end{abstract}

\section{Introduction}
\label{sec:intro}

\lang{Findings from
randomized trials are routinely transported} to target populations beyond the trial sample, a
practice known as \emph{generalizability}. Examples range from
extending oncology-trial results to broader patient cohorts
\citep{ColeStuart2010, Westreich2017transport} to transferring
policy-evaluation findings across regions
\citep{HotzImbensMortimer2005, StuartColeBradshaw2011}\lang{, and in each case
the goal is to estimate treatment effects that remain valid for the target population. This goal is harder to reach with survival outcomes, which accrue over time for individuals who
may be lost to follow-up, and researchers have therefore extended generalizability methods to transport}
time-to-event effects \citep{Dahabreh2020transport}, \lang{estimate}
heterogeneous effects under censoring \citep{CuiKosorok2023}, \lang{fuse}
survival data across sources \citep{liu2025targeted}, and \lang{incorporate}
machine-learning-based estimators
\citep{RytgaardvanderLaan2022, Westling2023survival}.

\lang{As a running example, we use} the National Job Training Partnership Act (JTPA) Study \citep{Bloom1997JTPA, AbadieAngristImbens2002}\lang{, in which eligible adult applicants were randomized} to be offered or
denied \lang{federal job-training services, so that its estimated effect on time-to-employment has} strong internal validity. To extend the
JTPA findings to a broader policy target, we consider a population
sample from the Current Population Survey (CPS), which represents
JTPA-eligible adults nationally \citep{FloodIPUMS, CEPRCPS}. Under
the assumptions in Section~\ref{sec:setup}, JTPA training is
estimated to increase the probability of employment at any point
during the $24$ months by $2.7$ percentage points
among U.S. JTPA-eligible adults. \lang{Despite} the large trial sample
size, \lang{however, the} $95\%$
confidence interval (CI) \lang{of} $[-10.5,\, +15.9]$ percentage points \lang{is wide enough to include} both substantial benefit and substantial harm\lang{, and so too wide to inform labor policy.}

\lang{Generalizability} of
time-to-event trial findings can fail \lang{through two mechanisms:} \emph{baseline
underrepresentation}\lang{, which} occurs when a subgroup's covariate profile is
sparsely observed in the trial relative to the target\lang{, and} \emph{time-varying underrepresentation}\lang{, which} occurs when
differential censoring depletes a subgroup's trial sample over
follow-up\lang{.} Both \lang{mechanisms are at work in} the JTPA-to-CPS
generalization (Section~\ref{sec:jtpa})\lang{, where they widen the interval above in ways that cannot be detected from the point estimate alone.}

To our knowledge, no existing diagnostic \lang{handles} both mechanisms\lang{: the}
closest, \citet{Parikh2025-ck}, \lang{targets} baseline mismatch alone
in cross-sectional settings and conflates the two when extended to
time-to-event \lang{outcomes} (Section~\ref{sec:related}). \lang{A policymaker using the JTPA findings therefore cannot tell} \emph{which} target subgroups the
trial can be reliably generalized to, \emph{why} generalization
fails for the rest, \lang{or} \emph{when} during follow-up precision is
lost. We propose \textbf{TRACE} (\textbf{T}emporal \textbf{R}ashomon \textbf{A}nalysis of \textbf{CE}nsored data), \lang{an} interpretable diagnostic \lang{built around these three questions, which, when applied to JTPA-to-CPS,} identifies the \lang{reliably generalizable} population, attributes failures elsewhere to baseline mismatch or censoring depletion, and \lang{narrows} the $95\%$ confidence interval \rev{for the refined estimand} by $60\%$.

\paragraph{Contributions.} We propose TRACE, a general framework for diagnosing subgroup-level
underrepresentation in time-to-event generalizability settings.
\begin{enumerate}[label=\arabic*., leftmargin=*, itemsep=1pt, topsep=1pt]
   \item \lang{We derive} a variance decomposition that isolates
   baseline and time-varying underrepresentation as two unbounded
   factors, reducing subgroup diagnosis to identifying where those
   factors inflate variance (Theorem~\ref{thm:var}; Section~\ref{sec:setup-var}).

  \item \lang{To answer} the \emph{which}, \emph{why}, and \emph{when}
  questions\lang{, we cast} diagnosis as a variance-minimization
  problem with a self-regulating constraint\lang{, whose optimum trims units and intervenes on person-intervals in the subgroups} most affected by each failure mechanism (Theorem~\ref{thm:Bh-variance}; Section~\ref{sec:trace-output}).

  \item \lang{We solve this optimization with} a \emph{scalable} parametric \lang{dynamic program} nested inside an \emph{interpretable} Rashomon ensemble of near-optimal trees (\cite{Breiman2001twocultures}; Algorithm~\ref{alg:nested}).
\end{enumerate}

\section{Related Work}
\label{sec:related}
\textbf{Generalizability: cross-sectional and survival.} Generalizability and transportability of trial findings are well-studied in biomedical research \citep{HotzImbensMortimer2005, ColeStuart2010, StuartColeBradshaw2011, Tipton2014generalize, Westreich2017transport, Dahabreh2020transport, Colnet, generalizabilityreview} and in the machine learning community under the names domain shift/adaptation \citep{Sugiyama2007covariateshift, QuiononeroCandela2009, zhou2023domain}, with parallel developments in semi-supervised learning \citep{testa2025semiparametric, hou2025efficient}\lang{. Across} these literatures, weighted, outcome-regression, and doubly robust estimators dominate practice. Extensions to time-to-event outcomes are more recent: \citet{wen2025multi} derive semiparametric inference for a multi-source PATE on time-to-event outcomes\lang{, which is the basis} for our framework\lang{,} and \lang{doubly robust and targeted estimators that accommodate machine-learning nuisances have been developed in parallel} \citep{lee2022doubly, liu2025targeted}\rev{, including orthogonal survival learners for heterogeneous effects under censoring \citep{frauen2025orthogonal}}, with applied demonstrations in oncology \citep{ramagopalan2022transportability, zuo2022transportability}. Although a few of these \lang{authors note} that heterogeneity inflates variance, none target subgroup underrepresentation as a primary methodological object, and none distinguish baseline from censoring-driven depletion.

\textbf{Underrepresentation and refined estimands.} \lang{Methods that correct for distributional shift, such as group distributionally robust optimization (DRO)} \citep{Sagawa2020groupdro}\lang{, are well developed, whereas less is known about where a shift reduces precision, and we are not aware of work on when during follow-up this happens.} The closest precedent, ROOT \citep{Parikh2025-ck}, uses a sparse-tree Rashomon set to diagnose baseline underrepresentation in cross-sectional generalizability settings via examination of variance. TRACE \lang{builds on this tree-based approach and extends it to} (i) decompose variance into baseline and censoring channels, (ii) attribute each underrepresented subgroup to \lang{its} mechanism, and (iii) \lang{locate} the intervals where censoring drives the loss of precision in settings with longitudinal outcome follow-up.

\section{Setup and Variance Decomposition}
\label{sec:setup}

In this section, we identify two unbounded factors from the asymptotic variance of the efficient generalizability
estimator that correspond, respectively, to the baseline and time-varying failure mechanisms of Section~\ref{sec:intro}. The decomposition is the foundation of the diagnostic in Section~\ref{sec:method}.

\subsection{Data and Estimand}
\label{sec:setup-data}
Data are observed at discrete
follow-up intervals $t = 0, 1, \ldots, T$. A binary source indicator
$S$ marks each unit as trial ($S=1$) or target ($S=0$); baseline
covariates $X$ are observed on every unit. Trial units additionally
carry a binary treatment $A \in \{0,1\}$ and a longitudinal
censoring-and-survival history. \rev{For each $t$, $C_t = 1$ if the unit is censored (lost to follow-up) by interval $t$ and $0$ otherwise, and $Y_{t+1} = 1$ if the unit is still event-free (at risk) at interval $t+1$ and $0$ once the event has occurred; $Y_{t+1}$ is observed only when $C_t = 0$.} Each trial unit $i$ contributes
$
O_i = \bigl(S_i = 1,\, X_i,\, A_i,\, C_{i,0},\, (1 - C_{i,0}) Y_{i,1},\, C_{i,1},\, \ldots,\, C_{i,T-1},\, (1 - C_{i,T-1}) Y_{i,T}\bigr),
$
while target units contribute only $(S_i = 0, X_i)$. \rev{That is, treatment, censoring, and event histories are observed only for the trial cohort, and target units contribute baseline covariates alone; the censoring and survival quantities introduced below are therefore defined on the trial population ($S = 1$) and transported to the target through $X$.} \rev{For any generic time-indexed variable $Z_t$, set}
$\bar Z_t = (Z_0, \ldots, Z_t)$, $\ubar Z_t = (Z_t, \ldots, Z_T)$,
and let $Y_T^a$ denote the potential survival outcome at horizon
$T$ under $A = a \in \{0,1\}$.

Our estimand is the \emph{target
population average treatment effect (PATE)} on the survival
outcome at horizon $T$, $\tau_T = \tau_T(1) - \tau_T(0)$ with
$\tau_T(a) = \mathbb{E}[Y_T^a \mid S = 0]$. \rev{Equivalently, $\tau_T(a) = \Pr(Y_T^a = 1 \mid S = 0)$, so $\tau_T$ is a contrast of survival probabilities under $A=0$ and $A=1$ at horizon $T$ in the target population of interest.}

\subsection{Identification}
\label{sec:setup-id}

Following \cite{wen2025multi}, assuming consistency,
treatment-assignment positivity, and treatment exchangeability, we require the following identification assumptions:
\begin{enumerate}[label=\textbf{A\arabic*.}, leftmargin=*, itemsep=1pt, topsep=1pt]
  \item \textit{Censoring positivity in the trial.} For each $a$ and
  each $t \le T-1$,
  $\Pr(C_t=0 \mid \bar C_{t-1}=0,\, \bar Y_t=1,\, A=a,\, X,\, S=1) > 0$.
  \item \textit{Conditional exchangeability in censoring.} For each
  $a$ and each $t \le T-1$,
  $\ubar Y_{t+1} \perp C_t \mid (\bar C_{t-1}=0,\, \bar Y_t=1,\,
  A=a,\, X,\, S=1)$.
  \item \textit{Conditional exchangeability over $S$.} For each $a$
  and each $t \le T$,
  $Y_t^a \perp S \mid (\bar C_{t-1}=0,\, \bar Y_{t-1}=1,\, A=a,\, X)$.
  \item \textit{Trial-participation positivity.} If
  $\Pr(X=x \mid S=0) > 0$, then $\Pr(X=x \mid S=1) > 0$.
\end{enumerate}
\rev{Together with the standard trial assumptions, A1--A2 identify the within-trial treatment effect on survival and A3--A4 transport it to the target \citep{Dahabreh2020transport, Colnet, generalizabilityreview}, so} A1--A4 suffice for \emph{point} identification of $\tau_T$ (formal
statements and discussion in Appendix~\ref{ap: id}). They do not,
however, suffice for \emph{uniform reliability across the covariate
space and across follow-up}: A4 may hold globally yet be
near-violated for specific covariate profiles, and A1 may hold
globally yet be near-violated at specific covariate values and
follow-up intervals. These near-violations are the two
failure mechanisms of Section~\ref{sec:intro}, and they \lang{appear} in
two identifiable quantities:
\[
\ell(X) \equiv \frac{\Pr(S=0\mid X)}{\Pr(S=1\mid X)},
\qquad
\pi^a_{C,j}(X) \equiv \Pr(C_j = 0 \mid \bar C_{j-1}=0,\, \bar Y_j=1,\, A=a,\, X,\, S=1).
\]$\ell(X)$ is large at covariate profiles common in the target but
rare in the trial, a near-violation of A4 that drives \emph{baseline
underrepresentation}. $\pi^a_{C,j}(X)$ is small at person-intervals
where remaining uncensored is unlikely; its inverse, the IPCW weight
$1/\pi^a_{C,j}(X)$, is correspondingly large, a near-violation of
A1 that drives \emph{time-varying underrepresentation}.

\subsection{Variance Decomposition}
\label{sec:setup-var}

Under A1--A4 and standard randomization, $\tau_T$ admits an EIF $\chi_{\tau_T}(O)$ whose variance attains the
semiparametric efficiency bound; the EIF and the associated TMLE,
which attains this bound under nuisance-rate conditions analogous to
\citet{wen2025multi}, are derived in Appendices~\ref{ap: eif} and ~\ref{ap: tmle}. We now state the central diagnostic identity of TRACE.

\begin{theorem}[Variance decomposition]
\label{thm:var}
Given A1--A4 and standard randomization assumptions, under a
nonparametric model on the observed data, the asymptotic variance attaining the semiparametric efficiency bound is
\begin{equation}
\begin{aligned}
& \mathrm{Var}\{\chi_{\tau_T}(O)\} = \sum_{a \in \{0,1\}}\sum_{t=1}^T \mathbb{E}\biggl[ \;
  \colorbox{red!25}{$\displaystyle
    \bigl(\underbrace{\ell(X)}_{\text{\scriptsize baseline underrep.}}\bigr)^{\!2}
  $}
  \cdot
  \colorbox{red!25}{$\displaystyle
    \biggl(\underbrace{\prod\nolimits_{j < t} \pi^a_{C,j}(X)}_{\text{\scriptsize time-varying underrep.}}\biggr)^{\!\!-2}
  $} \\[2pt] 
& \hspace{3em} \cdot
  \underbrace{
    \colorbox{green!25}{$\displaystyle
      \bigl(W^a_{t-1}(O)\bigr)^2 \cdot \left( \left[Q^a_{t}(X)\right]^2 - \left[Q^a_{t-1}(X)\right]^2 \right)
    $}
  }_{\text{\scriptsize $\le 1/[\,\Pr(S=0)^2\, \Pr(A=a \mid X, S=1)^2]$}}
\biggr] + \mathbb{E}\!\biggl[ \;
  \underbrace{
    \colorbox{green!25}{$\displaystyle
      \frac{\mathbf{1}\{S=0\}}{\Pr(S=0)^2}\bigl(\tau_T(X) - \tau_T\bigr)^2
    $}
  }_{\text{\scriptsize $\le 4/\Pr(S=0)^2$}}
\; \biggr] \;.
\end{aligned}
\label{eq:varderomp}
\end{equation}
\rev{where $W^a_{t-1}(O) = \mathbf{1}\{S = 1\}\,\mathbf{1}\{A = a\}\,\mathbf{1}\{\bar C_{t-1} = 0\} / \{\Pr(S = 0)\,\pi^a(X)\}$ denotes the trial-at-risk weight with $\pi^a(X) = \Pr(A = a \mid X, S = 1)$,} $\{Q^a_j(X)\}_{j=0}^T$ the
iterated conditional survival probabilities, and $\tau_T(X)$ the
conditional average treatment effect (CATE) at horizon $T$. Formal definitions, variance derivation, and proof of
the worst-case bounds are in Appendix~\ref{ap: var}.
\end{theorem}

\lang{Equation}~\eqref{eq:varderomp} \lang{has} two \lang{kinds of factors}. The \textcolor{green!50!black}{green factors}, i.e.,
bounded survival probabilities, the randomization probability, and
design constants, are \lang{bounded} (worst-case bounds beneath each box)\lang{, whereas} the \textcolor{red!70!black}{red factors} are not:
$\ell(X)$ inflates at baseline-underrepresented covariate profiles, and $1/\prod_{j<t}\pi^a_{C,j}(X)$ inflates
at time-varying-underrepresented person-intervals. The two
red factors map onto the diagnostic questions of
Section~\ref{sec:intro}\lang{, in that large values of} $\ell(X)$ \lang{mark} \emph{which}
\lang{subgroups are underrepresented at baseline} (the baseline \emph{why})\lang{, and large values of} $1/\prod_{j<t}\pi^a_{C,j}(X)$ \lang{mark}
\emph{when} \lang{censoring depletes them} (the temporal \emph{why}). Characterizing
underrepresented subgroups therefore reduces to localizing where the
red factors inflate the variance most, a search problem that
\lang{we formulate as a tractable optimization in} Section~\ref{sec:method}. 
\begin{revblock}
\textbf{What counts as underrepresented.} \lang{We call} a subgroup underrepresented when the trial cannot support a precise generalized estimate for it, i.e., when its contribution to the efficiency-bound variance \lang{in Theorem}~\ref{thm:var} is large. This decision-relevant notion \citep{Rudolph2022underrep, Chen2023genscore, Parikh2025-ck} couples how rare a profile is in the trial with how much remains to be learned about it, so a large $\ell(X)$ alone is not enough: $\ell(X)^2$ enters \eqref{eq:varderomp} only multiplied by the survival-increment variance $[Q^a_t(X)]^2 - [Q^a_{t-1}(X)]^2$, and a profile that is rare in the trial but whose treatment effects are nearly deterministic (homogeneous) is estimated precisely from the few trial units available and is not flagged. A rare profile with variable treatment effects (heterogeneous) is flagged, because the generalized estimate there is imprecise and can change the treatment decision.
\end{revblock}

\section{The TRACE Framework}
\label{sec:method}

\lang{Diagnosis therefore amounts to finding} the
covariate profiles and follow-up intervals at which the two red factors inflate the variance most\lang{, which we do} in three steps. We
first introduce two interventions, one per red factor, that contract
the variance along its baseline and time-varying axes
(Section~\ref{sec:trace-interventions}). We then pose a bilevel
optimization problem that minimizes the variance jointly over the two
interventions, and provide the analytical and empirical objectives that
make it tractable (Section~\ref{sec:trace-objective}). Finally, we read
the optimum as a \rev{four-region} partition of the data \lang{and use it to answer} the
\emph{which}, \emph{why}, and \emph{when} questions of
Section~\ref{sec:intro} (Section~\ref{sec:trace-output}).

\subsection{Two interventions: \texorpdfstring{$B$}{B} and \texorpdfstring{$h$}{h}}
\label{sec:trace-interventions}

\textbf{Baseline trim $B$.} Let $B : \mathcal{X} \to \{0,1\}$ be a
measurable function applied symmetrically to trial and target. Setting
$B(x) = 0$ refines the estimand from the target PATE to the
\emph{$B$-restricted PATE}
$\tau_T(B) \equiv \mathbb{E}[Y_T^{1} - Y_T^{0} \mid S = 0,\, B(X) = 1]$,
with $p_B \equiv \Pr(S = 0,\, B(X) = 1) > 0$ and identified under
A1--A4 restricted to $\{B(X) = 1\}$. $B$ contracts the $\ell(X)^2$
red factor across covariate space\lang{, so that the retained set $\{B(X)=1\}$ is the} subpopulation \lang{to which the trial generalizes reliably (the} \emph{which} \lang{question).}

\textbf{Censoring intervention $h$.} Let $h_{i,t} \in \{0,1\}$ be
a trial-side, person-interval indicator. Setting $h_{i,t} = 1$
replaces unit $i$'s observed law with the counterfactual law in
which $\Pr(C_t = 0 \mid \cdot) = 1$ at interval $t$, i.e., \emph{as if} censoring had not occurred there. $h$ contracts the
$1 / \prod_{j<t}\pi^a_{C,j}(X)$ red factor across
person-intervals\lang{, and the intervals with $h_{i,t}=1$ are those at which} censoring depletion \lang{inflates the variance most (the} \emph{when} \lang{question).} \rev{$h$ is thus both a diagnostic and a guide for future trials\lang{, since better retention or follow-up in the flagged subgroups and intervals would most improve precision.}}

Together, $(B, h)$ map each red factor to a distinct intervention,
a correspondence we make precise in Theorem~\ref{thm:Bh-variance} below.

\subsection{The TRACE objective}
\label{sec:trace-objective}

\textbf{Bilevel optimization.} The diagnostic we want is the joint
$(B, h)$ that contracts the variance most. Because $B$ refines the
estimand and $h$ refines the trial sample, we couple the two through
a profiled bilevel problem:
\begin{equation}
\label{eq:trace-bilevel}
B^{*} \;=\; \arg\min_{B}\, \widehat V\!\bigl(B,\, h^{*}(B)\bigr),
\qquad
h^{*}(B) \;=\; \arg\min_{h}\, \widehat V(B, h) \;+\; \lambda_h \|h\|_1.
\end{equation}
The outer level chooses the baseline trim; the inner level returns,
for each candidate $B$, the censoring intervention that minimizes a
sparsity-penalized variance. The remainder of the section gives the
analytical form of $V(B,h)$, the empirical estimator $\widehat V$ that
plugs into~\eqref{eq:trace-bilevel}, and the calibration of
$\lambda_h$.

\textbf{Analytical objective.} The next theorem expresses the
variance under $(B, h)$ as a parametric family containing
Theorem~\ref{thm:var}'s expression as the no-intervention special case,
and \lang{implies} that $B$ and $h$ act on \lang{separate} red factors.
\begin{theorem}[Variance under $(B, h)$]
\label{thm:Bh-variance}
Given A1--A4 and standard randomization, with $p_B > 0$, the
variance of the efficient estimator of $\tau_T(B)$ under intervention
$(B, h)$ is
\begin{equation}
\label{eq:Bh-variance}
\begin{aligned}
& V(B, h)
\equiv \mathbb{E}_{\mathcal P}\!\left[\,\mathbb{E}_{\mathcal P^*_i}\!\left[\chi_{\tau_T}(O; B)^2\right]\,\right]
= \sum_{a \in \{0,1\}}\sum_{t=1}^T \mathbb{E}_{\mathcal P}\!\biggl[ \;
  \colorbox{red!25}{$\displaystyle
    \bigl(\underbrace{B(X) \; \ell(X)}_{\text{\scriptsize contracted by }B}\bigr)^{\!2}
  $}
  \\[-4pt]
& \hspace{2em} \cdot\;
  \colorbox{red!25}{$\displaystyle
    \biggl(\underbrace{\textstyle\prod\nolimits_{j<t} \pi^a_{C,j}(X)^{1 - h_{i,j}/2}}_{\text{\scriptsize contracted by }h}\biggr)^{\!\!-2}
  $}
  \;\cdot\;
  \colorbox{green!25}{$\displaystyle
    \bigl(W^a_{t-1}(O; B)\bigr)^2 \cdot \bigl(\,[Q^a_t(X)]^2 - [Q^a_{t-1}(X)]^2\,\bigr)
  $}
\biggr]
  \\[-10pt]
& \hspace{15em} +\, \mathbb{E}_{\mathcal P}\!\biggl[\,
  \colorbox{green!25}{$\displaystyle
    \frac{\mathbf{1}\{S=0,\, B(X)=1\}}{p_B^2}\bigl(\tau_T(X) - \tau_T(B)\bigr)^2
  $}
\,\biggr].
\end{aligned}
\end{equation}
where $\chi_{\tau_T}(O; B)$ is the EIF for $\tau_T(B)$
(Appendix~\ref{ap: eifb}), $\mathcal{P}$ is the observed-data law, and
$\mathcal{P}^*_i$ is the counterfactual law for unit $i$ in which
$\Pr(C_j = 0 \mid \cdot) = 1$ at every interval $j$ with
$h_{i,j} = 1$. In particular, $V(\mathbf{1}, \mathbf{0})$ recovers
$\mathrm{Var}\{\chi_{\tau_T}(O)\}$ from Theorem~\ref{thm:var}. Proof in Appendix~\ref{ap: varbh}.
\end{theorem}

\lang{The} red/green structure of
Theorem~\ref{thm:var} \lang{carries over to Theorem}~\ref{thm:Bh-variance}: the green factors remain bounded, while the
two red factors are now under explicit control of the interventions.
Setting $B(X_i) = 0$ removes unit $i$'s contribution from the trial
sum, attenuating the average of $\ell(X)^2$ over the retained
subpopulation; setting $h_{i,j} = 1$ halves the exponent on
$\pi^a_{C,j}(X)$ in unit $i$'s contribution, attenuating the
censoring product. Minimizing $V(B,h)$ over $(B,h)$ therefore
\lang{works by shrinking} the two unbounded factors \lang{we want to localize}.
\begin{revblock}
\textbf{Attribution by intervention.} The two red factors enter the variance multiplicatively, so the variance alone does not identify the baseline-vs-censoring split; TRACE attributes by intervention instead, using the variance only as the objective. \lang{Because} $B$ \lang{changes} only $\ell(X)$ and $h$ only the censoring product \lang{(Theorem}~\ref{thm:Bh-variance}\lang{), we attribute} a region's inflation to baseline underrepresentation if trimming with $B$ removes it\lang{,} and to censoring if $h$ does. Regions inflated on both axes are resolved by \emph{$R$-gating}: before any trimming, TRACE computes for each trial unit the ratio of its baseline to its censoring-driven variance inflation,
\begin{equation}
\label{eq:rgate-main}
R_i \;=\; \hat\ell(X_i)^2 / F_i ,
\end{equation}
where $F_i \ge 1$ is the unit's censoring-amplification factor (Appendix~\ref{ap:rgating}); large $R_i$ means baseline is the dominant cause and $h$ is suppressed for that unit; small $R_i$ means censoring dominates.
\end{revblock}

\textbf{Empirical objective.} Plugging \eqref{eq:Bh-variance} into
\eqref{eq:trace-bilevel} requires an estimator of $V(B, h)$. We consider the \emph{block-self-normalized} (mixed-Hájek)
estimator
\begin{align}
\label{eq:mixed-hajek}
\widehat V(B, h)
=
\underbrace{\frac{1}{n\,(\hat p_B^{\text{trl}})^{2}}\!\!\sum_{i:\,S_i=1}\!\! B(X_i)\,\Psi(O_i; h_i)}_{\text{trial block}}+ 
\underbrace{\frac{1}{n\,(\hat p_B^{\text{tgt}})^{2}}\!\!\sum_{i:\,S_i=0}\!\! B(X_i)\,\bigl(\hat\tau(X_i) - \hat\tau_T(B)\bigr)^{2}}_{\text{target block}},
\end{align}
with block-specific Hájek denominators $\hat p_B^{\text{trl}} = n^{-1}\!\sum_{i:S_i=1} \hat\ell(X_i)\,B(X_i)$ and $\hat p_B^{\text{tgt}} = n^{-1}\!\sum_{i:S_i=0} B(X_i)$, where $\Psi(O_i; h_i)$ is a functional \lang{of the} trial data and $h_i$\lang{,} defined in Appendix~\ref{ap: empirical}.

Two features of the proposed $\widehat V(B,h)$ make the
bilevel search practical. \emph{(i) Self-regulation.} The squared
H\'ajek denominator $\hat p_B^{-2}$ grows superlinearly as $B$
excludes more units, so each additional trim carries an internally
generated cost. The optimum $B^{*}$ therefore removes a region only
when its variance contribution exceeds that cost, \lang{so no}
explicit penalty on $B$ \lang{is needed} (formal statement in Theorem~\ref{thm:trim},
Appendix~\ref{ap:trim}). \emph{(ii) Decoupling.} The nuisance
functions in \eqref{eq:mixed-hajek} are fitted once from the full
observed data, independently of $(B,h)$: changing $B$ only zeroes
$\ell(X)$ and rescales $\hat p_B$, and changing $h_{i,t}$ updates a
single interval's $\pi_{C,j}^a(X)$ contribution. The bilevel search
therefore avoids nuisance refitting altogether and can be decomposed into sub-optimization problems in Section~\ref{sec:alg}.

\textbf{Calibration of $\lambda_h$.} The inner penalty $\lambda_h$ is calibrated to a sparsity
level $\eta \in (0, 1)$ rather than chosen as a free hyperparameter.
At the no-trim baseline $(B, h) = (\mathbf{1}, \mathbf{0})$, define
each $h_{i,k}$'s \emph{standalone variance reduction}
\[
\Delta \widehat V_{i,k} \;\coloneqq\; \widehat V(\mathbf{1}, \mathbf{0}) - \widehat V(\mathbf{1},\, h_{i,k}=1),
\]
i.e., the drop in $\widehat V$ from activating $h_{i,k}$ alone.
Set $\lambda_h$ equal to the $\eta$-quantile of $\Delta\widehat V_{i,k}$.
\lang{An} $h_{i,k}$ \lang{is then kept by} the inner penalty iff $\Delta\widehat V_{i,k} > \lambda_h$,
which by construction holds for exactly the top $1 - \eta$ fraction
of person-intervals. With the default $\eta = 0.95$, $h^{*}(\mathbf{1})$ activates the $5\%$
of person-intervals whose standalone contribution to the
censoring-driven variance is largest; the remaining $95\%$ are
penalized out. \lang{We specify} $\eta$ rather than $\lambda_h$ \lang{because
$1-\eta$ is simply the fraction of person-intervals flagged.}

\subsection{Reading the optimum: a \rev{four-region} partition}
\label{sec:trace-output}
The optimum $(B^*, h^*)$ \rev{and the gate $R_i$} partition the target into \rev{four}
diagnostic regions, each with a distinct interpretation:
\begin{enumerate}[label=(\roman*), leftmargin=*, itemsep=1pt, topsep=1pt]
  \item \textbf{Reliably generalizable} ($B^*=1$, $h^*=\mathbf{0}$\lang{; \emph{which}}):
  subgroups whose trial estimates transport precisely to the target.
  \item \textbf{Censoring-depleted} ($B^*=1$, $h^*\neq\mathbf{0}$\lang{; temporal \emph{why} and \emph{when}}):
  subgroups that are well-represented at baseline but lose precision
  through differential censoring; the indices $(i,t)$ at which
  $h^*_{i,t}=1$ \lang{give} the follow-up intervals responsible.
  \item \textbf{Baseline-underrepresented} ($B^*=0$\rev{, $R_i$ large}\lang{; baseline \emph{why}}): subgroups whose
  covariate profiles are too rare in the trial relative to the
  target\rev{, with censoring playing no material role, so that future trials could fix this region directly through recruitment}.
  \item \rev{\textbf{Doubly underrepresented} ($B^*=0$, $R_i$ small\lang{; baseline and temporal \emph{why}}): subgroups that are both rare at baseline and depleted by censoring\lang{. Both interventions reduce their variance contribution,} and $R_i$ \lang{measures} which \lang{mechanism} dominates.}
\end{enumerate}
\rev{Since $h$ is not computed for trimmed units, (iii) and (iv) are separated by $R_i$ rather than by $h^*$ (Appendix~\ref{ap:rgating:attr}).}

\lang{In region (i) the trial supports a precise generalized estimate, whereas regions} (ii)\rev{--(iv)} \lang{need} further data
collection\lang{, in the form of better retention for censoring depletion, more trial participants with the rare baseline profiles, or} \rev{both}\lang{. For censoring-depleted subgroups, $h^*$ also gives the follow-up intervals where retention matters most.}

\section{Nested Optimization}
\label{sec:alg}
The profiling identity in \eqref{eq:trace-bilevel}, along with the fact that the nuisance functions are independent of $(B,h)$, allows the bilevel optimization problem to split into two independent subproblems: an inner over $h \in \{0,1\}^{n_{\text{trl}} \times T}$, which we solve with a \emph{scalable} parametric dynamic program, and an outer over $B \in \{0,1\}^{\mathcal X}$, solved with an \emph{interpretable} Rashomon ensemble of trees.

\subsection{Inner: a per-unit affine envelope}
\label{sec:alg-inner}

For fixed $B$, the trial block of $\widehat V(B)$ decouples across
units, so the inner problem reduces to a per-individual
minimization. A short structural argument
(Appendix~\ref{ap:inner-dp}) further collapses the $2^T$-way binary
search over $h_i \in \{0,1\}^T$ to a $(T{+}1)$-way search over the
integer $\ell_1$-budget $j = \|h_i\|_1$:
\begin{equation}
\label{eq:envelope}
\min_{h_i \in \{0,1\}^T}\!\bigl[\widehat V_i(B, h_i) + \lambda_h\,\|h_i\|_1\bigr]
\;=\;
\min_{j \in \{0, \ldots, T\}}\!\bigl[\alpha_i(B)\, f_{i,j}
\,+\, \lambda_h\, j\bigr],
\qquad
\alpha_i(B) \coloneqq \frac{B(X_i)}{n\,(\hat p_B^{\text{trl}})^{2}}.
\end{equation}

\textbf{Intuition.} Two structural facts drive~\eqref{eq:envelope}.
\emph{(i) Budget sufficiency.} Within the class of $h_i$ sharing a
common $\ell_1$-budget $j$, the penalty $\lambda_h j$ is constant
and only the variance contribution varies; partitioning
$\{0,1\}^T$ by budget reduces $2^T$ candidate configurations to
$T{+}1$ representative slopes $\{f_{i,j}\}_{j=0}^{T}$, each the
budget-$j$ minimum of the variance contribution alone. The
right-hand side of~\eqref{eq:envelope} is then a lower envelope of
$T{+}1$ affine functions of the scalar $\alpha_i(B)$.
\emph{(ii) Decoupling.} The slopes $f_{i,j}$ are functions of the
fitted nuisances and the budget alone, \lang{i.e.,} $B$ enters
\eqref{eq:envelope} only through $\alpha_i(B)$, so the slopes
are independent of every candidate $B$ the outer ensemble visits.

\textbf{Computational gain.} Each $f_{i,j}$ is the optimum of a
$\binom{T}{j}$-way subset selection, but the additive cumulative
state of the censoring product (Appendix~\ref{ap:inner-dp}) admits
a budget-augmented Bellman recursion that returns all $T{+}1$
slopes in $\mathcal O(T^{2} G)$ time per unit on a state grid of
size $G$, with grid error vanishing as $G \to \infty$
(Theorem~\ref{thm:dp}). Decoupling lets us precompute the slopes
once at $\mathcal O(n_{\text{trl}}\, T^{2}\, G)$ total cost; each
subsequent outer query of $B$ then reduces to walking the per-unit
envelopes at $\mathcal O(n_{\text{trl}}\, T)$, a factor of
$\mathcal O(TG)$ faster than a non-parametric DP that absorbs
$\alpha_i(B)$ into the recursion and rebuilds at every $B$. \lang{This precomputation makes the joint
search over an ensemble of $M$ trees practical.} \rev{In total, growing $M$ depth-$D$ trees over $F$ covariates costs $\mathcal O(n_{\mathrm{trl}}\,T^{2}G + M\,2^{D}F\,n_{\mathrm{trl}}\,T)$: the expensive $T^{2}G$ build is paid once and never compounds with the $2^{D}$ tree term (Appendix~\ref{ap:compute}).}
\subsection{Outer problem: Rashomon ensemble of trees}

Given $h^*(B)$, we optimize $B$ via the Rashomon Set of Optimal
Trees (ROOT) algorithm of \citet{Parikh2025-ck}. ROOT is
\lang{designed} for variance-style objectives that depend globally on
the current partition rather than additively on its leaves\lang{, a
structure that, to our knowledge, other tree methods do not accommodate.} The
obstacle for off-the-shelf alternatives is that $\widehat V$ is not
branch-additive: flipping a single leaf of $B$ rescales the Hájek
denominators $\hat p_B^{\text{trl}}$ and $\hat p_B^{\text{tgt}}$,
which multiply every leaf's contribution, so any greedy learner
that scores splits against branch-local quantities (CART
\citep{Breiman1984CART}, MOB \citep{Zeileis2008MOB}, causal forests
\citep{WagerAthey2018}) \lang{optimizes a different} objective. ROOT instead
scores every candidate split against $\widehat V(B)$ on the full
data: $M$ randomized trees are grown over the union of trial and
target individuals, each split is accepted only if it lowers the
global $\widehat V(B)$ on the current partition, the top-$K$ trees
within tolerance of the best objective form the \emph{Rashomon set}
\citep{Breiman2001twocultures}, and $B^*$ is aggregated by majority
vote across that set\lang{. The aggregated $B^*$ can be summarized as a decision tree over baseline covariates, so} the \emph{which} answer of Section~\ref{sec:trace-output}
\lang{can be audited} directly.

We summarize the full nested procedure in Algorithm~\ref{alg:nested} (Appendix~\ref{ap:alg}).

\section{Experiments}
\label{sec:experiments}

We apply TRACE to synthetic settings with baseline and time-varying underrepresentation\lang{, in which} Experiment~1 combines both mechanisms with ten covariates \lang{and} Experiments~2a--2d isolate each mechanism and include an adversarial case \lang{(details in} Appendix~\ref{ap:dgp}\lang{)}.

\begin{figure}[t]
\centering
\includegraphics[width=0.9\textwidth]{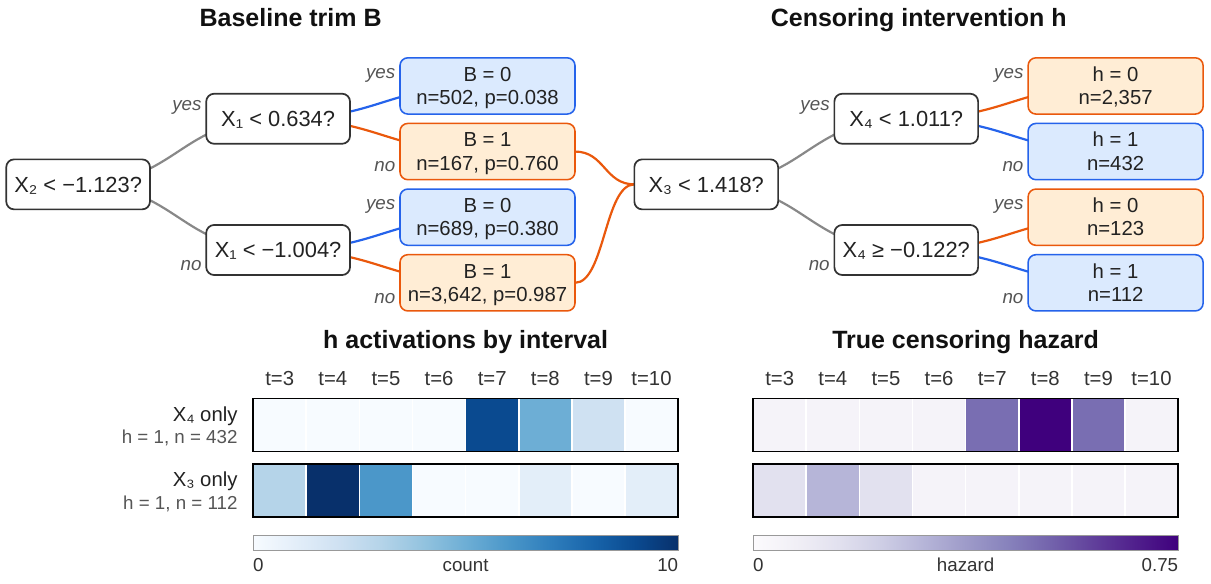}
\caption{\lang{Underrepresented subgroups found by} TRACE in Experiment 1.}
\label{fig:simres-exp1}
\end{figure}

\textbf{Experiment 1.} Ten i.i.d.\ $\mathcal{N}(0,1)$ covariates are generated, where $X_1$ and $X_2$ affect trial-participation and survival, $X_3$ and $X_4$ affect censoring, and $X_5, \ldots, X_{10}$ 
are noise. The censoring structure creates two subgroups: high $X_3$ (but not $X_4$) produces an early modest 
spike in censoring at $t \in \{3,4,5\}$ and high $X_4$ (but not $X_3$) produces a 
late heavy spike at $t \in \{7,8,9\}$. TRACE recovers the underrepresented subgroups (Figure~\ref{fig:simres-exp1}): $B^*$ 
trims on $\{X_1, X_2\}$, while $h^*$ 
activates almost exclusively on individuals in the two censoring-spike subgroups defined by $X_3$ and $X_4$, with the noise covariates unused 
by either intervention. The two spikes are recovered at the 
correct intervals, and $h^*$ activates comparable numbers of 
person-intervals in both subgroups despite the late spike's larger 
peak hazard, \lang{consistent with} the greater downstream impact of early censoring.

\begin{figure}[t]
\centering
\includegraphics[width=0.8\textwidth]{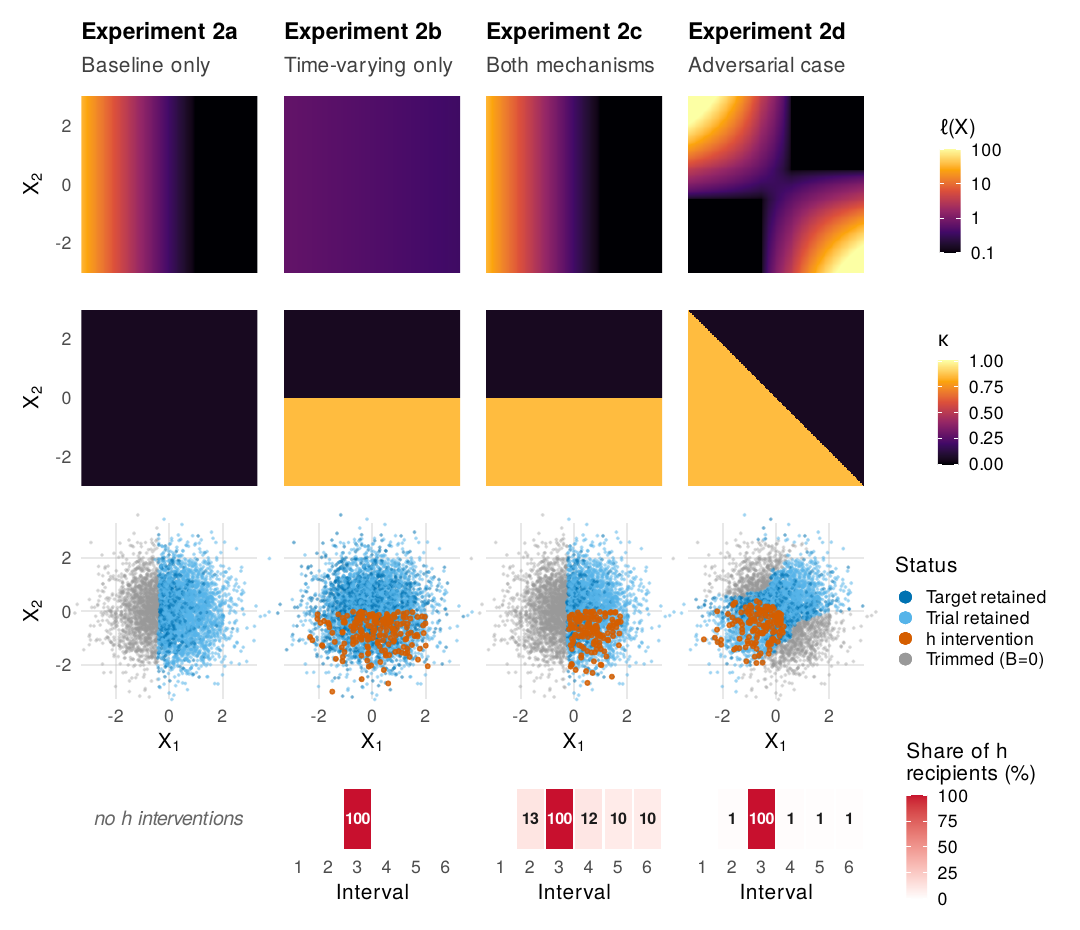}
\caption{\lang{Underrepresented subgroups found by} TRACE in
Experiments 2a--2d. \textbf{Top row}: generalizability
weight $\ell(X)$ on the
$(X_1, X_2)$ plane; \textbf{Second row}: peak censoring probability
at the spike interval; \textbf{Third row}: per-subject $B$-trim and $h$-intervention status based on TRACE; \textbf{Fourth row}:
proportion of $h$ recipients intervened at each interval.}
\label{fig:simres-exp2}
\end{figure}

\textbf{Experiment 2a: Baseline underrepresentation only.}
$\ell(X)$ depends on $X_1$ and the censoring probability is negligible in this experiment. TRACE correctly trims the region with high values of $\ell$ and does not invoke $h$ (Figure~\ref{fig:simres-exp2}, column~1).

\textbf{Experiment 2b: Time-varying underrepresentation only.} Here $\ell(X)$ is negligible and censoring spikes at $t = 3$ when $X_2 < 0$\lang{, so} TRACE trims only a thin region \lang{and intervenes on} every $h$ recipient at $t = 3$ and nowhere else (Figure~\ref{fig:simres-exp2}, column~2).

\textbf{Experiment 2c: Baseline and time-varying underrepresentation.}
This experiment combines 2a and 2b\lang{, with} trial participation \lang{depending} on $X_1$ and censoring \lang{spiking} at $t = 3$ when $X_2 < 0$. Because the two mechanisms act on independent covariates, \lang{both} trimming and $h$-targeting \lang{are needed, and} TRACE trims the high-$\ell$ tail on $X_1$ \lang{while} $h^*$ activates on individuals with $X_2 < 0$ at and around $t = 3$ (Figure~\ref{fig:simres-exp2}, column~3).

\textbf{Experiment 2d: Adversarial case for TRACE.}
In this experiment, the trial-participation odds are \rev{low (so that $\ell(X)$ is high)} in two off-diagonal corners of the
$(X_1, X_2)$ plane, a structure no single axis-aligned split can exactly
recover, while censoring spikes when $X_1 + X_2 < 0$ at time $3$. TRACE trims the two high-$\ell$ corners rather than a half-plane\lang{, and} $h^*$ activates almost exclusively at $t = 3$ among 
the retained trial individuals (Figure~\ref{fig:simres-exp2}, 
column~4)\lang{, so} $h$ remains temporally precise even when 
$B$ incompletely recovers the underrepresented region.
\begin{revblock}
\textbf{Quantitative recovery and baselines.} Appendix~\ref{ap:metrics} reports recovery metrics for $B$ and $h$ across seeds (precision, recall, interval localization, and noise-covariate split rates), and \lang{in} Appendix~\ref{ap:baselines}\lang{, we compare} TRACE with $\ell$-thresholding, censoring thresholding, and ROOT applied to $\ell$ alone. TRACE recovers the regions \lang{where each DGP induces baseline or censoring underrepresentation} and is the only method that attributes both mechanisms correctly.
\end{revblock}

\section{Generalizing JTPA Job-Training Effects to a National Target}
\label{sec:jtpa}

\lang{Applied to} the JTPA-to-CPS generalization \lang{of} Section~\ref{sec:intro}\lang{, TRACE reduces} the variance objective from
$\widehat V(\mathbf 1, \mathbf 0) = 81.58$ to
$\widehat V(B^*, h^*) = 12.68$ ($-84.5\%$)\lang{, with Rashomon-set objectives between} $13.13$ \lang{and} $13.77$ \lang{and} a stable $2/3$-vote supermajority \lang{that}
trims $46.3\%$ of the target and $13.8\%$ of the trial\lang{. Because}
$h^*$ activates only $98 / 249{,}096$ trial person-intervals
($0.04\%$, vs.\ $1{,}796$ without R-gating), baseline mismatch\lang{, rather than}
censoring depletion\lang{, is the main} source of underrepresentation.
The $95\%$ CI for the 24-month employment effect is $[-10.5,\,+15.9]$ percentage points for the original PATE and $[-1.9,\,+8.6]$ for the refined estimand
(\rev{$-\widehat\tau_{24}$}$ = +0.034$, $\widehat{\mathrm{SE}} = 0.027$), a
$60.6\%$ reduction in width\lang{,} \rev{\lang{although} the two intervals \lang{are for different estimands, the first for} the full CPS target \lang{and the second for} the reliably generalizable subpopulation, \lang{which} is more credible but less general \citep{Parikh2025-ck}.} \rev{Because the event in JTPA is first employment ($Y_t = 1$ while not yet employed; Appendix~\ref{ap:jtpa-outcome}), all JTPA effects and intervals are reported on the employment scale, i.e., as $-\tau_{24}$ in the notation of Section~\ref{sec:setup-data}, so that positive values mean training raises the probability of having become employed by month 24.}

\begin{figure}[t]
\centering
\includegraphics[width=0.9\textwidth]{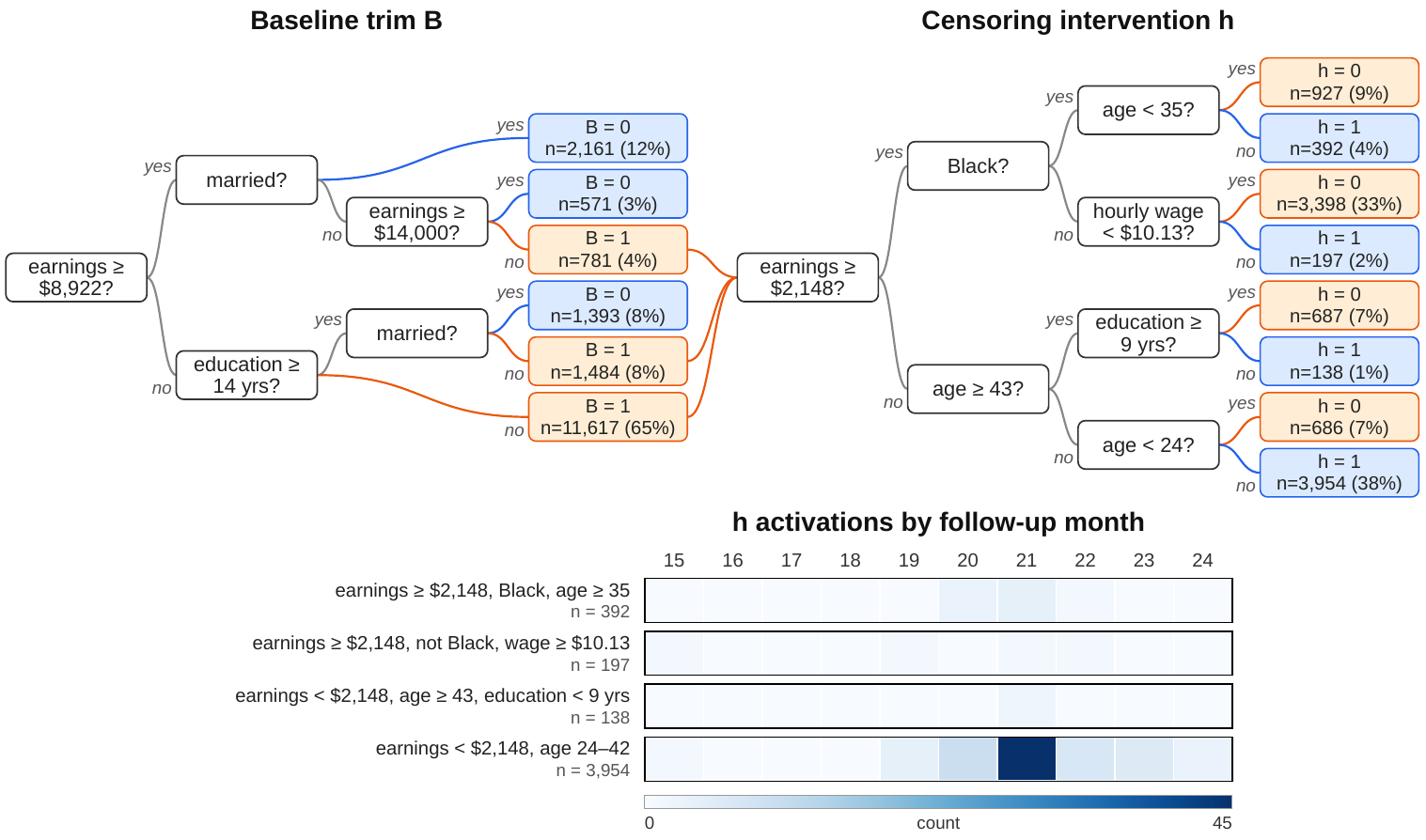}
\caption{\lang{Underrepresented subgroups found by} TRACE in \lang{the} JTPA-to-CPS generalization.}
\label{fig:jtpa}
\end{figure}

\textbf{Localizing JTPA underrepresentation (Figure~\ref{fig:jtpa}).} The $B$-tree trims higher socioeconomic status branches common in CPS but rare
among the disadvantaged-targeted JTPA enrollees (earnings
$\ge \$8{,}922$ and married; earnings $\ge \$14{,}000$ unmarried;
earnings $< \$8{,}922$ with $\ge 14$ years of education and
married), retaining the \lang{typical} low-earnings, low-education
profile as the reliably generalizable region, which is consistent with the JTPA selection rule \citep{Bloom1997JTPA}. Within the (much
smaller) censoring-depleted region, the dominant $h$-active leaf
(earnings $<\$2{,}148$ and ages $24$--$42$; $n = 3{,}954$)
\lang{accounts for} $\sim\!80\%$ of activations and is sharply peaked at
$t = 21$ ($45/50$ leaf interventions), \lang{consistent with} a
late-second-year loss-to-follow-up shock\rev{ in the window where JTPA's attrition is documented \citep{Bloom1997JTPA}}. In short, the trial over-represents the most disadvantaged \lang{adults} relative to the CPS target (\emph{where}, baseline \emph{why}), and late attrition \lang{depletes} low-earnings young adults (\emph{when}, temporal \emph{why}). Details\rev{, including sensitivity analyses,} are in Appendix~\ref{ap:jtpa}.

\section{Discussion}
\label{sec:discussion}

\lang{We developed TRACE to find} \emph{which} \lang{target subgroups a trial generalizes to,} \emph{why} \lang{generalization fails for the others, and} \emph{when} \lang{during follow-up precision is lost.} \lang{The method rests on Theorem}~\ref{thm:var}\lang{, by which} $\ell(X)$ and $1/\prod_{j<t}\pi^a_{C,j}(X)$ are the only channels through which baseline misalignment and differential censoring can inflate variance, \lang{and on} the coupled interventions $(B,h)$ of Theorem~\ref{thm:Bh-variance}\lang{, which} contract each channel separately \lang{while} the self-normalized objective in~\eqref{eq:mixed-hajek} \lang{penalizes} over-trimming. \lang{Nesting a} parametric DP \lang{inside} a Rashomon ensemble (Algorithm~\ref{alg:nested}) keeps the search scalable and yields an interpretable \rev{four-region} partition.

\textbf{Limitations, future work, and broader impacts.}
TRACE is diagnostic \lang{rather than corrective:} it identifies subgroups for whom generalization is unreliable but does not adjust the point estimator $\hat\tau_T$\lang{, and} the $B$-restricted estimand changes the inferential question rather than recovering the original PATE\lang{, a trade-off that needs to be described and discussed in any presentation of results}. \rev{When both mechanisms are large on the same units, how credit is split between them depends on the objective. Likewise, $h$ \lang{flags} where and when censoring \lang{reduces} precision\lang{, so its practical use is} to guide retention and follow-up in future trials. The point estimate and CI for the $B$-restricted PATE are \lang{secondary outputs that condition} on the selected $(B^{*}, h^{*})$\lang{, and inference that accounts for this selection is outside our diagnostic aim. As in} the convention for data-defined estimands \citep{Crump2009, Li2019overlap, Parikh2025-ck}\lang{, we leave such post-selection inference to future work.}} \lang{Extensions} include continuous-time survival via counting-process EIFs \citep{RytgaardvanderLaan2022, wen2025multi}, cross-fit or sample-splitting Rashomon variants that propagate nuisance uncertainty, multi-source generalizations that connect to data fusion \citep{liu2025targeted}, and corrective \lang{methods that use TRACE output for} adaptive trial design or targeted oversampling. \lang{Reporting which subgroups a trial underrepresents can make evidence-based policy more transparent and equitable.} \lang{TRACE can, however,} be misused if its outputs are read as causal claims about excluded subgroups or if the $B$-restricted estimand is conflated with the PATE\lang{, and we therefore} recommend reporting trimmed-region characteristics with the refined estimand so that the retained population is auditable.

\begin{ack}
This work was supported by the National Institute on Drug Abuse of the National Institutes of Health under award number R01DA056407. The content is solely the responsibility of the authors and does not necessarily represent the official views of the National Institutes of Health. The authors declare no competing interests.
\end{ack}

\bibliographystyle{apalike}
\bibliography{reference}
\newpage
\appendix

\section{Identification and Discussion of Assumptions}
\label{ap: id}

We require the following assumptions for identifying PATE.

\begin{enumerate}[leftmargin=*, itemsep=2pt, topsep=3pt]
  \item \emph{(Consistency).} If $A = a$, then $\bar{Y}_t = \bar{Y}_t^a$ for all $t \leq T$.
  \item \emph{(Positivity of treatment assignment in the trial).} For each treatment $a$, if $\Pr(X = x \mid S = 1) > 0$, then $\Pr(A = a \mid X = x, S = 1) > 0$.
  \item \emph{(Exchangeability over treatment assignment in the trial).} $Y_t^a \perp A \mid (X, S = 1), \quad \forall t \leq T$.
  \item \emph{(A1: Censoring positivity in the trial).} For each treatment $a$, $\Pr(C_t = 0 \mid \bar C_{t-1} = 0,\, \bar Y_t = 1,\, A = a,\, X,\, S = 1) > 0$ for all $t \leq T-1$.
  \item \emph{(A2: Conditional exchangeability in censoring).} $\ubar{Y}_{t+1} \perp C_t \mid (\bar C_{t-1} = 0,\, \bar Y_t = 1,\, A = a,\, X,\, S = 1)$ for all $t \leq T-1$.
  \item \emph{(A3: Conditional exchangeability over $S$).} For each treatment $a$, $Y_t^a \perp S \mid (\bar{C}_{t - 1} = 0,\; \bar{Y}_{t - 1} = 1,\; A = a,\; X), \quad \forall t \leq T$.
  \item \emph{(A4: Trial-participation positivity).} If $\Pr(X = x \mid S = 0) > 0$, then $\Pr(X = x \mid S = 1) > 0$.
\end{enumerate}

\paragraph{Discussion of Assumptions.} Consistency, positivity of treatment assignment in the trial, and exchangeability over treatment assignment in the trial are standard conditions assumed in randomized trials \lang{and} extend to the survival setting. A1 (censoring positivity in the trial) ensures that no subgroup is fully censored at any interval $t$, and A2 (conditional exchangeability in censoring) states that, conditional on the history, censoring is missing at random and ignorable. Together with the standard randomization assumptions, A1 and A2 identify the within-trial treatment effect on the survival outcome. A3 (conditional exchangeability over $S$) is the generalizability assumption adapted to the time-to-event setting, allowing trial-derived risk-set hazards to be transported to the target conditional on the history; A4 (trial-participation positivity) ensures that every covariate profile prevalent in the target has a corresponding trial counterpart. Under A1--A4, $\tau_T(a)$ is identified by an observed-data functional that integrates the within-trial risk-set hazards at each interval, reweighted to the target covariate distribution, as proved below.

By the law of iterated expectations, we first rewrite $\tau_t(a)$ as
\[
\tau_t(a) = \mathbb{E}(Y_t^a \mid S = 0)
= \mathbb{E}\!\left( \, \mathbb{E}(Y_t^a \mid X, S = 0) \,\middle|\, S = 0 \right).
\]
From A3, we have
\begin{align*}
\mathbb{E}(Y_t^a \mid \{Y_{t - 1} = 1, C_{t - 1} = 0\}, X)
=& \;\mathbb{E}(Y_t^a \mid \{Y_{t - 1} = 1, C_{t - 1} = 0\}, X, S = 0) \\
=& \; \mathbb{E}(Y_t^a \mid \{Y_{t - 1} = 1, C_{t - 1}
= 0\}, X, S = 1) \;.
\end{align*}
Consequently,
\begin{align*}
\mathbb{E}(Y_t^a \mid X) = &\;\mathbb{E}(Y_t^a \mid  X, S = 0)
= \mathbb{E}(Y_t^a \mid X, S = 1) \;.
\end{align*}
Thus, the PATE can be estimated using the observed data from the trial:
\[
\mathbb{E}(Y_t^a \mid S = 0)
= \mathbb{E}\!\left( \, \mathbb{E}(Y_t^a \mid X, S = 1) \,\middle|\, S = 0 \right).
\]
We now focus on the identification of $\ee(Y^a_t \mid X = x, S = 1)$. By definition, all individuals are at risk at baseline (i.e., $Y_0 = 1$), and by convention we set
$\bar{C}_{-1} = 0$. From consistency, treatment-assignment positivity, treatment exchangeability, A1, and A2, we have
\begin{align*}
    & \quad \; \mathbb{E}(Y_t^a \mid X = x, S = 1) \\
    &= \ee (Y_t^a \mid A = a, X = x, S = 1) \tag*{\hfill ((Exchangeability over treatment assignment)} \\
     &= \ee (Y_t \mid A = a, X = x, S = 1) \tag*{\hfill (Consistency)} \\
    & = \ee (Y_t \mid A = a, X = x, S = 1, C_0 = 0) \tag*{\hfill (A2)} \\
    & = \ee (Y_t \mid Y_1 = 1, A = a, X = x, S = 1, C_0 = 0) \ee(Y_1 \mid A = a, X = x, S = 1, C_0 = 0)\\
    & = \ee (Y_t \mid Y_1 = 1, A = a, X = x, S = 1, C_1 = 0, C_0 = 0) \tag*{\hfill (A2)} \\
    & \quad \quad \quad \quad \quad \quad \quad \quad \ee(Y_1 \mid A = a, X = x, S = 1, C_0 = 0) \\
    & = \underbrace{\ee (Y_t \mid Y_1 = 1, A = a, X = x, S = 1, C_1 = 0)}_{\text{{Iterate with previous three steps}}} \ee(Y_1 \mid A = a, X = x, S = 1, C_0 = 0) \\
    & \qquad \qquad \qquad \vdots \\
    & = \ee (Y_t \mid Y_{t - 1} = 1, A = a, X = x, S = 1, C_{t - 2} = 0) \\
    & \quad \quad \quad \quad \quad \quad \quad \quad  \prod_{k = 1}^{t - 1} \ee(Y_k \mid Y_{k - 1} = 1,A = a, X = x, S = 1, C_{k - 1} = 0) \\
    & = \ee (Y_t \mid Y_{t - 1} = 1, A = a, X = x, S = 1, C_{t - 1} = 0)  \tag*{\hfill (A2)} \\
    & \quad \quad \quad \quad \quad \quad \quad \quad \prod_{k = 1}^{t - 1}  \ee(Y_k \mid Y_{k - 1} = 1,A = a, X = x, S = 1, C_{k - 1} = 0)\\
&=\prod_{k=1}^{t}
\ee\!\left( Y_k \mid Y_{k-1} = 1, C_{k-1} = 0, X = x, A = a, S = 1 \right) \;. \tag*{\hfill (A3)}
\end{align*}
Therefore, $\tau_t(a)$ can be identified by the observed data functional
\begin{align*}
\psi_t(a)
&= \mathbb{E}\!\left( \, \mathbb{E}(Y_t^a \mid X, S = 1) \,\middle|\, S = 0 \right) \\
&= \mathbb{E}\!\left( \, \prod_{k=1}^{t}
    \mathbb{E}\!\left( Y_k \,\middle|\, Y_{k-1} = 1,\; C_{k-1} = 0,\; X,\; A = a,\; S = 1 \right)
    \,\middle|\, S = 0 \right) \\
&= \mathbb{E}\Bigg(
    \frac{\Pr(S=1)}{\Pr(S=0)} \cdot
    \frac{\Pr(S=0 \mid X)}{\Pr(S=1 \mid X)} \\
&\hspace{10em}\cdot
    \prod_{k=1}^{t}
    \mathbb{E}\!\left( Y_k \,\middle|\, Y_{k-1} = 1,\; C_{k-1} = 0,\; X,\; A = a,\; S = 1 \right)
    \,\Bigg|\, S = 1 \Bigg) \\
&= \mathbb{E}\Bigg(
    \frac{\mathbf{1}\{S=1\}}{\Pr(S=0)} \cdot
    \frac{\Pr(S=0 \mid X)}{\Pr(S=1 \mid X)} \\
&\hspace{10em}\cdot
    \prod_{k=1}^{t}
    \mathbb{E}\!\left( Y_k \,\middle|\, Y_{k-1} = 1,\; C_{k-1} = 0,\; X,\; A = a,\; S = 1 \right)
    \Bigg) \;.
\end{align*}
The PATE at time $T$ becomes
\[\tau_T = \psi_T(1) - \psi_T(0) \;.\]

\section{Derivation of EIF}
\label{ap: eif}

\begin{theoremapp}[Efficient influence function for $\tau_T(a)$]
\label{thm:eif}
Given A1--A4 and standard randomization assumptions, under a nonparametric model on the observed data,
\begin{equation}
\label{eq:eif}
\begin{aligned}
\chi_{\psi_T(a)}(O) &= \sum_{t=1}^T \underbrace{\frac{\mathbf{1}\{\bar C_{t-1}=0,\, A=a,\, S=1\}}{\Pr(S=0)\,\pi^a(X)}}_{\equiv\, W^a_{t-1}(O)} \cdot \underbrace{\frac{\Pr(S=0\mid X)}{\Pr(S=1\mid X)}}_{\equiv\, \ell(X)} \\
& \quad \cdot \frac{1}{\bar\pi^a_{C,t-1}(X)} \cdot \underbrace{\bigl[Q^a_t(X) - Q^a_{t-1}(X)\bigr]}_{\equiv\, \Delta Q^a_t(X)} \; +   \; \frac{\mathbf{1}\{S=0\}}{\Pr(S=0)}\bigl[Q^a_0(X) - \tau_T(a)\bigr]
\end{aligned}
\end{equation}
is the EIF for $\psi_T(a) = \mathbb{E}[Y_T^a \mid S=0]$, and $\chi_{\tau_T}(O) = \chi_{\psi_T(1)}(O) - \chi_{\psi_T(0)}(O)$ is the EIF for the contrast. 
\end{theoremapp}
\begin{proof}
    
Define the weight
\[
w(X) = \frac{\mathbf{1}(S = 1)}{\Pr(S = 0)} \cdot
        \frac{\Pr(S = 0 \mid X)}{\Pr(S = 1 \mid X)} \;.
\]
For $a \in \{0, 1\}$, define the treatment probability in the trial as
$$\pi^a(X) \equiv \Pr(A=a \mid X,S=1) \;,$$
the uncensored history up to time $k$ as
$$H_k^a \equiv (\bar Y_k=1,\ \bar C_k=0,\ A=a,\ S=1) \;,$$
the probability of being uncensored at time $k$ given previous history as
$$\pi_{C_k}^a(X) \equiv \Pr(C_k=0 \mid Y_k=1,\ X,\ H^a_{k-1}) \;,$$
and
$\bar\pi^a_{C_k}(X) \equiv \prod_{j=0}^{k-1} \pi^a_{C_j}(X)$.
Moreover, let $Q_t^a \equiv Y_t^a$, and for $k=0,\ldots,t-1$, define
\[
Q^a_{k-1}(X) \equiv \ee \,\!\bigl(Q_k^a \mid X,H^a_{k-1}\bigr) = \mathbb{E}(Y_t^a \mid X, H_{k-1}^a)
= \prod_{j=k}^t \Pr(Y_j=1 \mid X,H^a_{j-1}) \; .
\]
Note that, in the identification formula, we have a series of concatenated conditional expectations. Specifically, the $\sigma$-algebra satisfies
\[\sigma(H^a_0, X) \subset \sigma(H^a_1, X) \subset \cdots  \subset \sigma(H^a_T, X) \;.\]
Therefore, by the iterated law of conditional expectation, the identification formula can be rewritten as
\[\psi_t(a) \equiv \ee(Y_t^a \mid S = 0) = \ee [w(X) Q^a_0(X)] \;.\]
Let
\[
p(S, X,A,C_0,(1-C_0)\times Y_1,\ldots,C_{T-1},(1-C_{T-1})\times Y_T;\theta)
\]
denote a regular parametric submodel of $\mathcal{M}_{NP}$ indexed by $\theta$.

The efficient influence function in $\mathcal{M}_{NP}$ is defined as the unique mean zero, finite variance random variable $\chi_{\psi_t(a)}(O)$ such that
\[
\left.\frac{\partial}{\partial \theta}\right|_{\theta=0} \psi_t(a;\theta)
= \ee\!\left\{ \chi_{\psi_t(a)}(O) \cdot \mathbb{S}(O) \right\},
\]
where
\[
\mathbb{S}(O) = \mathbb{S}(S,X,A,C_0,(1-C_0)\times Y_1,\ldots,C_{T-1},(1-C_{T-1})\times Y_T)
\]
is the score function of the regular parametric submodel at $\theta=0$.

By Leibniz rule, for any expectation of function $g_\theta(O)$ under the parametric model $p(O;\theta)$, we have
\[
\left. \frac{d}{d\theta} \,\mathbb{E}_{\theta} \!\left[ g_\theta(O) \right] \right|_{\theta=0}
= \ee\!\left[ \left. \partial_\theta g_\theta(O) \right|_{\theta=0} \right]
  + \ee\!\left[ g_0(O)\, \mathbb{S}(O) \right].
\]
Now, let $g_\theta(O) = w(X) Q^a_0(X)$, we have
\begin{align}
\label{eq:pathwise}
\begin{aligned}
   &\quad  \left. \partial_\theta \psi(a;\theta) \right|_{\theta=0} \\
&= \mathbb{E}\!\left[ \left. \partial_\theta \big( w_\theta(X)\, Q^a_{0,\theta}(X) \big) \right|_{\theta=0} \right]
   + \mathbb{E}\!\left[ w(X)\, Q^a_0(X)\, \mathbb{S}(O) \right] \\
&= \mathbb{E}\!\left[ \left. w(X)\, \partial_\theta Q^a_{0,\theta}(X) \right|_{\theta=0} \right] + \mathbb{E}\!\left[
   \left. Q^a_0(X)\, \partial_\theta w_\theta(X) \right|_{\theta=0} \right]
   + \mathbb{E}\!\left[ w(X)\, Q^a_0(X)\, \mathbb{S}(O) \right].
\end{aligned}
\end{align}
We first focus on the first term, specifically $ \partial_\theta Q^a_{0,\theta}(X)|_{\theta=0}$, and introduce the following lemma to help with the iterated conditional expectation.
\begin{lemma}
\label{lem: score}
Let $X$ be a random variable with conditional density $p_\theta(x \mid H=h)$
depending smoothly on a scalar parameter $\theta$, and let
$\mathbb{S}_\theta(x \mid H) = \partial_\theta \log p_\theta(x \mid H)$ denote the conditional score.
For any measurable function $g$ with $\mathbb{E}[|g(X)|]<\infty$, we have
\[
\left.\frac{\partial}{\partial\theta}\right|_{\theta=0}
\mathbb{E}_\theta\!\big[g(Y) \mid H=h\big]
\;=\;
\mathbb{E}\!\big[g(Y)\,\mathbb{S}(Y \mid H) \mid H=h\big],
\]
where $\mathbb{S}(Y \mid H) := \mathbb{S}_{\theta}(Y \mid H)\big|_{\theta=0}$.
\end{lemma}
Proof of Lemma~\ref{lem: score} can be found in Appendix~\ref{ap: lemma1}.

Apply Lemma~\ref{lem: score} for $Q^a_{0,\theta}(X)$, we have
\begin{align*}
    \left. \partial_\theta Q^a_{0,\theta}(X) \right|_{\theta=0}
=& \left. \partial_\theta \mathbb{E}\!\left[ Q^a_{1,\theta}(X) \,\middle|\, H^a_{0} \right] \right|_{\theta=0} \\
=& \mathbb{E}\!\left[ Q^a_1(X)\, \mathbb{S}(Y_1 \mid H^a_{0}) \,\middle|\, H^a_{0} \right]
  + \mathbb{E}\!\left[ \left. \partial_\theta Q^a_{1,\theta}(X) \right|_{\theta=0} \,\middle|\, H^a_{0} \right].
\end{align*}
Note that the second term allows us to iterate over the conditional expectations
\begin{align*}
&\left.\partial_\theta Q^a_{0,\theta}(X)\right|_{\theta=0} \\
=& \mathbb{E}\!\left[\, Q^a_1(X)\, \mathbb{S}\big(Y_1 \mid H^a_{0}\big) \,\middle|\, H^a_{0} \right]
  + \mathbb{E}\!\left[\, \left.\partial_\theta Q_{1,\theta}(X)\right|_{\theta=0} \,\middle|\, H^a_{0} \right] \\
=& \mathbb{E}\!\left[\, Q^a_1(X)\, \mathbb{S}\big(Y_1 \mid H^a_{0}\big) \,\middle|\, H^a_{0} \right]
  + \mathbb{E}\!\Big[\, \mathbb{E}\!\left\{ Q^a_2(X)\, \mathbb{S}\big(Y_2 \mid H^a_{1}\big) \,\middle|\, H^a_{1} \right\} \\
  &\qquad + \mathbb{E}\!\left\{ \left.\partial_\theta Q^a_{2,\theta}(X)\right|_{\theta=0} \,\middle|\, H^a_{1} \right\}
      \,\Big|\, H^a_{0} \Big] \\
=&\mathbb{E}\!\left[\, Q^a_1(X)\, \mathbb{S}\big(Y_1 \mid H^a_{0}\big) \,\middle|\, H^a_{0} \right]
  + \mathbb{E}\!\left[\, Q^a_2(X)\, \mathbb{S}\big(Y_2 \mid H^a_{1}\big) \,\middle|\, H^a_{0} \right]
  + \mathbb{E}\!\left[\, \left.\partial_\theta Q^a_{2,\theta}(X)\right|_{\theta=0} \,\middle|\, H^a_{0} \right] \\
&\;\;\vdots \\
=& \sum_{k=1}^{t-1} \mathbb{E}\!\left[\, Q^a_k(X)\, \mathbb{S}\big(Y_k \mid H^a_{k-1}\big) \,\middle|\, H^a_{0} \right] \\
  &+ \mathbb{E}\!\left[\, \left.\partial_\theta Q^a_{t,\theta}(X)\right|_{\theta=0} \,\middle|\, H^a_{0} \right]\\
=& \sum_{k=1}^{t-1} \mathbb{E}\!\left[\, Q^a_k(X)\, \mathbb{S}\big(Y_k \mid H^a_{k-1}\big) \,\middle|\, H^a_{0} \right] \\
  &+ \mathbb{E}\!\left[\, \left.\partial_\theta Y_t\right|_{\theta=0} \,\middle|\, H^a_{0} \right]\\
  =& \sum_{k=1}^{t} \mathbb{E}\!\left[\, Q^a_k(X)\, \mathbb{S}\big(Y_k \mid H_{k-1}\big) \,\middle|\, H^a_{0} \right].
\end{align*}
Therefore, the first term in (\ref{eq:pathwise}) becomes
\begin{align}
\label{eq:firstterm}
\begin{aligned}
    \mathbb{E}\!\left[\, w(X)\, \left.\partial_\theta Q^a_{0,\theta}(X)\right|_{\theta=0} \right]
&= \mathbb{E}\!\left\{ \sum_{k=1}^{t} w(X)\, \mathbb{E}\!\left[\, Q^a_k(X)\, \mathbb{S}\big(Y_k \mid H^a_{k-1}\big) \,\middle|\, H^a_{0} \right] \right\} \\
&= \mathbb{E}\!\left\{ \sum_{k=1}^{t} \mathbb{E}\!\left[\, w(X)\, Q^a_k(X)\, \mathbb{S} \big(Y_k \mid H^a_{k-1}\big) \,\middle|\, H^a_{0} \right] \right\} \\
&= \sum_{k=1}^{t} \mathbb{E}\!\left[\, w(X)\, Q^a_k(X)\, \mathbb{S}\big(Y_k \mid H^a_{k-1}\big) \right].
\end{aligned}
\end{align}
We now want to link the partial above with the full score $\mathbb{S}(O)$. First note that the parametric submodel $p_\theta$ can be factorized as
\begin{align*}
   p(o) \propto & \; p(s)\, p(x \mid s)\, p(a \mid x, s)^{\mathbf{1}(A = a)}  \\
      & \times \prod_{k=1}^{T} p(y_k \mid X, H^a_{k-1})^{\mathbf{1}(H^a_{k - 1}) = 1} \\
      & \times \prod_{k=0}^{T-1} p(c_k \mid y_k = 1, X, H^a_{k - 1} = 1)^{\mathbf{1}(y_k = 1, H^a_{k - 1}) = 1} \;.
\end{align*}
Some of the conditional densities are omitted due to censoring. Thus, the full score can also be factorized as
\begin{align*}
\mathbb{S}(O)
&= \mathbb{S}_S(S)
+ \mathbb{S}_X(X \mid S)
+ \mathbb{S}_A(A \mid S, X) \\
&+ \sum_{k=1}^{T}\mathbf{1}(H^a_{k - 1} = 1) \; \mathbb{S}_{Y_k}(Y_k \mid H^a_{k-1}) \\
& + \sum_{k=0}^{T-1} \mathbf{1}(y_k = 1, H^a_{k - 1} = 1) \; \mathbb{S}_{C_k}(C_k \mid Y_k = 1, H^a_{k - 1})
\end{align*}
Since from the identification formula, there is no randomness in $C_k$ for $k = 0, \cdots, T - 1$, we can focus on the submodels defined on the Hilbert space
\[\mathcal{H}^a = \mathcal{T}_{S} \oplus \mathcal{T}_{X \mid S} \oplus \mathcal{T}_{A \mid X,S} \oplus \mathcal{T}_{Y_1 \mid H^a_0} \oplus \cdots \oplus  \mathcal{T}_{Y_T \mid H^a_{T - 1}} \;.\]
Note that each component in (\ref{eq:firstterm}) lives in the tangent space $\mathcal{T}_{Y_k \mid H^a_{k - 1}}$. Thus, to map it with the full score, we center as a mean-zero function on $\mathcal{T}_{Y_k \mid H^a_{k - 1}}$:
\begin{align*}
  w(X)Q^a_k(X) - \ee [w(X) Q^a_k(X) \mid H^a_{k - 1}] &= w(X)Q^a_k(X) - w(X) \ee [Q^a_k(X) \mid H^a_{k - 1}] \\
&=  w(X)Q^a_k(X) - w(X)Q^a_{k - 1}(X) \;.
\end{align*}
Thus
\begin{align*}
    &\mathbb{E}\!\left[\, w(X)\, \left.\partial_\theta Q^a_{0,\theta}(X)\right|_{\theta=0} \right] \\
=& \sum_{k=1}^{t} \mathbb{E}\!\left[\, w(X)\, Q^a_k(X)\, \mathbb{S}\big(Y_k \mid H^a_{k-1}\big) \right] \\
=& \sum_{k=1}^{t} \mathbb{E}\!\left[\, w(X)\, Q^a_k(X)\, \mathbb{S}\big(Y_k \mid H^a_{k-1}\big) \right] - \underbrace{\mathbb{E}\!\left[\, w(X)\, Q^a_{k-1}(X)\, \mathbb{S}\big(Y_k \mid H^a_{k-1}\big) \right]}_{\text{Mean } 0} \\
=& \sum_{k=1}^{t} \mathbb{E}\!\left[\,\underbrace{ w(X)\, [Q^a_k(X) - Q^a_{k-1}(X)]}_{\in \mathcal{T}_{Y_k \mid H^a_{k-1}}}\, \mathbb{S}\big(Y_k \mid H^a_{k-1}\big) \right] \\
=& \sum_{k=1}^{t} \mathbb{E}\!\left[\frac{\mathbf{1}(A = a)\,\mathbf{1}(\bar{C}_{k-1} = 0)}{\pi^a(X) \bar{\pi}^a_{C_{k-1}}(X)}\,w(X)\, [Q^a_k(X) - Q^a_{k-1}(X)]\, \mathbb{S}\big(O\big) \right] \tag*{\hfill (By orthogonality)}
\end{align*}
Note that there is no need for restricting to $\bar{Y}_k = 1$, as $Q^a_k(X) - Q^a_{k-1}(X)$ is zero off-risk.

We now handle the rest of (\ref{eq:pathwise}). It is cumbersome to handle the derivative of $w(X)$. We \lang{instead} consider the identification formula before changing the measure to $S = 1$. In that case, we have directly
\begin{align*}
&\mathbb{E}\!\left[\, Q^a_0(X)\,\left.\partial_\theta w_\theta(X)\right|_{\theta=0} \right]
\;+\;
\mathbb{E}\!\big[\, w(X)\,Q^a_0(X)\,\mathbb{S}(O) \big] \\
&\quad=\;
\mathbb{E}\!\left[\, \frac{\mathbf{1}\{S=0\}}{\Pr(S=0)}\,\{ Q^a_0(X)-\psi_t(a) \}\,\mathbb{S}(O) \right].
\end{align*}

Thus, collectively, we have the EIF written as
\[
\begin{aligned}
\chi_{\psi_t(a)}(O)
&= \sum_{k=1}^{t}
\frac{\mathbf{1}\{S=1\}\,\mathbf{1}\{A=a\}\,\mathbf{1}\{\bar C^a_{k-1}=0\}}
     {\Pr(S=0)\,\pi^a(X)\,\bar{\pi}^a_{C,k-1}(X)}
\cdot \frac{\Pr(S=0\mid X)}{\Pr(S=1\mid X)}\,[Q^a_k(X)-Q^a_{k-1}(X)]\\
&\quad + \frac{\mathbf{1}\{S=0\}}{\Pr(S=0)}\,\{\,Q^a_0(X)-\psi_t(a)\,\} \;.
\end{aligned}
\]
And the EIF for the PATE at time $t$ is the linear difference
\[\chi_{\tau_t}(O) = \chi_{\psi_t(1)}(O) - \chi_{\psi_t(0)}(O) \;.\]

\end{proof}

\section{Proof of Lemma~\ref{lem: score}}
\label{ap: lemma1}
\begin{proof}
    Write the conditional expectation as
\[
E_\theta[g(Y)\mid H=h]
= \int g(y)\,p_\theta(y \mid h)\,dy
= \frac{\int g(y)\,p_\theta(y,h)\,dy}{p_\theta(h)}.
\]
Differentiating under the integral sign and applying the quotient rule gives
\[
\partial_\theta E_\theta[g(Y)\mid h]
= \frac{\int g(y)\,\partial_\theta p_\theta(y,h)\,dy}{p_\theta(h)}
- \frac{\int g(y)\,p_\theta(y,h)\,dy}{p_\theta(h)^2}\,\partial_\theta p_\theta(h).
\]
Since
\[
S_\theta(y,h) = \partial_\theta \log p_\theta(y,h)
= \frac{\partial_\theta p_\theta(y,h)}{p_\theta(y,h)},
\qquad
S_\theta(h) = \partial_\theta \log p_\theta(h)
= \frac{\partial_\theta p_\theta(h)}{p_\theta(h)} \,,
\]
we have
\[
\partial_\theta p_\theta(y,h) = p_\theta(y,h)\, S_\theta(y,h),
\qquad
\partial_\theta p_\theta(h) = p_\theta(h)\, S_\theta(h).
\]
The derivative becomes
\[
\partial_\theta E_\theta[g(Y)\mid h]
= \frac{\int g(y)\, p_\theta(y,h)\, S_\theta(y,h)\, dy}{p_\theta(h)}
\;-\;
\frac{\int g(y)\, p_\theta(y,h)\, S_\theta(h)\, dy}{p_\theta(h)}.
\]
Recognizing the first term as
$E_\theta[g(Y)S_\theta(Y,H)\mid H=h]$, the second term as $E_\theta[g(Y)S_\theta(H)\mid H=h]$, and using the identity
$S_\theta(Y\mid H)=S_\theta(Y,H)-S_\theta(H)$, we obtain
\[
\partial_\theta E_\theta[g(Y)\mid h]
= E_\theta\!\big[g(Y)(S_\theta(Y,H)-S_\theta(H))\mid H=h\big]
= E_\theta\!\big[g(Y)S_\theta(Y\mid H)\mid H=h\big].
\]
Evaluating at $\theta=0$ yields the result.
\end{proof}

\newpage 
\section{TMLE Estimator}
\label{ap: tmle}

To construct an estimator that achieves the semiparametric efficiency bound characterized by the EIF, we employ the doubly robust Targeted Maximum Likelihood Estimator (TMLE), detailed in Algorithm \ref{alg:tmle}. The TMLE proceeds through a backward recursive targeting procedure that iteratively updates pseudo estimates of the sequential conditional expectations, $Q_k^a(X)$. At each step of the backward iteration, the estimator targets the interval-specific component of the EIF. Because these components are sequentially orthogonal, accumulating these targeted updates recursively ensures that the final estimator solves the global EIF equation. Under
standard nuisance-rate conditions ($o_p(n^{-1/4})$ rates on
$\hat\pi^a$, $\hat\pi^a_C$, $\hat Q^a_k$, $\hat\ell$, and
$\hat\Pr(S=0)$, by Donsker class argument or cross-fitting), the resulting TMLE is asymptotically linear and
attains the semiparametric efficiency bound.

\begin{algorithm}
\caption{Targeted Maximum Likelihood Estimator (TMLE) for $\psi_t(a)$}
\label{alg:tmle}
\begin{algorithmic}[1]
    \Require Nuisance estimates $\hat{\pi}^a(X), \hat{\pi}_0, \hat{\pi}_{C,k-1}^a(X), \hat{l}(X)$ for the mechanisms defined as $\pi^a(X) = \Pr(A = a \mid X, S = 1)$, $\pi_0 = \Pr(S = 0)$, $\pi_{C,k-1}^a(X) = \Pr(C_{k} = 0 \mid Y_k = 1, C_{k-1} = 0, X, S = 1)$, and $l(X) = \Pr(S = 0 \mid X) / \Pr(S = 1 \mid X)$.
    \State \textbf{Initialize:} $\hat{Q}^a_T(X) = Y_T \mid A = a$.
    \For{$k = T, T-1, \dots, 1$}
        \State \textbf{Initial Regression:} Among subjects with $S = 1$ and $\bar{C}_{k-1}=0$, fit the initial outcome model $\eta_{k-1}(A,X;\theta_{k-1})$ by regressing $\hat{Q}^a_k(X)$ on $(A,X)$ using the inverse-logit link $g^{-1}(x)=\{1+\exp(-x)\}^{-1}$.
        \State \textbf{Targeting Step:} Among subjects with $S = 1$, $\bar{C}_{k-1}=0$, and $A=a$, compute the generalizability and IPW weights $\hat \omega_k(X) = \hat{l}(X) / \{\hat{\pi}_0 \hat{\pi}^a(X) \prod_{j = 1}^k \hat{\pi}_{C, j - 1}^a\}$. Solve for the fluctuation parameter $\nu_{k-1}$ by fitting a logistic regression of $\hat{Q}^a_k(X)$ on an intercept using $\omega_k(X)$ as weights and $g(\eta_{k-1}(a,X;\hat\theta_{k-1}))$ as the offset, i.e., solve for $\nu_{k-1}$ in:
        $$
        \mathbb{P}_n\!\left[ \hat \omega_k(X) \left\{ \hat{Q}^a_k(X) - g^{-1}\!\big( g\{\eta_{k-1}(a,X;\hat\theta_{k-1})\} + \nu_{k-1} \big) \right\} \right] = 0.
        $$
        \State \textbf{Update:} For subjects who remain uncensored at $k - 2$ (i.e., $\bar{C}_{k-2} = 0$), predict
        $$
        \hat{Q}^a_{k-1}(X) = g^{-1}\!\big( g\{\eta_{k-1}(a,X;\hat\theta_{k-1})\} + \hat\nu_{k-1} \big).
        $$
    \EndFor
    \State \textbf{Marginalize:} Compute the final empirical mean over the target population $S=0$.
    \State \Return $\hat\psi_{\mathrm{TMLE}} = \mathbb{P}_{S = 0}\!\left\{\hat{Q}_0(X)\right\}$.
\end{algorithmic}
\end{algorithm}

\section{Efficiency Bound}
\label{ap: var}

\begin{theorem*}[Variance decomposition]
Given A1--A4 and standard randomization assumptions, under a
nonparametric model on the observed data, the asymptotic variance attaining the semiparametric efficiency bound is
\begin{equation*}
\begin{aligned}
& \mathrm{Var}\{\chi_{\tau_T}(O)\} = \sum_{a \in \{0,1\}}\sum_{t=1}^T \mathbb{E}\biggl[ \;
  \colorbox{red!25}{$\displaystyle
    \bigl(\underbrace{\ell(X)}_{\text{\scriptsize baseline underrep.}}\bigr)^{\!2}
  $}
  \cdot
  \colorbox{red!25}{$\displaystyle
    \biggl(\underbrace{\prod\nolimits_{j < t} \pi^a_{C,j}(X)}_{\text{\scriptsize time-varying underrep.}}\biggr)^{\!\!-2}
  $} \\[2pt] 
& \hspace{3em} \cdot
  \underbrace{
    \colorbox{green!25}{$\displaystyle
      \bigl(W^a_{t-1}(O)\bigr)^2 \cdot \left( \left[Q^a_{t}(X)\right]^2 - \left[Q^a_{t-1}(X)\right]^2 \right)
    $}
  }_{\text{\scriptsize $\le 1/[\,\Pr(S=0)^2\, \Pr(A=a \mid X, S=1)^2]$}}
\biggr] + \mathbb{E}\!\biggl[ \;
  \underbrace{
    \colorbox{green!25}{$\displaystyle
      \frac{\mathbf{1}\{S=0\}}{\Pr(S=0)^2}\bigl(\tau_T(X) - \tau_T\bigr)^2
    $}
  }_{\text{\scriptsize $\le 4/\Pr(S=0)^2$}}
\; \biggr] \;.
\end{aligned}
\end{equation*}
where $W^a_{t-1}$
denotes the trial-at-risk weight, $\{Q^a_j(X)\}_{j=0}^T$ the
iterated conditional survival probabilities, and $\tau_T(X)$ the
conditional average treatment effect (CATE) at horizon $T$. 
\end{theorem*}
\proof{
The variance of the TMLE estimator that attains the semiparametric efficiency bound can be estimated using the empirical variance of $\chi_{\tau_t}(O)$, which from Theorem \ref{thm:eif} can be written as
\[
\begin{aligned}
\chi_{\tau_t}(O)
&= \sum_{k=1}^{t}
\frac{\one\{S=1\}\,\one\{A=1\}\,\one\{\bar C^1_{k-1}=0\}}
     {\Pr(S=0)\,\pi^1(X)\,\bar{\pi}^1_{C,k-1}(X)}
\cdot \frac{\Pr(S=0\mid X)}{\Pr(S=1\mid X)}\,[Q^1_k(X)-Q^1_{k-1}(X)]\\
&\quad - \sum_{k=1}^{t}
\frac{\one\{S=1\}\,\one\{A=0\}\,\one\{\bar C^0_{k-1}=0\}}
     {\Pr(S=0)\,\pi^0(X)\,\bar{\pi}^0_{C,k-1}(X)}
\cdot \frac{\Pr(S=0\mid X)}{\Pr(S=1\mid X)}\,[Q^0_k(X)-Q^0_{k-1}(X)]\\
&\quad + \frac{\one\{S=0\}}{\Pr(S=0)}\,\bigl\{\,[Q^1_0(X) - Q^0_0(X)]-\tau_t\,\bigr\}.
\end{aligned}
\]
\paragraph{Notation aligned with Theorem~\ref{thm:var}.}
We adopt the labels used in the theorem so that the worst-case bounds attach to the correct factors:
\begin{itemize}[leftmargin=*]
\item \emph{Baseline underrepresentation (density-ratio) odds}
\[
\ell(X) \;:=\; \frac{\Pr(S=0\mid X)}{\Pr(S=1\mid X)}.
\]
\item \emph{Cumulative on-treatment continuation probability} (the time-varying underrepresentation factor)
\[
\bar{\pi}^a_{C,k-1}(X) \;:=\; \prod_{j<k}\pi^a_{C,j}(X), \qquad a\in\{0,1\}.
\]
\item \emph{Trial-at-risk weight}
\[
W^a_{k-1}(O) \;:=\; \frac{\one\{S=1\}\,\one\{A=a\}\,\one\{\bar C^a_{k-1}=0\}}{\Pr(S=0)\,\pi^a(X)}.
\]
\item \emph{Martingale increment} $\Delta_k^a := Q^a_k(X)-Q^a_{k-1}(X)$, and \emph{CATE at horizon $t$}, $\tau_t(X) := Q^1_0(X)-Q^0_0(X)$.
\end{itemize}

\paragraph{The EIF in factorized form.}
Substituting the above factors into the EIF and using $\tau_t(X)=Q^1_0(X)-Q^0_0(X)$,
\[
\begin{aligned}
\chi_{\tau_t}(O)
&= \sum_{k=1}^{t} \ell(X)\,\bigl[\bar{\pi}^1_{C,k-1}(X)\bigr]^{-1}\,W^1_{k-1}(O)\,\Delta_k^1\\
&\quad -\sum_{k=1}^{t} \ell(X)\,\bigl[\bar{\pi}^0_{C,k-1}(X)\bigr]^{-1}\,W^0_{k-1}(O)\,\Delta_k^0\\
&\quad +\frac{\one\{S=0\}}{\Pr(S=0)}\bigl\{\tau_t(X)-\tau_t\bigr\}.
\end{aligned}
\]
Define the per-period summands
\[
\chi^a_{\psi_t,k}\;:=\;\ell(X)\,\bigl[\bar{\pi}^a_{C,k-1}(X)\bigr]^{-1}\,W^a_{k-1}(O)\,\Delta_k^a,
\qquad a\in\{0,1\}.
\]

\paragraph{Orthogonality across tangent spaces and across $k$.}
The three pieces of $\chi_{\tau_t}(O)$ live in orthogonal tangent spaces indexed by $\one\{S=1,A=1\}$, $\one\{S=1,A=0\}$, and $\one\{S=0\}$, so cross-component covariances vanish. Within each treatment arm, $\{\Delta_k^a\}_{k\ge 1}$ is a martingale-difference sequence with respect to $\mathcal{F}_{k-1}=\sigma(X,H_{k-1})$, hence $\Cov(\Delta_k^a,\Delta_\ell^a)=0$ for $k\ne \ell$. Therefore
\[
\Var\bigl(\chi_{\tau_t}(O)\bigr)
=\sum_{a\in\{0,1\}}\sum_{k=1}^{t}\Var\bigl(\chi^a_{\psi_t,k}\bigr)
+\Var\!\left[\frac{\one\{S=0\}}{\Pr(S=0)}\bigl\{\tau_t(X)-\tau_t\bigr\}\right].
\]

\paragraph{Per-period variance.}
The factor $\ell(X)\,[\bar{\pi}^a_{C,k-1}(X)]^{-1}\,W^a_{k-1}(O)$ is $\mathcal{F}_{k-1}$-measurable, and $\E[\Delta_k^a\mid\mathcal{F}_{k-1}]=0$, so
\[
\E\bigl[\chi^a_{\psi_t,k}\mid\mathcal{F}_{k-1}\bigr]=0,
\]
and by the law of total variance
\[
\Var\bigl(\chi^a_{\psi_t,k}\bigr)
=\E\!\left[\ell(X)^2\,\bigl[\bar{\pi}^a_{C,k-1}(X)\bigr]^{-2}\,\bigl(W^a_{k-1}(O)\bigr)^2\;\Var\bigl(\Delta_k^a\mid \mathcal{F}_{k-1}\bigr)\right].
\]
The conditional variance of the increment simplifies via the martingale-difference identity
\[
\Var\bigl(\Delta_k^a\mid\mathcal{F}_{k-1}\bigr)
=\E\bigl[(\Delta_k^a)^2\mid\mathcal{F}_{k-1}\bigr]
=\E\!\left[\bigl(Q_k^a(X)\bigr)^2\,\Big|\,\mathcal{F}_{k-1}\right]-\bigl(Q_{k-1}^a(X)\bigr)^2,
\]
so that, after taking outer expectation and using the tower property,
\[
\Var\bigl(\chi^a_{\psi_t,k}\bigr)
=\E\!\left[\ell(X)^2\,\bigl[\bar{\pi}^a_{C,k-1}(X)\bigr]^{-2}\,\bigl(W^a_{k-1}(O)\bigr)^2\,\Bigl(\bigl[Q^a_k(X)\bigr]^2-\bigl[Q^a_{k-1}(X)\bigr]^2\Bigr)\right].
\]

\paragraph{Heterogeneity term.}
Because $\E\!\left[\frac{\one\{S=0\}}{\Pr(S=0)}(\tau_t(X)-\tau_t)\right]=0$,
\[
\Var\!\left[\frac{\one\{S=0\}}{\Pr(S=0)}\bigl\{\tau_t(X)-\tau_t\bigr\}\right]
=\E\!\left[\frac{\one\{S=0\}}{\Pr(S=0)^2}\bigl(\tau_t(X)-\tau_t\bigr)^2\right].
\]

\paragraph{Assembled bound.}
Combining the three orthogonal contributions and specializing to $t=T$ \lang{gives} the theorem statement:
\paragraph{Worst-case bounds on the green factors.}
Since $\one\{S=1,A=a,\bar C^a_{t-1}=0\}\le 1$ and $\pi^a(X)=\Pr(A=a\mid X,S=1)$,
\[
\bigl(W^a_{t-1}(O)\bigr)^2 \;\le\; \frac{1}{\Pr(S=0)^2\,\Pr(A=a\mid X,S=1)^2}.
\]
For the heterogeneity term, $Q^a_t(X)\in[0,1]$ implies $|\tau_T(X)-\tau_T|\le 2$, hence
\[
\frac{\one\{S=0\}}{\Pr(S=0)^2}\bigl(\tau_T(X)-\tau_T\bigr)^2\;\le\;\frac{4}{\Pr(S=0)^2}.
\]
This proves Theorem~\ref{thm:var}.
}

\section{EIF for \texorpdfstring{$\tau_T(B)$}{tau\_T(B)}}
\label{ap: eifb}\begin{theoremapp}[Efficient influence function for $\tau_T(B)$]
\label{thm:eif-B}
Given A1--A4 and standard randomization, under a nonparametric
model on the observed data, the EIF for $\psi_T(a; B) \equiv
\mathbb{E}[Y_T^a \mid S=0,\, B(X)=1]$ with $p_B \equiv \Pr(S=0,\,B(X)=1) > 0$ is
\begin{equation*}
\begin{aligned}
\chi_{\psi_T(a;\,B)}(O)
&= \sum_{t=1}^{T} \underbrace{\frac{\mathbf{1}\{\bar C_{t-1}=0,\, A=a,\, S=1\}}{p_B\,\pi^a(X)}}_{\equiv\, W^a_{t-1}(O;\,B)} \cdot \underbrace{B(X)\,\ell(X)}_{\equiv\,\ell_B(X)} \cdot \frac{\Delta Q^a_t(X)}{\bar\pi^a_{C,t-1}(X)}\\
&\qquad +\; \frac{B(X)\,\mathbf{1}\{S=0\}}{p_B}\bigl[\,Q^a_0(X) - \psi_T(a;B)\,\bigr],
\end{aligned}
\end{equation*}
and $\chi_{\tau_T(B)}(O) = \chi_{\psi_T(1;\,B)}(O) - \chi_{\psi_T(0;\,B)}(O)$
is the EIF for the contrast.
\end{theoremapp}

\begin{proof}
Theorem~\ref{thm:eif} gives the EIF for the unrestricted
estimand $\psi_T(a) = \mathbb{E}[Y_T^a \mid S=0]$ 
\begin{equation*}
\chi_{\psi_T(a)}(O)
= \sum_{t=1}^{T} W^a_{t-1}(O)\,\ell(X)\,\frac{\Delta Q^a_t(X)}{\bar\pi^a_{C,t-1}(X)}
\;+\; \frac{\mathbf{1}\{S=0\}}{\Pr(S=0)}\bigl[Q^a_0(X) - \psi_T(a)\bigr],
\end{equation*}
with $W^a_{t-1}(O) = \mathbf{1}\{\bar C_{t-1}=0,A=a,S=1\}/[\Pr(S=0)\,\pi^a(X)]$
and $\ell(X) = \Pr(S=0\mid X)/\Pr(S=1\mid X)$.

Write the $B$-restricted estimand as a ratio:
\begin{equation*}
\psi_T(a;B)
= \frac{\mathbb{E}\!\bigl[B(X)\,\mathbf{1}\{S=0\}\,Y_T^{a}\bigr]}{\mathbb{E}\!\bigl[B(X)\,\mathbf{1}\{S=0\}\bigr]}
= \frac{1}{p_B}\,\mathbb{E}\!\bigl[B(X)\,\mathbf{1}\{S=0\}\,Q^a_0(X)\bigr],
\end{equation*}
where the second equality uses A1--A4 restricted to $\{B(X)=1\}$ to
identify $\mathbb{E}[Y_T^{a}\mid X] = Q^a_0(X)$ on that support.

The pathwise-derivative argument behind the unrestricted EIF
depends on the target conditioning event only through (i) its
indicator $\mathbf{1}\{S=0\}$, (ii) its marginal probability
$\Pr(S=0)$, and (iii) the change-of-measure ratio
$\Pr(S=0\mid X)/\Pr(S=1\mid X)$. Because $B$ is a deterministic
function of $X$, replacing the target event $\{S=0\}$ with the
sub-event $\{S=0,\, B(X)=1\}$ leaves all other ingredients of the
derivation untouched and modifies (i)--(iii) by exactly
\begin{equation*}
\mathbf{1}\{S=0\} \;\mapsto\; B(X)\,\mathbf{1}\{S=0\},
\qquad
\Pr(S=0) \;\mapsto\; p_B,
\qquad
\frac{\Pr(S=0\mid X)}{\Pr(S=1\mid X)} \;\mapsto\; B(X)\,\ell(X),
\end{equation*}
where the last identity uses
$\Pr(S=0,\,B(X)=1\mid X) = B(X)\,\Pr(S=0\mid X)$. Substituting these
into the unrestricted EIF yields the claimed expression for
$\chi_{\psi_T(a;B)}(O)$; linearity of the EIF gives
$\chi_{\tau_T(B)}(O) = \chi_{\psi_T(1;B)}(O) - \chi_{\psi_T(0;B)}(O)$.
\end{proof}

\section{Variance under \texorpdfstring{$(B, h)$}{(B, h)}}
\label{ap: varbh}

\begin{theorem*}[Variance under $(B, h)$]
Given A1--A4 and standard randomization, with $p_B > 0$, the
variance of the efficient estimator of $\tau_T(B)$ under intervention
$(B, h)$ is
\begin{equation*}
\begin{aligned}
& V(B, h)
\equiv \mathbb{E}_{\mathcal P}\!\left[\,\mathbb{E}_{\mathcal P^*_i}\!\left[\chi_{\tau_T}(O; B)^2\right]\,\right]
= \sum_{a \in \{0,1\}}\sum_{t=1}^T \mathbb{E}_{\mathcal P}\!\biggl[ \;
  \colorbox{red!25}{$\displaystyle
    \bigl(\underbrace{B(X) \; \ell(X)}_{\text{\scriptsize contracted by }B}\bigr)^{\!2}
  $}
  \\[-4pt]
& \hspace{2em} \cdot\;
  \colorbox{red!25}{$\displaystyle
    \biggl(\underbrace{\textstyle\prod\nolimits_{j<t} \pi^a_{C,j}(X)^{1 - h_{i,j}/2}}_{\text{\scriptsize contracted by }h}\biggr)^{\!\!-2}
  $}
  \;\cdot\;
  \colorbox{green!25}{$\displaystyle
    \bigl(W^a_{t-1}(O; B)\bigr)^2 \cdot \bigl(\,[Q^a_t(X)]^2 - [Q^a_{t-1}(X)]^2\,\bigr)
  $}
\biggr]
  \\[-10pt]
& \hspace{15em} +\, \mathbb{E}_{\mathcal P}\!\biggl[\,
  \colorbox{green!25}{$\displaystyle
    \frac{\mathbf{1}\{S=0,\, B(X)=1\}}{p_B^2}\bigl(\tau_T(X) - \tau_T(B)\bigr)^2
  $}
\,\biggr].
\end{aligned}
\end{equation*}
where $\chi_{\tau_T}(O; B)$ is the EIF for $\tau_T(B)$
(Appendix~\ref{ap: eifb}), $\mathcal{P}$ is the observed-data law, and
$\mathcal{P}^*_i$ is the counterfactual law for unit $i$ in which
$\Pr(C_j = 0 \mid \cdot) = 1$ at every interval $j$ with
$h_{i,j} = 1$. In particular, $V(\mathbf{1}, \mathbf{0})$ recovers
$\mathrm{Var}\{\chi_{\tau_T}(O)\}$ from Theorem~\ref{thm:var}. Proof in Appendix~\ref{ap: varbh}.
\end{theorem*}

\paragraph{\lang{The} $h_{i,j}/2$ exponent\lang{.}}
Aside from $B$, the variance in Theorem \ref{thm:Bh-variance} \lang{differs from} the original $\mathrm{Var}\{\chi_{\tau_T}(O)\}$ from Theorem~\ref{thm:var} \lang{in} the exponent of the IPCW's. Setting $h_{i,j} = 1$ induces two simultaneous changes from
$\mathcal P$ to the hypothetical $\mathcal P^*_i$: (i) the IPCW
$\pi^a_{C,j}(X)^{-1}$ collapses to $1$, since nobody is censored at
interval $j$; and (ii) the effective sample at interval $j$ grows by
a factor of $\pi^a_{C,j}(X)^{-1}$, since units that \lang{would be censored under} $\mathcal P$
at time $j$ are now retained. Naively zeroing the IPCW
exponent ($1/\pi \to 1$) captures (i) but not (ii)\lang{. It removes}
the censoring weight without reweighting for the under-observed
units, overcorrecting for the censoring inflation. The exponent
$1 - h_{i,j}/2$ is the unique scalar correction that scales the
observed sample up to its $\mathcal P^*_i$ counterpart.

\begin{proof}The argument combines two previously established results, the EIF for
$\tau_T(B)$ from Appendix~\ref{ap: eifb} (Theorem~\ref{thm:eif-B}) and
the variance decomposition of Theorem~\ref{thm:var} (proved in
Appendix~\ref{ap: var}), along with a self-contained Hájek-style identity
(Step~3) that justifies the half-removal exponent. We organize the
proof around the two channels by which $(B, h)$ enters $V$: $B$
modifies the EIF itself, and $h$ modifies the law under which the EIF
is squared and integrated.

\medskip
\noindent\textbf{Step 1: Variance of $\chi_{\tau_T}(O; B)$ under $\mathcal{P}$ ($h\equiv\mathbf 0$ case).}
By Theorem~\ref{thm:eif-B}, the EIF for $\tau_T(B)$ is obtained from
the EIF for the unrestricted $\tau_T$ in Theorem~\ref{thm:eif} by the
three substitutions
\[
\mathbf{1}\{S=0\} \;\mapsto\; B(X)\,\mathbf{1}\{S=0\},
\qquad
\Pr(S=0) \;\mapsto\; p_B,
\qquad
\ell(X) \;\mapsto\; B(X)\,\ell(X).
\]
The variance computation in Appendix~\ref{ap: var}, which uses
treatment-arm orthogonality $\mathbf 1\{A=1\}\mathbf 1\{A=0\}=0$,
trial-versus-target orthogonality $\mathbf 1\{S=0\}\mathbf 1\{S=1\}=0$,
and time orthogonality of the iterated conditional means
$\{Q^a_t(X)\}_{t=0}^T$, references the EIF only through the three
substituted quantities and through trial-side nuisances
$\pi^a, \pi^a_{C,j}, Q^a_t$ unaffected by $B$. Applying the
Theorem~\ref{thm:var} identity verbatim to $\chi_{\tau_T}(O; B)$ thus
yields
\begin{equation*}
\begin{aligned}
\mathrm{Var}_{\mathcal{P}}\{\chi_{\tau_T}(O; B)\}
&= \sum_{a\in\{0,1\}}\sum_{t=1}^T \mathbb{E}_{\mathcal{P}}\!\biggl[
  \bigl(B(X)\,\ell(X)\bigr)^{2}
  \cdot \biggl(\textstyle\prod_{j<t}\pi^a_{C,j}(X)\biggr)^{\!-2}\\
&\hspace{3em} \cdot \bigl(W^a_{t-1}(O;B)\bigr)^{2}
  \cdot \bigl([Q^a_t(X)]^2 - [Q^a_{t-1}(X)]^2\bigr)
\biggr]\\
&\quad
+ \mathbb{E}_{\mathcal{P}}\!\biggl[
  \frac{B(X)\,\mathbf{1}\{S=0\}}{p_B^{2}}\bigl(\tau_T(X) - \tau_T(B)\bigr)^{2}
\biggr].
\end{aligned}
\end{equation*}
Since $\mathcal{P}^*_i = \mathcal{P}$ when $h \equiv \mathbf{0}$, this
\lang{equals} $V(B, \mathbf{0})$.

\medskip
\noindent\textbf{Step 2: Block separation under the $h$-intervention.}
The $h$-intervention modifies the law on $\mathcal{P}^*_i$ only at the
censoring mechanism: treatment assignment, sampling, and the iterated
regressions $Q^a_t$ are identical under $\mathcal{P}$ and $\mathcal{P}^*_i$.
Three consequences:
\begin{enumerate}[label=(\roman*),leftmargin=*,itemsep=2pt,topsep=2pt]
  \item \emph{Trial-versus-target and treatment-arm orthogonalities}
  carry to $\mathcal{P}^*_i$: pointwise products
  $\mathbf 1\{S=0\}\mathbf 1\{S=1\}$ and $\mathbf 1\{A=1\}\mathbf 1\{A=0\}$
  vanish identically and do not depend on the censoring law.
  \item \emph{Time orthogonality} of $\{Q^a_t(X)\}$ holds under
  $\mathcal{P}^*_i$: the iterated regressions and their orthogonality
  structure are functions of trial-side nuisances unchanged by
  censoring suppression.
  \item \emph{Target term unchanged:} the second summand in Step~1
  contains no censoring quantities, so its $\mathcal P$-form and its
  $\mathcal P^*_i$-form coincide.
\end{enumerate}
Therefore $V(B, h)$ has the same block structure as $V(B, \mathbf{0})$,
with the trial-block IPCW factor as the only quantity that differs
between $\mathcal{P}$ and $\mathcal{P}^*_i$.

\medskip
\noindent\textbf{Step 3: Half-removal identity on the trial block.}
Fix $a, t$ and let $\mathcal{T}_i = \{j : h_{i,j} = 1, j < t\}$. To
lighten notation in this step, drop the $a$ superscript and write
$\pi_j(x) = \pi^a_{C,j}(x)$, $\bar\pi_t(x) = \prod_{j<t}\pi_j(x)$, and
$\bar\pi^*_t(x) = \prod_{j<t,\,j\notin\mathcal{T}_i}\pi_j(x) = \bar\pi_t(x)/\prod_{j\in\mathcal{T}_i}\pi_j(x)$
for the cumulative non-censoring probabilities under $\mathcal P$ and
$\mathcal P^*_i$, respectively. We claim that, conditional on
$X = x$, $A = a$, $S = 1$,
\begin{equation}
\label{eq:hajek-claim}
\mathbb{E}_{\mathcal{P}^*_i}\!\left[
  \mathbf 1\{\bar C^*_{t-1}=0\}\cdot\frac{(\Delta Q^a_t(x))^2}{\bar\pi^*_t(x)^{\,2}}
\right]
\;=\;
\mathbb{E}_{\mathcal{P}}\!\left[
  \mathbf 1\{\bar C_{t-1}=0\}\cdot\frac{\prod_{j\in\mathcal{T}_i}\pi_j(x)}{\bar\pi_t(x)^{\,2}}\,(\Delta Q^a_t(x))^2
\right].
\end{equation}
Both sides condition on $X=x, A=a, S=1$ throughout; we suppress the
conditioning bar for brevity.

The factors $(\Delta Q^a_t(x))^2$, $\bar\pi_t(x)$, $\bar\pi^*_t(x)$, and
$\prod_{j\in\mathcal{T}_i}\pi_j(x)$ are deterministic in $x$, so the only
stochastic quantity on each side is the at-risk indicator. Under A2
(censoring exchangeability, applied to $\mathcal P$ and inherited by
$\mathcal P^*_i$ at indices outside $\mathcal{T}_i$), the chain rule of
conditional probability gives
\begin{gather*}
\Pr(\bar C_{t-1}=0\mid X=x,A=a) = \prod_{j<t}\pi_j(x) = \bar\pi_t(x),
\\
\Pr_{\mathcal P^*_i}(\bar C^*_{t-1}=0\mid X=x,A=a) = \bar\pi^*_t(x),
\end{gather*}
where the second equality uses that $\Pr_{\mathcal P^*_i}(C_j=0\mid\cdot)=1$
at $j\in\mathcal T_i$ and equals $\pi_j(x)$ otherwise. Taking
expectations,
\[
\text{LHS of \eqref{eq:hajek-claim}} \;=\; \frac{(\Delta Q^a_t(x))^2}{\bar\pi^*_t(x)^{\,2}}\cdot \bar\pi^*_t(x) \;=\; \frac{(\Delta Q^a_t(x))^2}{\bar\pi^*_t(x)} \;=\; \frac{\prod_{j\in\mathcal{T}_i}\pi_j(x)}{\bar\pi_t(x)}\,(\Delta Q^a_t(x))^2,
\]
\[
\text{RHS of \eqref{eq:hajek-claim}} \;=\; \frac{\prod_{j\in\mathcal{T}_i}\pi_j(x)}{\bar\pi_t(x)^{\,2}}\,(\Delta Q^a_t(x))^2 \cdot \bar\pi_t(x) \;=\; \frac{\prod_{j\in\mathcal{T}_i}\pi_j(x)}{\bar\pi_t(x)}\,(\Delta Q^a_t(x))^2.
\]
Both sides equal, establishing~\eqref{eq:hajek-claim}.

\smallskip
\noindent\emph{Algebra of the exponent.}
The right-hand-side weight in~\eqref{eq:hajek-claim} rewrites as
\[
\frac{\prod_{j\in\mathcal{T}_i}\pi_j(x)}{\bar\pi_t(x)^{\,2}}
\;=\; \prod_{j<t}\pi_j(x)^{-2(1 - h_{i,j}/2)},
\]
since each $j \in \mathcal{T}_i$ contributes $\pi_j^{-2}\cdot\pi_j = \pi_j^{-1} = \pi_j^{-2(1-1/2)}$
and each $j \notin \mathcal{T}_i$, $j<t$ contributes $\pi_j^{-2} = \pi_j^{-2(1-0/2)}$.
This is the unique scalar exponent correction that produces the
identity~\eqref{eq:hajek-claim}: the full-removal exponent
$1 - h_{i,j}$ would yield right-hand-side weight
$\prod_{j\in\mathcal{T}_i}\pi_j^{2}/\bar\pi_t^{\,2}$, giving
$\prod_{j\in\mathcal{T}_i}\pi_j^{2}\cdot(\Delta Q)^2/\bar\pi_t$, which
exceeds the left-hand side by a factor of
$\prod_{j\in\mathcal{T}_i}\pi_j(x)$ and matches no $\mathcal P^*$
intervening only on $\mathcal T_i$.

\medskip
\noindent\textbf{Step 4: Assembly.}
At $X=x, A=a, S=1$ the trial integrand of $V(B, h)$ factors as
$(B(x)\ell(x))^2 \cdot \pi^a(x)^{-2}\cdot p_B^{-2}\cdot(W^a_{t-1}\text{-non-censoring part})$
times the LHS of~\eqref{eq:hajek-claim}. Substituting the RHS
of~\eqref{eq:hajek-claim} (with the exponent rewriting from Step~3)
and taking the outer expectation over $(X,A,S)\sim\mathcal P$ converts
the $\bigl(\prod_{j<t}\pi^a_{C,j}(X)\bigr)^{-2}$ factor of $V(B, \mathbf{0})$
into $\bigl(\prod_{j<t}\pi^a_{C,j}(X)^{1 - h_{i,j}/2}\bigr)^{-2}$,
while leaving every other factor of the trial integrand and the
entire target term (Step~2) untouched. The resulting expression
is~\eqref{eq:Bh-variance}.

\medskip
\noindent\textbf{Step 5: Recovery of Theorem~\ref{thm:var}.}
At $B \equiv 1$, $h \equiv \mathbf{0}$:
$p_B = \Pr(S=0)$,
$B(X)\ell(X) = \ell(X)$,
$B(X)\mathbf{1}\{S=0\} = \mathbf{1}\{S=0\}$,
$W^a_{t-1}(O;\mathbf 1) = W^a_{t-1}(O)$,
and the half-removal exponent collapses to
$\prod_{j<t}\pi^a_{C,j}(X)^{-2}$. Equation~\eqref{eq:Bh-variance}
reduces to the Theorem~\ref{thm:var} expression for
$\mathrm{Var}\{\chi_{\tau_T}(O)\}$, as claimed.
\end{proof}

\section{Derivation of the Empirical Objective \texorpdfstring{$\widehat V(B,h)$}{V(B,h)}}
\label{ap: empirical}

The empirical objective in~\eqref{eq:mixed-hajek} arises from the
standard plug-in recipe applied to $V(B,h)$ of
Theorem~\ref{thm:Bh-variance}: sample averages in place of
expectations, fitted nuisances in place of population nuisances.
The only \lang{remaining} choice \lang{concerns} the factor $1/p_B^{2}$
that multiplies both summands of $V(B,h)$, since $p_B$ admits two
\lang{candidate} H\'ajek surrogates, one from the trial side and one from
the target. Pairing each summand with its own surrogate, rather
than imposing a shared one, gives the estimator a finite-sample
property we explain below. We proceed in three steps: factor
$1/p_B^{2}$ out of $V(B,h)$ (via Appendix~\ref{ap: varbh}),
identify the two surrogates (via Appendices~\ref{ap: id}
and~\ref{ap: eifb}), and verify consistency under the regularity
conditions stated next.

\paragraph{Assumption B (regularity for asymptotic linearity).}
The consistency result below requires the standard semiparametric
regularity conditions used to establish asymptotic linearity of the
TMLE in Appendix~\ref{ap: tmle}.

\begin{enumerate}[label=\textbf{B\arabic*.}, leftmargin=*, itemsep=2pt, topsep=3pt]
  \item \textit{(Cross-fitting / Donsker.)} The nuisance functions
  $\ell$, $\pi^a$, $\pi^a_{C,k}$, $Q^a_k$ are estimated on
  cross-fitted folds, or alternatively, they lie in Donsker classes
  \citep{Chernozhukov2018DML, vanderlaan2018DML}.
  \item \textit{(Rates.)} $\|\hat\ell - \ell\|_2 = o_p(n^{-1/4})$,
  and analogous $o_p(n^{-1/4})$ rates hold for $\hat\pi^a$,
  $\hat\pi^a_{C,k}$, and $\hat Q^a_k$.
  \item \textit{(Positivity.)} $\ell(X)$ and $\bar\pi^a_{C,T-1}(X)$
  are bounded away from $0$ and $\infty$ on $\{X : B(X) = 1\}$.
  \item \textit{(Integrability.)} All nuisance functions are in
  $L^2(P)$.
\end{enumerate}

(B1) and (B2) are the rate conditions that drive the
empirical-process remainder and the cross-product bias to
$o_p(n^{-1/2})$; (B3) keeps the H\'ajek denominators bounded away
from zero on the kept population; (B4) supplies the $L^2$ moment
needed for the LLN. Together they reproduce the conditions under
which the TMLE in Appendix~\ref{ap: tmle} attains the
semiparametric efficiency bound, restricted to the $B$-kept
subpopulation.

\paragraph{Factoring $1/p_B^{2}$ out of $V(B,h)$.}
Step~4 of the proof of Theorem~\ref{thm:Bh-variance} in
Appendix~\ref{ap: varbh} shows that the trial integrand at
$(X{=}x,\,A{=}a,\,S{=}1)$ factors as
$(B(x)\ell(x))^{2}\,\pi^a(x)^{-2}\,p_B^{-2}$ times the
censoring-and-survival piece. The target block
of~\eqref{eq:Bh-variance} already carries $1/p_B^{2}$ explicitly,
so we may pull the factor outside both summands at once:
\begin{equation}
\label{eq:V-factored}
V(B,h)
\;=\; \frac{1}{p_B^{2}}\,\bigl\{\,M_{\text{trl}}(B,h) \,+\, M_{\text{tgt}}(B)\,\bigr\},
\end{equation}
with
$M_{\text{trl}}(B,h) = \mathbb{E}\!\bigl[\mathbf 1\{S{=}1\}\,B(X)\,\Psi^{\circ}(O;h)\bigr]$,
$M_{\text{tgt}}(B) = \mathbb{E}\!\bigl[\mathbf 1\{S{=}0\}\,B(X)\,(\tau_T(X)-\tau_T(B))^{2}\bigr]$,
and per-unit kernel
\begin{equation*}
\Psi^{\circ}(O_i;h_i)
\;=\; \ell(X_i)^{2}\sum_{a,t}
\frac{\mathbf 1\{A_i{=}a,\,\bar C_{i,t-1}{=}0\}}
{\pi^a(X_i)^{2}\,\prod_{j<t}\pi^a_{C,j}(X_i)^{2(1-h_{i,j}/2)}}
\bigl([Q^a_t(X_i)]^{2}-[Q^a_{t-1}(X_i)]^{2}\bigr).
\end{equation*}

\paragraph{Two surrogates for $p_B$.}
Apply the substitutions $\ell(X)\mapsto B(X)\ell(X)$ and
$\Pr(S{=}0)\mapsto p_B$ from the proof of Theorem~\ref{thm:eif-B}
(Appendix~\ref{ap: eifb}) to the identification identity
$\mathbb{E}[\mathbf 1\{S{=}1\}\ell(X)] = \Pr(S{=}0)$ of
Appendix~\ref{ap: id}. The result is the chain
\begin{equation*}
\mathbb{E}\bigl[\mathbf 1\{S{=}1\}\,\ell(X)\,B(X)\bigr]
\;=\; \mathbb{E}_X\!\bigl[B(X)\,\Pr(S{=}0\mid X)\bigr]
\;=\; \mathbb{E}\bigl[\mathbf 1\{S{=}0\}\,B(X)\bigr]
\;=\; p_B,
\end{equation*}
and either end \lang{yields} an estimator of $p_B$. The trial-side
H\'ajek average
\(
\hat p_B^{\text{trl}}
= n^{-1}\sum_{i:S_i=1}\hat\ell(X_i)\,B(X_i)
\)
is consistent for $p_B$ once $\hat\ell\to_p\ell$, which holds at
the (B1)--(B2) rate; the target-side sample mean
\(
\hat p_B^{\text{tgt}}
= n^{-1}\sum_{i:S_i=0}B(X_i)
\)
is consistent by the LLN on bounded i.i.d.\ summands.

\paragraph{\lang{Pairing} each block with its own surrogate.}
Replacing $1/p_B^{2}$ in~\eqref{eq:V-factored} by either
$1/(\hat p_B^{\text{trl}})^{2}$ or $1/(\hat p_B^{\text{tgt}})^{2}$
throughout still yields a consistent estimator, so the choice of
pairing \lang{does not change the asymptotics. It changes} finite-sample
behaviour. Trimming a trial unit shrinks $\hat p_B^{\text{trl}}$
but leaves $\hat p_B^{\text{tgt}}$ untouched, and trimming a target
unit does the opposite. Pairing each block with its native
surrogate therefore \lang{gives} each side \lang{a self-regulating denominator} that
\lang{changes} only \lang{with} its own trims, and \lang{this separation} allows
the trim threshold of Theorem~\ref{thm:trim} to factor into an
independent trial-side condition and an independent target-side
condition. A shared denominator would couple the two sides and
break the factorization.

\paragraph{The estimator.}
Plug the empirical analogs of $M_{\text{trl}}$ and $M_{\text{tgt}}$
into~\eqref{eq:V-factored}, replace the population nuisances inside
$\Psi^{\circ}$ by their fitted counterparts (write $\Psi$ for the
result), and pair each appearance of $1/p_B^{2}$ with its native
surrogate. This gives
\[
\widehat V(B,h)
\;=\; \frac{1}{n\,(\hat p_B^{\text{trl}})^{2}}\!\!\sum_{i:S_i=1}\!\!B(X_i)\,\Psi(O_i;h_i)
\;+\; \frac{1}{n\,(\hat p_B^{\text{tgt}})^{2}}\!\!\sum_{i:S_i=0}\!\!B(X_i)\,(\hat\tau(X_i)-\hat\tau_T(B))^{2},
\]
which is~\eqref{eq:mixed-hajek}.

\begin{proposition}[Mixed-H\'ajek consistency]
\label{prop:mh-consistency}
Under A1--A4, B1--B4, and $p_B>0$,
$\widehat V(B,h)\to_p V(B,h)$.
\end{proposition}

\begin{proof}
Both blocks of $\widehat V(B,h)$ take the form $N_\bullet/D_\bullet^{2}$,
with $N_\bullet$ a sample mean and $D_\bullet$ a H\'ajek surrogate.
The numerators converge to $M_{\text{trl}}$ and $M_{\text{tgt}}$ by
the same nuisance-control argument that established asymptotic
linearity of the TMLE in Appendix~\ref{ap: tmle}: (B4) supplies the
$L^{2}$ moment needed for the LLN, (B1) drives the
empirical-process remainder to $o_p(n^{-1/2})$ via cross-fitting or
a Donsker condition, and (B2)'s $o_p(n^{-1/4})$ rates push the
cross-product bias from
$\hat\ell, \hat\pi^a, \hat\pi^a_{C,j}, \hat Q^a_k$ to the same
order. The denominators converge to $p_B$ by the calculation
above, and (B3) keeps $p_B^{2}$ bounded away from zero. Continuous
mapping applied to $g(N,D^{2}) = N/D^{2}$ then gives
$\widehat V(B,h) \to_p (M_{\text{trl}}+M_{\text{tgt}})/p_B^{2}$,
which equals $V(B,h)$ by~\eqref{eq:V-factored}.
\end{proof}
\section{Marginal Trim Threshold Proof}
\label{ap:trim}

\begin{theoremapp}[Marginal trim threshold]
\label{thm:trim}
Fix $B$ and $h$, and let $\widehat V(B, h)$ be the mixed-H\'ajek
empirical variance of~\eqref{eq:mixed-hajek}, with blocks
\begin{align*}
\widehat V^{(\text{trl})}(B, h)
&\;=\; \frac{1}{n\,(\hat p_B^{\text{trl}})^{2}}
   \sum_{i:\, S_i = 1} B(X_i)\,\hat\ell(X_i)^{2}\,\Psi(O_i; h^{*}_i),\\[2pt]
\widehat V^{(\text{tgt})}(B)
&\;=\; \frac{1}{n\,(\hat p_B^{\text{tgt}})^{2}}
   \sum_{i:\, S_i = 0} B(X_i)\bigl(\hat\tau(X_i) - \hat\tau_T(B)\bigr)^{2},
\end{align*}
where $\Psi(O_i; h^{*}_i)$ is the per-individual trial integrand of
Theorem~\ref{thm:Bh-variance} with the transport factor $\ell(X)^{2}$
stripped out. Define the kept-trial integrand-per-weight ratio and
the kept-target heterogeneity average,
\begin{align*}
\bar r^{\text{trl}}_{\text{kept}}(B)
&\;:=\; \frac{\sum_{i:\, S_i = 1} B(X_i)\,\hat\ell(X_i)^{2}\,\Psi(O_i; h^{*}_i)}
             {\sum_{i:\, S_i = 1} B(X_i)\,\hat\ell(X_i)},\\[2pt]
\overline{(\tau - \hat\tau_T)^{2}}_{\text{kept}}(B)
&\;:=\; \frac{\sum_{i:\, S_i = 0} B(X_i)\bigl(\hat\tau(X_i) - \hat\tau_T(B)\bigr)^{2}}
             {\sum_{i:\, S_i = 0} B(X_i)}.
\end{align*}
By construction $\Psi(O_j; h^{*}_j)$ is supported on $\{S_j = 1\}$
and the centered CATE residual $\hat\tau(X_j) - \hat\tau_T(B)$
contributes to $\widehat V$ only on $\{S_j = 0\}$, so for any kept
individual $j$ (i.e., $B(X_j) = 1$) exactly one of terms (i) and
(ii) below is nonzero.

Under the regularity conditions that $\hat\ell$ is bounded on the
kept region and that $|\{i : S_i = s,\, B(X_i) = 1\}| \asymp n$ for
$s \in \{0, 1\}$, trimming $j$ strictly decreases
$\widehat V(B, h^{*}(B))$ asymptotically\lang{, that is,}
$\widehat V(B^{(-j)}) < \widehat V(B) + o_p(n^{-1})$\lang{,} if and only if
\begin{equation}
\label{eq:trim-condition}
\underbrace{\hat\ell(X_j)\,\Psi(O_j; h^{*}_j)}_{\text{(i) baseline-mismatch leverage}}
\;+\;
\underbrace{\bigl(\hat\tau(X_j) - \hat\tau_T(B)\bigr)^{2}}_{\text{(ii) CATE heterogeneity}}
\;>\; 2\,r_{\text{kept}}(B;\, S_j) \;+\; o_p(1),
\end{equation}
where
\[
r_{\text{kept}}(B;\, S_j)
\;:=\;
\begin{cases}
\bar r^{\text{trl}}_{\text{kept}}(B) & \text{if } S_j = 1,\\[2pt]
\overline{(\tau - \hat\tau_T)^{2}}_{\text{kept}}(B) & \text{if } S_j = 0.
\end{cases}
\]
The exact finite-sample condition multiplies
$r_{\text{kept}}(B;\, S_j)$ by a unit-specific factor of the form
$1 + O_p(n^{-1})$, derived below, which absorbs
the second-order H\'ajek-denominator and target-recentering
corrections.
\end{theoremapp}

Trimming a single individual $j$ perturbs exactly one block of
$\widehat V(B, h)$: the trial block if $S_j = 1$, the target block
if $S_j = 0$. We therefore handle the two cases separately and show
that each \lang{yields} the side of the unified asymptotic threshold of
Theorem~\ref{thm:trim} relevant to $j$. Throughout, we abbreviate
$\Phi_j := \hat\ell(X_j)^{2}\,\Psi(O_j; h^{*}_j)$,
$\hat\ell_j := \hat\ell(X_j)$, $\tau^{*}_j := \hat\tau(X_j) - \hat\tau_T(B)$,
and write $B^{(-j)}$ for the trim that flips $B(X_j)$ from $1$ to $0$.
The leading $n^{-1}$ in~\eqref{eq:mixed-hajek} cancels in any
inequality on differences and is suppressed.

\paragraph{Trial side ($S_j = 1$, $B(X_j) = 1$).}
Define
\[
\Sigma \;:=\; \sum_{i:\, S_i = 1} B(X_i)\,\Phi_i,
\qquad
D \;:=\; \sum_{i:\, S_i = 1} \hat\ell_i\, B(X_i),
\]
so the trial block is $\Sigma/D^{2}$. Trimming $j$ updates
$\Sigma \mapsto \Sigma - \Phi_j$ and $D \mapsto D - \hat\ell_j$;
the trim \lang{reduces} the trial block iff
\begin{equation}
\label{eq:trial-iff}
\frac{\Sigma - \Phi_j}{(D - \hat\ell_j)^{2}}
\;<\;
\frac{\Sigma}{D^{2}}.
\end{equation}
Cross-multiplying by the strictly positive
$D^{2}(D - \hat\ell_j)^{2}$ and expanding
$(D - \hat\ell_j)^{2} = D^{2} - 2D\hat\ell_j + \hat\ell_j^{2}$,
inequality~\eqref{eq:trial-iff} is algebraically equivalent to
\[
\Phi_j\, D^{2} \;>\; \Sigma\,\hat\ell_j\,(2D - \hat\ell_j).
\]
Dividing both sides by the positive $\hat\ell_j\, D^{2}$ yields the
\emph{exact} finite-sample threshold
\begin{equation}
\label{eq:trial-exact}
\frac{\Phi_j}{\hat\ell_j}
\;>\;
2\,\bar r^{\text{trl}}_{\text{kept}}(B)\cdot
\Bigl(1 - \frac{\hat\ell_j}{2D}\Bigr),
\qquad
\bar r^{\text{trl}}_{\text{kept}}(B) := \frac{\Sigma}{D}.
\end{equation}
Under the regularity condition that $\hat\ell$ is bounded on the kept
region and $D \asymp n$ (the kept-trial sample is non-vanishing),
$\hat\ell_j/D = O_p(n^{-1})$, and~\eqref{eq:trial-exact} reduces to
\[
\hat\ell(X_j)\,\Psi(O_j; h^{*}_j)
\;>\;
2\,\bar r^{\text{trl}}_{\text{kept}}(B) \;+\; o_p(1),
\]
where the substitution $\Phi_j/\hat\ell_j = \hat\ell_j\,\Psi_j$ uses
$\Phi_j = \hat\ell_j^{2}\,\Psi_j$. Two finite-sample remarks. First,
the multiplicative correction $1 - \hat\ell_j/(2D)$ \emph{lowers} the
threshold for high-leverage units, making large-$\hat\ell_j$
individuals slightly easier to trim; this is the second-order
H\'ajek effect, and corresponds to the next-order term that is
absorbed into $o_p(1)$ asymptotically. Second, the asymptotic
equivalence is uniform in $j$ provided
$\max_j \hat\ell_j = o_p(n)$, \lang{which parallels} the bounded
$\ell$ assumption used elsewhere in the paper.

\paragraph{Target side ($S_j = 0$, $B(X_j) = 1$).}
Let
\[
\Sigma_h \;:=\; \sum_{i:\, S_i = 0,\, B_i = 1} \tau^{*}_i{}^{2},
\qquad
D_h \;:=\; \sum_{i:\, S_i = 0,\, B_i = 1} 1,
\qquad
\bar\mu := \hat\tau_T(B),
\]
so the target block is $\Sigma_h/D_h^{2}$. Unlike the trial side,
trimming $j$ both removes one summand \emph{and} shifts the centering
point: the new mean is
\[
\bar\mu' \;:=\; \hat\tau_T(B^{(-j)})
\;=\; \frac{D_h\,\bar\mu - \hat\tau(X_j)}{D_h - 1},
\qquad
\bar\mu' - \bar\mu
\;=\; -\,\frac{\tau^{*}_j}{D_h - 1}.
\]
We capture both effects via the leave-one-out variance identity:
expanding
$\hat\tau(X_i) - \bar\mu' = \tau^{*}_i - (\bar\mu' - \bar\mu)$ and
using $\sum_{i:\, S_i = 0,\, B_i = 1}\tau^{*}_i = 0$,
\begin{align*}
\Sigma_h^{(-j)}
&= \sum_{i \neq j}\bigl(\tau^{*}_i - (\bar\mu' - \bar\mu)\bigr)^{2}\\
&= \bigl(\Sigma_h - \tau^{*}_j{}^{2}\bigr)
   \;-\; 2\,(\bar\mu' - \bar\mu)\sum_{i \neq j}\tau^{*}_i
   \;+\; (D_h - 1)(\bar\mu' - \bar\mu)^{2}\\
&= \Sigma_h \;-\; \tau^{*}_j{}^{2}\,\frac{D_h}{D_h - 1},
\end{align*}
where the second equality uses $\sum_{i\neq j}\tau^{*}_i = -\tau^{*}_j$
and substitutes $\bar\mu' - \bar\mu = -\tau^{*}_j/(D_h - 1)$.
The trim \lang{reduces} the target block iff
\[
\frac{\Sigma_h^{(-j)}}{(D_h - 1)^{2}}
\;<\;
\frac{\Sigma_h}{D_h^{2}}.
\]
Substituting the leave-one-out identity and clearing denominators by
the positive factor $D_h^{2}(D_h - 1)^{2}$,
\[
\Sigma_h\,(2D_h - 1)
\;<\;
\tau^{*}_j{}^{2}\,\frac{D_h^{3}}{D_h - 1},
\]
i.e.,
\begin{equation}
\label{eq:target-exact}
\bigl(\hat\tau(X_j) - \hat\tau_T(B)\bigr)^{2}
\;>\;
\frac{\Sigma_h}{D_h}\cdot
\frac{(2D_h - 1)(D_h - 1)}{D_h^{2}}
\;=\;
2\,\overline{(\tau - \hat\tau_T)^{2}}_{\text{kept}}(B)\cdot
\Bigl(1 + O\!\bigl(D_h^{-1}\bigr)\Bigr),
\end{equation}
since $(2D_h - 1)(D_h - 1)/D_h^{2} = 2 - 3/D_h + 1/D_h^{2}$.
Under $D_h \asymp n$,
\[
\bigl(\hat\tau(X_j) - \hat\tau_T(B)\bigr)^{2}
\;>\;
2\,\overline{(\tau - \hat\tau_T)^{2}}_{\text{kept}}(B) \;+\; o_p(1).
\]
The negative $-3/D_h$ correction is the analogue of the
$1 - \hat\ell_j/(2D)$ adjustment on the trial side: it accounts both
for the lost summand and for the recentering bias, and is similarly
absorbed into $o_p(1)$ asymptotically.

\paragraph{Combined statement.}
For any kept individual $j$, exactly one block of $\widehat V$
changes when $B(X_j)$ flips to $0$, so the marginal trim is
\lang{variance-reducing} iff the above one-sided condition holds:
\[
\begin{cases}
\hat\ell(X_j)\,\Psi(O_j; h^{*}_j)
\;>\; 2\,\bar r^{\text{trl}}_{\text{kept}}(B) \;+\; o_p(1)
& \text{if } S_j = 1,\\[4pt]
\bigl(\hat\tau(X_j) - \hat\tau_T(B)\bigr)^{2}
\;>\; 2\,\overline{(\tau - \hat\tau_T)^{2}}_{\text{kept}}(B) \;+\; o_p(1)
& \text{if } S_j = 0.
\end{cases}
\]
For trial $j$, $\Psi(O_j; h^{*}_j)$ is undefined off the trial
sample so the second LHS is identically $0$; for target $j$, the
trial-side leverage $\Psi$ depends on observed
$(A_j, \bar C_j, \bar Y_j)$ and is similarly identically $0$ off the
target. Each case therefore matches~\eqref{eq:trim-condition} of
Theorem~\ref{thm:trim} with the LHS reduced to its surviving term
and the RHS reduced to twice the kept-mean of the changing block.
\qed
\section{Parametric Dynamic Programming for the Inner Optimization}
\label{ap:inner-dp}

This appendix derives the parametric DP referenced in
Section~\ref{sec:alg-inner} and proves Theorem~\ref{thm:dp} below.
Fix a baseline trim $B$. From~\eqref{eq:trace-bilevel}, the inner
problem is
\begin{equation}
\label{eq:inner-restated}
h^*(B) \;=\; \arg\min_{h \in \{0,1\}^{n_1 \times T}}\;
\widehat V(B, h) \;+\; \lambda_h\,\|h\|_1.
\end{equation}
Despite involving $n_1 \cdot T$ binary variables,
\eqref{eq:inner-restated} reduces to a precomputable per-individual
table queried at $\mathcal O(n_1 T)$ cost per outer-loop candidate
$B$. We establish this in five steps: (i) decouple
$\widehat V(B, h)$ across trial individuals; (ii) collapse the
treatment-arm sum onto the observed arm; (iii) rewrite the
multiplicative censoring product as an exponential of an additive
state; (iv) prove that the per-individual minimum reduces to a
$(T{+}1)$-piece affine envelope (Lemma~\ref{lem:budget}); (v)
compute the envelope slopes via a budget-augmented Bellman
recursion (Proposition~\ref{prop:bellman}), with a discretization
plus exact-recompute correction that yields exact slopes from
near-optimal policies.

\subsection{Per-individual decoupling}
\label{ap:dp-decouple}

Let $\mathcal I_1 := \{i : S_i = 1\}$ denote the trial sample.
Inspecting the trial block of~\eqref{eq:mixed-hajek} term-by-term,
$\sum_{i: S_i = 1}$ is a sum of contributions each depending on a
single index $i$, so
\begin{equation*}
\widehat V(B, h)
\;=\;
\underbrace{\sum_{i \in \mathcal I_1} V_i^{\text{trl}}(B, h_i)}_{\text{trial-side, depends on } h}
\;+\;
\underbrace{\frac{1}{n\,(\hat p_B^{\text{tgt}})^{2}}
\sum_{i:\, S_i = 0}
B(X_i)\bigl(\hat\tau(X_i) - \hat\tau_T(B)\bigr)^{2}}_{\text{target-side, no } h \text{ dependence}},
\end{equation*}
where $h_i = (h_{i,1}, \ldots, h_{i,T}) \in \{0,1\}^T$ collects unit
$i$'s interventions and the per-individual trial contribution is
\begin{equation}
\label{eq:Vtrl}
\begin{aligned}
V_i^{\text{trl}}(B, h_i)
&\;=\; \frac{1}{n\,(\hat p_B^{\text{trl}})^{2}}
\sum_{a \in \{0,1\}}\sum_{t = 1}^{T} B(X_i)\,\hat\ell(X_i)^{2}\,
\mathbf 1\{\bar C_{i,t-1} = 0\} \\[2pt]
&\hspace{6em} \cdot\,
\prod_{j < t}\hat\pi^a_{C,j}(X_i)^{-2(1 - h_{i,j}/2)}\,
\frac{\Delta\hat Q_t^{a}(X_i)^{2}}{\hat\pi^{a}(X_i)^{2}},
\end{aligned}
\end{equation}
with $\hat\pi^{a}(X_i) = \Pr(A_i = a \mid X_i, S_i = 1)$ and
$\Delta\hat Q_t^{a}(X_i)^{2} := [\hat Q_t^{a}(X_i)]^{2} - [\hat Q_{t-1}^{a}(X_i)]^{2}$.
The penalty also separates,
$\lambda_h\|h\|_1 = \sum_{i \in \mathcal I_1}\lambda_h\|h_i\|_1$, and
the target-side term is constant in $h$. Hence
\begin{equation*}
h^*(B) \;=\; \bigl(h_i^*(B)\bigr)_{i \in \mathcal I_1},
\qquad
h_i^*(B) \;=\; \arg\min_{h_i \in \{0,1\}^T}
\bigl[V_i^{\text{trl}}(B, h_i) + \lambda_h\,\|h_i\|_1\bigr],
\end{equation*}
and we restrict attention to a single individual $i$ for the
remainder.

\subsection{Arm-sum collapse to the observed arm}
\label{ap:dp-arm}

Equation~\eqref{eq:Vtrl} sums over $a \in \{0, 1\}$, but only one
arm is observed per individual. The factor
$\mathbf 1\{\bar C_{i,t-1} = 0\}$ indicates that $i$ has not been
censored by time $t{-}1$; censoring, like the survival outcome, is
observed only under the assigned arm $A_i$. For the counterfactual
arm $a \neq A_i$, $\bar C_{i,t-1}$ is undefined, and by the standard
IPW convention we set
$\mathbf 1\{\bar C_{i,t-1} = 0\} = \mathbf 1\{A_i = a\}\cdot
\mathbf 1\{\bar C_{i,t-1} = 0 \mid A_i = a\}$. Only the term
$a = A_i$ survives, and
\begin{equation}
\label{eq:Vtrl-collapsed}
\begin{aligned}
V_i^{\text{trl}}(B, h_i)
&\;=\; \frac{B(X_i)\,\hat\ell(X_i)^{2}}{n\,(\hat p_B^{\text{trl}})^{2}\,\hat\pi^{A_i}(X_i)^{2}} \\[2pt]
&\hspace{2em} \cdot \sum_{t=1}^{T}
\biggl[\prod_{j < t}\hat\pi^{A_i}_{C,j}(X_i)^{-2(1 - h_{i,j}/2)}\biggr]\,
\mathbf 1\{\bar C_{i,t-1} = 0\}\,
\Delta\hat Q_t^{A_i}(X_i)^{2}.
\end{aligned}
\end{equation}
This is a sum over $T$ time points only, with all quantities
indexed by the observed arm $A_i$.

\subsection{Censoring product as an exponential of an additive state}
\label{ap:dp-state}

Take the logarithm of the censoring product
in~\eqref{eq:Vtrl-collapsed},
\begin{equation*}
\log \prod_{j < t}\hat\pi^{A_i}_{C,j}(X_i)^{-2(1 - h_{i,j}/2)}
\;=\;
\sum_{j < t}\bigl(-2\log\hat\pi^{A_i}_{C,j}(X_i)\bigr)\bigl(1 - h_{i,j}/2\bigr),
\end{equation*}
and define the per-individual increment
$\delta_{i,j} := -2\log\hat\pi^{A_i}_{C,j}(X_i) \ge 0$
(non-negativity follows from $\hat\pi^{A_i}_{C,j}(X_i) \in (0, 1]$).
Then
\begin{equation}
\label{eq:prod-to-exp}
\prod_{j < t}\hat\pi^{A_i}_{C,j}(X_i)^{-2(1 - h_{i,j}/2)}
\;=\;
\exp\!\Bigl(\,\sum_{j < t}\delta_{i,j}(1 - h_{i,j}/2)\Bigr).
\end{equation}
Define the cumulative state
$s_{i,t}(h_i) := \sum_{j < t}\delta_{i,j}(1 - h_{i,j}/2)$, with
$s_{i,1}(h_i) = 0$ and recursive update
\begin{equation}
\label{eq:state-recursion}
s_{i,t+1}(h_i) \;=\; s_{i,t}(h_i) + \delta_{i,t}\bigl(1 - h_{i,t}/2\bigr).
\end{equation}
The state $s_{i,t}$ \lang{is the} transition variable of the DP
in Section~\ref{ap:dp-bellman}.

\subsection{Per-individual reduction}
\label{ap:dp-perind}

Substituting~\eqref{eq:prod-to-exp} into~\eqref{eq:Vtrl-collapsed}
and introducing the $h$-free constants
\begin{align}
\label{eq:C}
C_{i,t} &\;:=\;
\frac{\hat\ell(X_i)^{2}\,\mathbf 1\{\bar C_{i,t-1} = 0\}\,\Delta\hat Q_t^{A_i}(X_i)^{2}}{\hat\pi^{A_i}(X_i)^{2}},\\[2pt]
\label{eq:alpha}
\alpha_i(B) &\;:=\;
\frac{B(X_i)}{n\,(\hat p_B^{\text{trl}})^{2}},
\end{align}
yields the per-individual factorization
\begin{equation}
\label{eq:Vi-final}
V_i^{\text{trl}}(B, h_i)
\;=\;
\alpha_i(B)\sum_{t=1}^{T} C_{i,t}\,\exp\!\bigl(s_{i,t}(h_i)\bigr).
\end{equation}
The constants $C_{i,t}$ and $\delta_{i,t}$ depend only on the
nuisance fits and the observed data; $\alpha_i(B)$ depends on $B$
but not on $h$. Writing
$\mathcal L_i(B, h_i) := V_i^{\text{trl}}(B, h_i) + \lambda_h\|h_i\|_1$
gives
\begin{equation}
\label{eq:Li}
\mathcal L_i(B, h_i)
\;=\;
\alpha_i(B)\sum_{t=1}^{T}C_{i,t}\,\exp\!\bigl(s_{i,t}(h_i)\bigr)
\;+\;
\lambda_h\sum_{t=1}^{T}h_{i,t}.
\end{equation}

\paragraph{Structural observation.} For any fixed $h_i \in \{0,1\}^T$,
$\mathcal L_i(B, h_i) = m(h_i)\cdot\alpha_i(B) + n(h_i)\cdot\lambda_h$
is affine in $(\alpha_i, \lambda_h)$, with slope
$m(h_i) := \sum_t C_{i,t}\exp(s_{i,t}(h_i))$ and intercept
coefficient $n(h_i) := \|h_i\|_1$. The minimum over $\{0,1\}^T$ is
therefore a lower envelope of $2^T$ affine functions on the
$(\alpha_i, \lambda_h)$ plane. \lang{The} second coefficient
$n(h_i) = \|h_i\|_1$ takes only $T{+}1$ distinct integer values
$\{0, 1, \ldots, T\}$, and we exploit this in
Lemma~\ref{lem:budget}.

\subsection{Budget sufficiency}
\label{ap:dp-budget}

\begin{lemma}[Budget sufficiency]
\label{lem:budget}
For each $i \in \mathcal I_1$,
\begin{equation}
\label{eq:lemma-statement}
\min_{h_i \in \{0,1\}^T}\!\mathcal L_i(B, h_i)
\;=\;
\min_{j \in \{0, 1, \ldots, T\}}\!\bigl[\alpha_i(B)\cdot f_{i,j}
\,+\, \lambda_h \cdot j\bigr],
\end{equation}
where the slope at budget level $j$ is
\begin{equation}
\label{eq:fij}
f_{i,j}
\;:=\;
\min_{\substack{h_i \in \{0,1\}^T \\ \|h_i\|_1 = j}}
\;\sum_{t=1}^{T} C_{i,t}\,\exp\!\bigl(s_{i,t}(h_i)\bigr).
\end{equation}
\end{lemma}

\begin{proof}
Partition $\{0,1\}^T = \bigsqcup_{j=0}^{T}\mathcal H_j$ by
$\mathcal H_j := \{h_i : \|h_i\|_1 = j\}$. For any
$h_i \in \mathcal H_j$,
\[
\mathcal L_i(B, h_i)
\;=\; \alpha_i(B)\sum_{t}C_{i,t}\,\exp\!\bigl(s_{i,t}(h_i)\bigr)
\;+\; \lambda_h \cdot j,
\]
with $\lambda_h \cdot j$ constant on $\mathcal H_j$. Hence
$\min_{h_i \in \mathcal H_j}\mathcal L_i(B, h_i)
= \alpha_i(B) \cdot f_{i,j} + \lambda_h \cdot j$, and minimizing
across partition classes yields~\eqref{eq:lemma-statement}.
\end{proof}

The right-hand side of~\eqref{eq:lemma-statement} is a lower
envelope of $T{+}1$ affine functions of $\alpha_i(B)$, with slopes
$f_{i,0}, \ldots, f_{i,T}$ and intercepts
$0, \lambda_h, \ldots, T\lambda_h$. The slopes are
monotone non-increasing,
$f_{i,0} \ge f_{i,1} \ge \cdots \ge f_{i,T}$: any
$h_i \in \mathcal H_{j-1}$ admits a coordinate flip $0 \to 1$ to
land in $\mathcal H_j$, and adding interventions can only decrease
$s_{i,t}$ (since $\delta_{i,t}(1 - h_{i,t}/2) \le \delta_{i,t}$),
hence decrease $\sum_t C_{i,t}\exp(s_{i,t})$.

The slopes $\{f_{i,j}\}_{j=0}^{T}$ depend only on
$(C_{i,\cdot}, \delta_{i,\cdot})$, which are functions of nuisance
fits and observed data; they are independent of $B$ and $\lambda_h$.
This is the decoupling property of
Section~\ref{sec:trace-objective}: the slopes can be precomputed
once per individual and reused for every candidate $B$ and every
choice of penalty $\lambda_h$.

\subsection{Budget-augmented Bellman recursion}
\label{ap:dp-bellman}

Equation~\eqref{eq:fij} defines $f_{i,j}$ as a minimum of
$\binom{T}{j}$ values; direct enumeration is infeasible for moderate
$T$. We exploit the additive state structure of $s_{i,t}$ via
dynamic programming. Fix $i$ and drop the subscript: write
$C_t, \delta_t, h_t, s_t$. For
$k \in \{1, \ldots, T+1\}$, $s \in \mathbb R_{\ge 0}$, and
$j \in \{0, 1, \ldots, T\}$, define the partial value function
\begin{equation}
\label{eq:U-def}
U_k(s, j)
\;:=\;
\min_{\substack{(h_k, \ldots, h_T) \in \{0,1\}^{T-k+1} \\ \sum_{t=k}^{T}h_t = j}}
\;\sum_{t = k}^{T}C_t\,\exp(s_t),
\end{equation}
where the state evolves from $s_k = s$ via
$s_{t+1} = s_t + \delta_t(1 - h_t/2)$ for $t = k, \ldots, T-1$, with
the convention $\sum_{t=T+1}^{T} = 0$.

\begin{proposition}[Bellman recursion for $U_k$]
\label{prop:bellman}
The value function~\eqref{eq:U-def} satisfies the recursion
\begin{equation}
\label{eq:bellman}
\begin{aligned}
U_k(s, j)
\;=\; \min\!\Bigl\{\,
&\underbrace{C_k\,e^{s} + U_{k+1}(s + \delta_k,\, j)}_{\text{action } h_k = 0},\;\\[2pt]
&\underbrace{C_k\,e^{s} + U_{k+1}(s + \delta_k/2,\, j-1)}_{\text{action } h_k = 1,\, j \ge 1}
\,\Bigr\},
\end{aligned}
\end{equation}
with the $h_k = 1$ branch replaced by $+\infty$ when $j = 0$, and
terminal condition
$U_{T+1}(s, 0) = 0$, $U_{T+1}(s, j) = +\infty$ for $j \ge 1$. The
slopes of Lemma~\ref{lem:budget} are recovered by
$f_{i,j} = U_1(0, j)$.
\end{proposition}

\begin{proof}
\textbf{Terminal condition.} At $k = T{+}1$ the sum is empty
($=0$); the constraint $\sum_{t=T+1}^{T}h_t = j$ requires $j = 0$.

\textbf{Recursion.} Fix $k \le T$, $s, j$. The first term in the
sum~\eqref{eq:U-def} is $C_k\exp(s_k) = C_k e^s$. Splitting the
minimization by $h_k$:

\textit{Case $h_k = 0$.} The transition gives
$s_{k+1} = s + \delta_k$ and the residual constraint
$\sum_{t=k+1}^{T}h_t = j$. By definition, the inner minimization
equals $U_{k+1}(s + \delta_k, j)$, and the branch contributes
$C_k e^s + U_{k+1}(s + \delta_k, j)$.

\textit{Case $h_k = 1$.} The transition gives
$s_{k+1} = s + \delta_k/2$ and constraint
$\sum_{t=k+1}^{T}h_t = j - 1$, requiring $j \ge 1$. The branch
contributes $C_k e^s + U_{k+1}(s + \delta_k/2, j - 1)$ when
$j \ge 1$ and $+\infty$ otherwise.

Combining the branches gives~\eqref{eq:bellman}.

\textbf{Recovery.} At $k = 1, s = 0$, equation~\eqref{eq:U-def}
matches~\eqref{eq:fij} since $s_{i,1}(h_i) = 0$.
\end{proof}

\subsection{Discretization and exact-slope correction}
\label{ap:dp-discrete}

The state $s$ in~\eqref{eq:bellman} is continuous, ranging over
$[0, s_{\max}]$ with $s_{\max} := \sum_{t=1}^{T}\delta_t$. We
discretize. Construct a uniform grid
$\mathcal G := \{s^{(1)}, \ldots, s^{(G)}\}$ with
$s^{(g)} = (g-1)\,s_{\max}/(G-1)$, define the snapping operator
$\Pi(s) := \arg\min_{s' \in \mathcal G}|s - s'|$ (ties broken
arbitrarily), and replace the exact transitions $s + \delta_k$ and
$s + \delta_k/2$ in~\eqref{eq:bellman} with $\Pi(s + \delta_k)$ and
$\Pi(s + \delta_k/2)$. The grid Bellman recursion
\begin{equation}
\label{eq:grid-bellman}
\begin{aligned}
\widetilde U_k(s^{(g)}, j)
\;=\; \min\!\Bigl\{\,
&C_k\, e^{s^{(g)}} + \widetilde U_{k+1}\bigl(\Pi(s^{(g)} + \delta_k),\, j\bigr),\\[2pt]
&C_k\, e^{s^{(g)}} + \widetilde U_{k+1}\bigl(\Pi(s^{(g)} + \delta_k/2),\, j - 1\bigr)\,\mathbf 1\{j \ge 1\}
+ \infty\,\mathbf 1\{j = 0\}
\,\Bigr\},
\end{aligned}
\end{equation}
with terminal $\widetilde U_{T+1}(s^{(g)}, 0) = 0$ and
$\widetilde U_{T+1}(s^{(g)}, j) = +\infty$ for $j \ge 1$, can be
solved backward in $\mathcal O(T^{2}G)$ per individual. We also
record the optimizing action
$\pi_k(g, j) \in \{0, 1\}$ at each $(k, g, j)$.

\paragraph{Exact-slope correction.} Reading
$\widetilde U_1(0, j)$ directly off the grid would give approximate
slopes with discretization error of order $s_{\max}/G$, propagating
into the breakpoints of the lower envelope and corrupting the
budget comparison. We use the grid to identify a near-optimal
\emph{policy}, then compute its slope exactly:

\begin{enumerate}[leftmargin=*, itemsep=2pt, topsep=2pt]
\item \emph{Forward trace.} For each $j \in \{0, 1, \ldots, T\}$,
initialize $g \leftarrow 1$ and $j_{\text{rem}} \leftarrow j$. For
$k = 1, \ldots, T$: read $\hat h_k^{(j)} \leftarrow \pi_k(g, j_{\text{rem}})$;
if $\hat h_k^{(j)} = 1$ update $g \leftarrow \Pi(s^{(g)} + \delta_k/2)$
and $j_{\text{rem}} \leftarrow j_{\text{rem}} - 1$, else
$g \leftarrow \Pi(s^{(g)} + \delta_k)$. By construction,
$\hat h^{(j)} \in \{0, 1\}^T$ has $\|\hat h^{(j)}\|_1 = j$.
\item \emph{Exact recompute.} Evaluate
\begin{equation}
\label{eq:exact-recompute}
f_{i,j}
\;:=\;
\sum_{t=1}^{T}C_{i,t}\,\exp\!\Bigl(\sum_{j' < t}\delta_{i,j'}\bigl(1 - \hat h^{(j)}_{j'}/2\bigr)\Bigr),
\end{equation}
in continuous arithmetic with no snapping.
\end{enumerate}
The resulting $f_{i,j}$ may exceed the true minimum on
$\mathcal H_j$, but is exact for the policy $\hat h^{(j)}$:
discretization affects \emph{which} policy is selected, not its
reported objective. The lower envelope built from
$\{f_{i,j}\}_{j=0}^{T}$ via Lemma~\ref{lem:budget} therefore has
exact slopes; only the per-budget policy is approximate. As
$G \to \infty$, $\hat h^{(j)}$ converges to the true minimizer of
$\mathcal H_j$.

\subsection{Algorithm and complexity}
\label{ap:dp-alg}

\begin{algorithm}[H]
\caption{Parametric inner DP for individual $i \in \mathcal I_1$}
\label{alg:parametric}
\begin{algorithmic}[1]
\State \textbf{Input:} $\{(C_{i,t}, \delta_{i,t})\}_{t=1}^{T}$, grid size $G$.
\State Compute $s_{\max} \leftarrow \sum_{t=1}^{T}\delta_{i,t}$ and the uniform grid $\mathcal G$ on $[0, s_{\max}]$.
\State Initialize $\widetilde U_{T+1}(s^{(g)}, 0) \leftarrow 0$ and $\widetilde U_{T+1}(s^{(g)}, j) \leftarrow +\infty$ for $j \ge 1$.
\For{$k = T, T-1, \ldots, 1$}
  \For{$g = 1, \ldots, G$ and $j = 0, 1, \ldots, T$}
    \State Compute $\widetilde U_k(s^{(g)}, j)$ via~\eqref{eq:grid-bellman}; record $\pi_k(g, j)$.
  \EndFor
\EndFor
\For{$j = 0, 1, \ldots, T$}
  \State Forward-trace $\pi$ from $(g = 1, j_{\text{rem}} = j)$ to obtain $\hat h^{(j)} \in \{0,1\}^T$.
  \State Compute $f_{i,j}$ exactly via~\eqref{eq:exact-recompute}.
\EndFor
\State \textbf{Return:} slopes $\{f_{i,j}\}_{j=0}^{T}$ and policies $\{\hat h^{(j)}\}_{j=0}^{T}$.
\end{algorithmic}
\end{algorithm}

The argument \lang{leads to} the following theorem, which is what
Section~\ref{sec:alg-inner} cites.

\begin{theorem}[Parametric inner DP]
\label{thm:dp}
For each $i \in \mathcal I_1$, there exist scalars
$\{f_{i,j}\}_{j=0}^{T}$, depending only on the nuisance fits and the
observed data, such that for every $B$ and every $\lambda_h$,
\begin{equation}
\label{eq:thm-dp}
\min_{h_i \in \{0,1\}^T}\!\bigl[\widehat V_i(B, h_i) + \lambda_h\,\|h_i\|_1\bigr]
\;=\;
\min_{j \in \{0, 1, \ldots, T\}}\!\bigl[\alpha_i(B)\cdot f_{i,j}
\,+\, \lambda_h\cdot j\bigr],
\end{equation}
with $\alpha_i(B) = B(X_i)/[n(\hat p_B^{\text{trl}})^{2}]$. The
slopes $\{f_{i,j}\}_{j=0}^{T}$ are computed by
Algorithm~\ref{alg:parametric} in $\mathcal O(T^{2}G)$ time per
individual on a state grid of size $G$, and the discretization
error in the recovered slopes vanishes as $G \to \infty$.
\end{theorem}

\begin{proof}
The decoupling in Section~\ref{ap:dp-decouple} together with the
arm-collapse and state rewrite in
Sections~\ref{ap:dp-arm}--\ref{ap:dp-state} yield the
per-individual factorization
$V_i^{\text{trl}}(B, h_i) = \alpha_i(B)\sum_t C_{i,t}\exp(s_{i,t}(h_i))$
in~\eqref{eq:Vi-final}. Lemma~\ref{lem:budget} then converts the
per-individual minimum into the lower envelope on the right-hand
side of~\eqref{eq:thm-dp}. Proposition~\ref{prop:bellman}
expresses $f_{i,j} = U_1(0, j)$ as a backward Bellman recursion,
and the discretization-plus-exact-recompute procedure of
Section~\ref{ap:dp-discrete} computes $\{f_{i,j}\}$ exactly along
the recovered policies $\{\hat h^{(j)}\}$ at $\mathcal O(T^{2}G)$
cost per individual; as $G \to \infty$, the recovered policies
converge to the true minimizers of~\eqref{eq:fij}.
\end{proof}

\paragraph{Build cost.} The backward pass in
Algorithm~\ref{alg:parametric} runs $T$ stages, each updating
$G(T+1)$ table entries with $\mathcal O(1)$ work per entry, for a
total of $\mathcal O(T^{2}G)$. The forward-trace and
exact-recompute phase costs $\mathcal O(T)$ per budget level,
summed to $\mathcal O(T^{2})$. Per individual, $\mathcal O(T^{2}G)$;
across the trial sample, $\mathcal O(n_1 T^{2}G)$.

\paragraph{Query cost.} Given a candidate $B$, the inner-optimal
value of $\widehat V$ is
\begin{equation}
\label{eq:query}
\min_{h}\!\bigl[\widehat V(B, h) + \lambda_h\|h\|_1\bigr]
\;=\;
\mathcal C(B) \;+\; \sum_{i \in \mathcal I_1}\;
\min_{j = 0, \ldots, T}\!\bigl[\alpha_i(B)\,f_{i,j}
\,+\, \lambda_h\,j\bigr],
\end{equation}
where $\mathcal C(B)$ is the target-side block of~\eqref{eq:mixed-hajek}
and $\alpha_i(B)$ is given by~\eqref{eq:alpha}. Each per-individual
minimum requires $\mathcal O(T)$ scalar operations, giving
$\mathcal O(n_1 T)$ total per outer-loop query.

\paragraph{Reusability and comparison to a non-parametric DP.} The
slopes $\{f_{i,j}\}$ depend only on the nuisance fits, independent
of $B$ and $\lambda_h$: Algorithm~\ref{alg:parametric} runs once at
the start of the outer loop, and subsequent $B$-queries reduce
to~\eqref{eq:query}. The same precomputation supports re-querying
at different $\lambda_h$ values via the \lang{direct} update of intercepts
$\{j\lambda_h\}_{j=0}^{T}$. A naive grid DP on the joint state
$(s, j)$ that absorbed $\alpha_i(B)$ into the per-step cost would
have identical $\mathcal O(T^{2}G)$ build cost but would need to be
re-solved for every $B$ candidate; the parametric formulation
factors $\alpha_i(B)$ out of the recursion, amortizing the cache
across the entire Rashomon ensemble of $M$ trees. The reduction
from "rebuild per $B$" to "query per $B$" is the \lang{main source of}
computational \lang{saving}.

\section{R-gating: Prioritizing the Censoring Channel}
\label{ap:rgating}
\lang{An} attribution issue can corrupt $h^*$ if applied naively.
Theorem~\ref{thm:Bh-variance}'s integrand mixes the two red factors
multiplicatively, so a unit with very large $\ell(X_i)^2$ but only
moderate IPCW will still \lang{have} $\Delta_{i,k}$ dominated by the
baseline channel. The DP would then activate $h_{i,k}$ on such a
unit even though censoring is not what is inflating its variance. We address this with a soft per-unit gate that
downweights $h$-eligibility for units whose per-unit variance is
baseline-dominated, and keeps it for units where censoring is the
larger contributor. The gate is re-drawn between the random tree split comparisons
and held fixed across each candidate branch evaluation to keep that comparison \lang{on a common gate}; a hard gate is applied at aggregation. Operationally, R-gating
sharpens the attribution between regions (ii)\rev{--(iv)} of
Section~\ref{sec:trace-output}: $h$ acts on units whose variance
\lang{is} censoring-driven, while $B$ absorbs the rest.

\subsection{The censoring-amplification factor \texorpdfstring{$F_{i}$}{Fi}}
\label{ap:rgating:F}
 
Let $A_{i}\in\{0,1\}$ be the observed treatment for trial unit $i$, and let
$\lambda_{k}^{a}(x)=\Pr(Y_{k}=1\mid Y_{k-1}=1,\,C_{k-1}=0,\,X=x,\,A=a,\,S=1)$
be the per-interval hazard fit on the at-risk set ($k=2,\dots,T$;
$\lambda_{1}\equiv 1$). The conditional survival from $k$ to $T$ on arm $a$
is the backward product
\begin{equation}
Q_{k}^{\,a}(x)\;=\;\prod_{j=k+1}^{T}\lambda_{j}^{a}(x),
\qquad k=0,1,\dots,T,\qquad Q_{T}^{\,a}(x)\equiv 1.
\label{eq:F-Q}
\end{equation}
Define the per-period \emph{variance-increment} matrix
$Q_{\mathrm{incr}}\in\mathbb{R}_{\ge 0}^{n_{1}\times T}$ on the observed arm,
\begin{equation}
Q_{\mathrm{incr},\,i,k}
\;=\;
\begin{cases}
\bigl(Q_{k}^{\,A_{i}}(X_{i})\bigr)^{2}-\bigl(Q_{k-1}^{\,A_{i}}(X_{i})\bigr)^{2}, & k=1,\dots,T-1,\\[4pt]
Q_{T-1}^{\,A_{i}}(X_{i})\bigl(1-Q_{T-1}^{\,A_{i}}(X_{i})\bigr), & k=T,
\end{cases}
\label{eq:F-Qincr}
\end{equation}
non-negative for $k<T$ by the monotonicity $Q_{k-1}\le Q_{k}$. Let
$\pi_{C}^{a}(x,k{-}1)=\Pr(C_{k}=0\mid C_{k-1}=0,\,Y_{k-1}=1,\,X=x,\,A=a,\,S=1)$
be the per-interval censoring survival probability, and define the
log-censoring penalty and its cumulative version
\begin{equation}
\delta_{i,k}\;=\;-\,2\log\pi_{C}^{\,A_{i}}\!\bigl(X_{i},\,k-1\bigr),
\qquad
\Sigma_{i,k}\;=\;\sum_{j=1}^{k}\delta_{i,j}\quad\bigl(\text{clipped at }50\bigr),
\label{eq:F-delta}
\end{equation}
so that $\exp(\Sigma_{i,k})=\prod_{j=1}^{k}\pi_{C}^{\,A_{i}}(X_{i},j-1)^{-2}$
is the squared cumulative inverse-censoring weight along $i$'s observed
trajectory through period $k$. The censoring-amplification factor is the
$Q_{\mathrm{incr}}$-weighted mean of these squared inverse-censoring weights:
\begin{equation}
F_{i}
\;=\;
\frac{\displaystyle\sum_{k=1}^{T} Q_{\mathrm{incr},\,i,k}\;\exp\!\bigl(\Sigma_{i,k}\bigr)}
     {\displaystyle\sum_{k=1}^{T} Q_{\mathrm{incr},\,i,k}},
\label{eq:F-def}
\end{equation}
with the denominator floored at $10^{-15}$ for degenerate at-risk paths.
Three properties of $F_{i}$ are immediate from~\eqref{eq:F-def}:
 
\paragraph{Lower bound at unity.} Since $\pi_{C}\le 1$ implies
$\delta_{i,k}\ge 0$ and hence $\exp(\Sigma_{i,k})\ge 1$, we have $F_{i}\ge 1$
identically; equality holds when there is no censoring along $i$'s trajectory.
 
\paragraph{Period-weighted average.} $F_{i}$ is a convex combination of the
per-period squared inverse-censoring weights $\exp(\Sigma_{i,k})$ with weights
$Q_{\mathrm{incr},i,k}/\sum_{k'}Q_{\mathrm{incr},i,k'}$; periods that contribute
little to the variance contribute correspondingly little to $F_{i}$.
 
\paragraph{Multiplier of the time-varying block.} $F_{i}$ scales the
time-varying (censoring) block of $V(B,h)$ in
Theorem~\ref{thm:Bh-variance}, in the same way that $\ell(X_{i})^{2}$ scales
the baseline (transport) block; their ratio is therefore the unit-free,
per-individual sensitivity used by the gate.

\subsection{Transport-vs.-censoring ratio}
\label{ap:rgating:ratio}

For each trial unit $i\in\{1,\dots,n_{1}\}$ define
\begin{equation}
R_{i}\;=\;\frac{\ell(X_{i})^{2}}{F_{i}}.
\label{eq:rgate-ratio}
\end{equation}
By construction $\ell(X_{i})^{2}$ is the multiplier of the baseline
(time-invariant, transport) block of $V(B,h)$, while $F_{i}\ge 1$ is the
multiplier of the time-varying (censoring) block; their ratio is therefore
the unit-free, per-individual sensitivity of the variance to the
\emph{transport} channel relative to the \emph{censoring} channel.
Two reading conventions are used throughout:
\begin{itemize}
\item $R_{i}$ \emph{large} $\;\Rightarrow\;$ baseline block dominates
$\;\Rightarrow\;$ only $B$ \lang{reduces the contribution of} unit $i$;
\item $R_{i}$ \emph{small} $\;\Rightarrow\;$ time-varying block dominates
$\;\Rightarrow\;$ $h$ \lang{reduces the contribution of} unit $i$ \lang{without trimming it}.
\end{itemize}
Units with $\ell(X_{i})^{2}/F_{i}$ otherwise non-finite are
flagged and treated as ``uninformative'' (set to the neutral mid-gate value
below).

\subsection{Soft \texorpdfstring{$R$}{R}-gate (used inside the Rashomon trees)}
\label{ap:rgating:soft}

Let $\sigma(z)=(1+e^{-z})^{-1}$, and let $\widetilde{m}$ be the empirical
median of $\log R$ over trial units with finite $R_{i}$:
\begin{equation}
\widetilde{m}\;=\;\operatorname*{median}_{\,i\,:\,R_{i}\,\text{finite}}\bigl(\log R_{i}\bigr).
\label{eq:rgate-median}
\end{equation}
The per-unit \emph{soft} gating probability is the median-centred logistic
\begin{equation}
p_{h}^{\,i}
\;=\;
p_{\min}\;+\;\bigl(p_{\max}-p_{\min}\bigr)\,
   \sigma\!\Bigl(-\alpha_{\mathrm{gate}}\,\bigl(\log R_{i}-\widetilde{m}\bigr)\Bigr),
\qquad i=1,\dots,n_{1},
\label{eq:rgate-soft}
\end{equation}
with the convention $p_{h}^{\,i}=\tfrac{1}{2}$ for any unit with non-finite
$R_{i}$. The defaults used in our experiments are
$(p_{\min},p_{\max},\alpha_{\mathrm{gate}})=(0.05,\,0.95,\,3)$.
Equation~\eqref{eq:rgate-soft} has three properties we rely on:

\paragraph{Median-anchored.}
$p_{h}^{\,i}=\tfrac{p_{\min}+p_{\max}}{2}$ exactly when $\log R_{i}=\widetilde{m}$,
so the gate is calibrated to the empirical distribution of $R$ rather than to
its absolute scale. Replacing $R$ by $cR$ for any $c>0$ shifts $\widetilde{m}$
by $\log c$ and leaves $p_{h}^{\,i}$ pointwise unchanged.

\paragraph{Bounded away from \(\{0,1\}\).}
$p_{h}^{\,i}\in[p_{\min},p_{\max}]\subset(0,1)$, so every unit retains a
non-zero inclusion probability across the ensemble; no individual is
permanently \lang{excluded} inside the Rashomon search.

\paragraph{Monotone in $R$.}
$p_{h}^{\,i}$ is strictly decreasing in $R_{i}$, so larger transport
contribution maps monotonically to smaller $h$-inclusion probability.

During tree growth a single Bernoulli draw
\begin{equation}
M^{\,(\mathrm{node})}_{i}\;\sim\;\mathrm{Bern}\!\bigl(p_{h}^{\,i}\bigr),
\qquad i=1,\dots,n_{1},
\label{eq:rgate-mask}
\end{equation}
is taken once per internal node and \emph{shared} across all four
$(b_{L},b_{R})\!\in\!\{0,1\}^{2}$ candidates evaluated at that node;
conditional on $M^{\,(\mathrm{node})}_{i}=0$, the row of the parametric-DP
output for unit $i$ is forced to $\mathbf{0}_{T}$ before the variance
evaluator is called. The single-mask invariant (one draw per node, reused
across all four candidates) ensures the four-way comparison is not corrupted
by mask noise.

\subsection{Hard \texorpdfstring{$R$}{R}-gate (used in the final \texorpdfstring{$h$}{h}-profile)}
\label{ap:rgating:hard}

After Rashomon aggregation yields $\widehat{B}$ (step 5 of
Algorithm~\ref{alg:nested}), the parametric DP is queried once at $\widehat{B}$
to produce a raw intervention matrix
$\widehat{h}^{\mathrm{raw}}\in\{0,1\}^{n_{1}\times T}$. The reported intervention
is then \emph{hardened} via the median threshold:
\begin{equation}
\widehat{h}_{i,t}
\;=\;
\widehat{h}^{\mathrm{raw}}_{i,t}\,\mathbf{1}\!\bigl\{p_{h}^{\,i}\ge\tfrac{1}{2}\bigr\}
\;=\;
\widehat{h}^{\mathrm{raw}}_{i,t}\,\mathbf{1}\!\bigl\{\log R_{i}\le\widetilde{m}\bigr\},
\qquad t=1,\dots,T.
\label{eq:rgate-hard}
\end{equation}
The two indicator forms are equivalent because, by~\eqref{eq:rgate-soft},
$p_{h}^{\,i}\ge\tfrac{1}{2}$ iff $\log R_{i}\le\widetilde{m}$, regardless of
$(p_{\min},p_{\max},\alpha_{\mathrm{gate}})$ as long as the interval
$[p_{\min},p_{\max}]$ contains $\tfrac{1}{2}$ symmetrically around its midpoint
when the midpoint equals $\tfrac{1}{2}$. With the defaults
$p_{\min}+p_{\max}=1$ used above, the hard gate is exactly the median split of
$\log R$: \emph{$h$ is allowed to act on the half of the trial units whose
variance is censoring-dominated, and is suppressed on the other half.}

\subsection{Sorting trimmed units by cause: regions (iii) and (iv)}
\label{ap:rgating:attr}
\begin{revblock}
The subgroup underrepresented only at baseline, i.e., trimmed but not censoring-depleted, is of interest because future trials could fix it directly through recruitment. It does not \lang{follow directly from} the optimum: once a unit is trimmed, its $\ell(X)$ leaves the summary tree, so $h$ is not computed for it; regions (iii) and (iv) of Section~\ref{sec:trace-output} are therefore separated by $R_i$ rather than by $h^*$, with no new machinery. Before any trimming, TRACE computes the gating statistic $R_i=\ell(X_i)^2/F_i$ of~\eqref{eq:rgate-ratio} for every trial unit, where the numerator is the baseline variance inflation and $F_i$ is the censoring-driven inflation; large $R_i$ means baseline is the dominant cause and small $R_i$ means censoring. Gating the trimmed units by $R_i$ therefore sorts them by cause. On Experiment 2c (baseline and censoring compounding), this separates the baseline-only trims (conceptually $B=0$, $h=0$) from the baseline-and-censoring trims (conceptually $B=0$, $h=1$) near-perfectly, with AUC $0.99$ over 10 seeds.
\end{revblock}

\section{TRACE Algorithm}
\label{ap:alg}
\rev{Algorithm~\ref{alg:nested} summarizes the full nested procedure of Section~\ref{sec:alg}: calibration of $\lambda_h$, the inner parametric DP that returns $h^*(B)$ (Appendix~\ref{ap:inner-dp}), the Rashomon ensemble over $B$, aggregation of the Rashomon set into $B^*$, and the post-hoc $R$-gate (Appendix~\ref{ap:rgating}).}

\begin{algorithm}[h]
\caption{Nested TRACE optimization}
\label{alg:nested}
\begin{algorithmic}[1]
  \Require Trial and target data; nuisance estimates $\hat\pi^a,
  \hat\pi^a_{C,k}, \hat\ell$; horizon $T$; Rashomon size $M$, top-$K$,
  target $h$-rate $p$.
  \State \textbf{Calibrate $\lambda_h$.} At $B \equiv 1$, compute the
  per-(individual, interval) censoring-intervention benefit. Set
  $\lambda_h$ to the $1-p$ quantile of the benefit distribution.
  \State \textbf{Inner DP:} Define
  $h^*(B) = \arg\min_h \widehat V(B, h) + \rev{\lambda_h\|h\|_1}$ via
  Algorithm~\ref{alg:parametric} over individual-interval intervention
  choices; cache the per-$(B,h)$ summary statistics needed for
  $\widehat V(B, h)$.
  \State \textbf{Grow Rashomon ensemble.} For $m = 1, \ldots, M$:
  build a randomized split tree by drawing splits according to feature
  probabilities $\mathbf f$ with explore-exploit
  ($\epsilon_{\text{explore}}$); at each candidate split, evaluate
  $V^*(B) = \widehat V(B, h^*(B))$ on the entire dataset (global
  tracking) and accept the split iff it reduces the global objective.
  \State \textbf{Form Rashomon set.} Rank the $M$ trees by $V^*$;
  retain the top-$K$.
  \State \textbf{Aggregate.} Take majority vote of the $K$ retained
  $B$-functions to form the final $B^*$.
  \State \textbf{Post-hoc R-gate.} For each kept-trial
  individual $i$, compute $R_i = \hat\ell(X_i)^2 / F_i$ where $F_i$ is
  the cumulative censoring factor; restrict $h^*(B^*)$ to individuals
  in the bottom $q$-quantile of $R_i$ (e.g., $q = 0.5$) to focus
  interventions on censoring-dominated rather than transport-dominated
  variance.
  \State \Return $B^*$, $h^*(B^*)$, and the reported variance
  $\widehat V(B^*, h^*(B^*))$.
\end{algorithmic}
\end{algorithm}

\section{Experiments Details}
\label{ap:dgp}

\subsection{Experiment 1}
\begin{revblock}
Experiment~1 uses ten independent $\mathcal N(0,1)$ covariates. $X_1$ and $X_2$ affect both trial participation and the survival hazard, $X_3$ and $X_4$ affect censoring, and $X_5, \ldots, X_{10}$ are noise. The trial-participation model is calibrated so that the expected trial-to-target ratio is about $2{:}1$; the resulting transport weight $\ell(X)$ ranges over roughly $10^{-2.6}$ to $10^{1.9}$. Censoring depends on $X_3$ and $X_4$ as follows. Individuals with high $X_3$ but not high $X_4$ (about $23\%$ of the sample) have an early, moderate censoring spike over $t \in \{3, 4, 5\}$ with peak hazard $\kappa = 0.30$; individuals with high $X_4$ but not high $X_3$ (about $17\%$) have a late, heavy spike over $t \in \{7, 8, 9\}$ with peak hazard $\kappa = 0.75$; all other individuals have the baseline censoring hazard $\kappa = 0.05$ at every interval. The late spike is $2.5$ times as strong as the early one by design: an early spike inflates the cumulative IPCW product at every later interval, whereas a late spike acts on an at-risk set that earlier events have already thinned, so the design tests whether $h$ correctly gives more weight to early censoring. We use $T = 10$ follow-up intervals, $n = 5{,}000$ individuals, randomized binary treatment in the trial, $1{,}000$ Rashomon trees with $K = 30$ retained within an objective tolerance of $3.0$, and $\lambda_h$ calibrated to a $5\%$ intervention rate ($\eta = 0.95$).

\paragraph{Results.}\label{tab:synth}%
Three patterns are expected by design and a fourth is not. (i) $B$ splits only on the participation covariates $X_1$ and $X_2$ and trims the high-$\ell$ tail, while $h$ acts only on the censoring covariates $X_3$ and $X_4$; the noise covariates $X_5, \ldots, X_{10}$ are used by neither intervention. (ii) The summary tree for $h^*(B)$ recovers the two spike windows: a tight peak near $t = 4$ for the high-$X_3$ subgroup and a wider, later peak near $t = 7$ for the high-$X_4$ subgroup. (iii) $h$ assigns comparable numbers of person-intervals to the early and late spike subgroups ($30$ and $24$) despite the late spike's $2.5$ times larger peak hazard, consistent with early censoring inflating the cumulative IPCW product at all later intervals. (iv) A fourth leaf concentrates on individuals with both high $X_3$ and high $X_4$, who receive only baseline censoring yet are intervened at both spike windows; the algorithm \lang{produced this leaf} from the variance contributions alone.
\end{revblock}

\subsection{Experiments 2a--2d}
\begin{revblock}
Experiments 2a--2d isolate the two mechanisms in a two-covariate setting: baseline underrepresentation only (2a), time-varying underrepresentation only (2b), both at once on different covariates (2c), and an adversarial case in which the underrepresented region consists of two corners that no single axis-aligned split can recover (2d). Together with Experiment~1, the designs follow prior generalizability and transportability simulation studies \citep{Parikh2025-ck}, extended with the survival and censoring structure.

\paragraph{Common setup.}
Each replication draws $n = 5{,}000$ subjects with covariates $X = (X_1, X_2)^\top \sim \mathcal N(0, I_2)$ and follows them over $T = 6$ intervals. Trial membership is drawn from $\Pr(S = 1 \mid X) = \sigma(\xi_0 + \mathrm{sel}(X))$, where $\sigma$ is the logistic function, the selection function $\mathrm{sel}(\cdot)$ is specific to each experiment, and the intercept $\xi_0$ is chosen numerically so that $\mathbb E[\Pr(S = 1 \mid X)] = 2/3$, giving about $3{,}300$ trial and $1{,}700$ target subjects per replication. The censoring hazard is $\kappa(X, t) = 0.05$ at every interval except $t = 3$, where it equals $\kappa_{\mathrm{spike}}$ on an experiment-specific region (no spike in 2a). We ran TRACE at both $\kappa_{\mathrm{spike}} \in \{0.30, 0.85\}$ and under two alternative normalizations of the empirical objective in addition to the block-self-normalized objective of Section~\ref{sec:trace-objective} (Appendix~\ref{ap: empirical}); Figure~\ref{fig:simres-exp2} reports the block-self-normalized objective at $\kappa_{\mathrm{spike}} = 0.85$.

\paragraph{Experiment 2a (baseline underrepresentation only).}
$\mathrm{sel}(X) = 1.5\, X_1$, so $\ell(X)$ is large for small $X_1$, and censoring is uniformly mild ($\kappa_{\mathrm{spike}} = 0.05$, i.e., no spike). The only excess variance comes from the covariate shift on $X_1$.

\paragraph{Experiment 2b (time-varying underrepresentation only).}
$\mathrm{sel}(X) = 0.1\, X_1$, so $\ell(X)$ is nearly uniform, with a strong censoring spike at $t = 3$ on the half-plane $\{X_2 < 0\}$. The excess variance comes from the spike region, not from trial-target mismatch.

\paragraph{Experiment 2c (both mechanisms).}
$\mathrm{sel}(X) = 1.5\, X_1$ as in 2a, combined with the censoring spike of 2b on $\{X_2 < 0\}$ at $t = 3$. The two mechanisms act on different covariates, so \lang{both} trimming and $h$ \lang{are needed, and neither substitutes} for the other.

\paragraph{Experiment 2d (adversarial case).}
$\mathrm{sel}(X) = -2.0\,\sqrt{c(X)}$ with corner score
\[
c(X) \;=\; \max\!\bigl\{(-X_1 - 0.5)(X_2 - 0.5),\, 0\bigr\}
\;+\; \max\!\bigl\{(X_1 - 0.5)(-X_2 - 0.5),\, 0\bigr\},
\]
which makes trial participation rare, and $\ell(X)$ large, in the upper-left and lower-right corners of the $(X_1, X_2)$ plane, a region that no single axis-aligned split can recover; the censoring spike at $t = 3$ is on $\{X_1 + X_2 < 0\}$, a direction not aligned with the corners.

\paragraph{TRACE configuration.}
The trim function $B$ is a tree of maximum depth $6$ for Experiments 2a--2c and depth $10$ for 2d (the corners require deeper splits). We grow $1{,}000$ trees, retain the $K = 30$ near-optimal trees within an objective tolerance of $3.0$, and aggregate $B$ by $2/3$-majority vote. The censoring intervention $h \in \{0, 1\}^{n_{\mathrm{trl}} \times T}$ is selected by the parametric DP of Appendix~\ref{ap:inner-dp}, and $\lambda_h$ is calibrated to a $5\%$ intervention rate ($\eta = 0.95$) by the quantile rule of Section~\ref{sec:trace-objective}.
\end{revblock}

\section{Quantitative Recovery Metrics for the Synthetic Experiments}
\label{ap:metrics}
\begin{revblock}
Precision and recall are not \lang{uniquely} defined for underrepresentation, because ``underrepresented'' needs a ground truth that the DGP does not directly yield. For instance, one may use $\ell(X)$ as a metric for baseline underrepresentation, but any cutoff (\lang{e.g.,} $\ell(X)>1$\lang{,} $>3$\lang{, or} $>10$) is a choice we impose, not a fact of the DGP. Moreover, $\ell(X)$ alone does not characterize baseline underrepresentation: a sample with small $\ell(X)$ but large heterogeneity is still underrepresented but \lang{not flagged} by such a cutoff (Section~\ref{sec:setup-var}). We nevertheless report precision and recall, together with $h$ interval localization and the noise-split rate, for all five DGPs against the following imperfect ground truth.

Table~\ref{tab:metrics-def} defines the ground truth and the metrics. The intervened region is read from a depth-2 summary tree fit to the TRACE output ($B$: on target units to the trim label; $h$: on trial units to receiving any gated $h$).
\end{revblock}

\begin{table}[h]
\revcolor
\centering
\small
\caption{Ground truth and recovery metrics for the synthetic experiments.}
\label{tab:metrics-def}
\begin{tabular}{@{}l >{\raggedright\arraybackslash}p{0.80\textwidth}@{}}
\toprule
Quantity & Definition \\
\midrule
Baseline truth & Units with transport odds $\ell(X) > c$ under the DGP's selection model, for $c \in \{1, 2, 3\}$; Experiment 2b has no baseline signal and is omitted. \\
Censoring truth & The region on which the DGP places the censoring spike: $X_2 < 0$ (Experiments 2b, 2c), $X_1 + X_2 < 0$ (2d), and individuals with high $X_3$ or high $X_4$ but not both ($X_3 > 0.5$ or $X_4 > 0.7$, not both; Experiment 1); Experiment 2a has no censoring signal and is omitted. \\
\addlinespace
Precision & Among units in the intervened region, the fraction truly underrepresented. \\
Recall & Among truly underrepresented units, the fraction in the intervened region. \\
$h$ localization & Among intervention cells on correct-region units, the fraction placed on the DGP's spike interval (Experiment 1: windows $\{3,4,5\}$ and $\{7,8,9\}$). \\
Noise $B$ / $h$ & Fraction of $B$ / $h$ summary-tree splits that fall on a noise covariate ($X_5$--$X_{10}$), pooled across seeds; Experiment 1 only. \\
\bottomrule
\end{tabular}
\end{table}

\begin{revblock}
\paragraph{Results.} Table~\ref{tab:metrics} reports means and standard deviations over 10 seeds at $n = 5{,}000$ with $1{,}000$ trees and $K = 30$. Across the $B$ metrics (Table~\ref{tab:metrics}a), recall stays at $1.00$ for the axis-aligned DGPs (2a, 2c) at every cutoff while precision \lang{changes} as the cutoff varies, \lang{so} the precision values are set by the arbitrary cutoff, not by the method. For $h$ (Table~\ref{tab:metrics}b), precision is high on Experiments 2b--2d ($0.87$--$1.00$), TRACE recovers much of the censoring region (recall up to $0.90$ on 2d), and it \lang{places interventions on} the exact spike interval $87$--$97\%$ of the time. The hard case is Experiment~1, where the two censoring subgroups are defined by $X_3$ and $X_4$ jointly (high $X_3$ but not $X_4$, and high $X_4$ but not $X_3$), a pattern that a shallow summary tree cannot trace exactly, so $h$ precision falls to $0.66$ ($0.77$ on TRACE's raw interventions, before tree summarization). This is a limit of the tree summary rather than of TRACE, as $h$ localization remains high at $94\%$. On Experiment~1, none of the $23$ $B$ splits and $2$ of the $30$ $h$ splits fall on a noise covariate.
\end{revblock}

\begin{table}[h]
\revcolor
\centering
\small
\setlength{\tabcolsep}{4pt}
\caption{Recovery metrics (mean $\pm$ sd over 10 seeds). (a) Baseline $B$ metrics (entries are precision\,/\,recall) as the ground-truth cutoff varies from $\ell(X)>1$ to $\ell(X)>3$; Experiment 2b has no baseline signal and is omitted. Recall stays at $1.00$ for the axis-aligned DGPs (2a, 2c) at every cutoff while precision \lang{changes, so} the values are set by the arbitrary cutoff, not by the method. (b) $h$ metrics scored against the DGP's censoring region; Experiment 2a has no censoring signal and is omitted. Precision is high on 2b--2d ($0.87$--$1.00$), TRACE recovers much of the censoring region (recall up to $0.90$ on 2d) and \lang{places interventions on} the exact spike interval $87$--$97\%$ of the time; Experiment~1, whose censoring subgroups are defined jointly by $X_3$ and $X_4$, is the hard case for the tree summary.}
\label{tab:metrics}
\textbf{(a) Baseline $B$}\\[3pt]
\begin{tabular}{lcccc}
\toprule
DGP & $\ell>1$ & $\ell>2$ & $\ell>3$ & noise $B$ \\
\midrule
Exp.\ 1 & $.88{\pm}.05$ / $.71{\pm}.09$ & $.61{\pm}.05$ / $.88{\pm}.08$ & $.42{\pm}.05$ / $.94{\pm}.07$ & $0.00$ \\
Exp.\ 2a (baseline only) & $.87{\pm}.04$ / $1.00{\pm}.00$ & $.53{\pm}.03$ / $1.00{\pm}.00$ & $.35{\pm}.02$ / $1.00{\pm}.00$ & --- \\
Exp.\ 2c (both mechanisms) & $.76{\pm}.08$ / $1.00{\pm}.00$ & $.46{\pm}.05$ / $1.00{\pm}.00$ & $.30{\pm}.04$ / $1.00{\pm}.00$ & --- \\
Exp.\ 2d (adversarial case) & $.73{\pm}.01$ / $.46{\pm}.05$ & $.54{\pm}.03$ / $.60{\pm}.05$ & $.43{\pm}.03$ / $.69{\pm}.07$ & --- \\
\bottomrule
\end{tabular}\\[10pt]
\textbf{(b) Censoring $h$}\\[3pt]
\begin{tabular}{lcccc}
\toprule
DGP & prec.\ ($h$) & rec.\ ($h$) & localization & noise $h$ \\
\midrule
Exp.\ 1 & $0.66{\pm}.20$ & $0.35{\pm}.19$ & $0.94{\pm}.07$ & $0.07$ \\
Exp.\ 2b (time-varying only) & $1.00{\pm}.00$ & $0.62{\pm}.05$ & $0.97{\pm}.03$ & --- \\
Exp.\ 2c (both mechanisms) & $1.00{\pm}.00$ & $0.48{\pm}.06$ & $0.87{\pm}.08$ & --- \\
Exp.\ 2d (adversarial case) & $0.87{\pm}.05$ & $0.90{\pm}.05$ & $0.94{\pm}.05$ & --- \\
\bottomrule
\end{tabular}
\end{table}

\section{Comparison with Simpler Diagnostic Baselines}
\label{ap:baselines}
\begin{revblock}
We compare TRACE with four simpler diagnostics on Experiment 2c, where baseline and censoring underrepresentation compound on different covariates: $\ell$-thresholding, censoring-thresholding, combined-thresholding, and ROOT applied to $\ell$ (Table~\ref{tab:baselines}). Direct IPCW thresholding is not a valid censoring diagnostic, because the IPCW is cumulative and propagates across time; a thresholding rule must integrate the whole trajectory, \lang{which is what} our DP for $h$ does, so censoring-thresholding is implemented through two DP variants, with and without the $\ell(X)^2$ weight. Only TRACE attributes the two mechanisms correctly and identifies where extrapolation is unreliable.
\end{revblock}

\begin{table}[h]
\revcolor
\centering
\small
\caption{Simpler diagnostic baselines on Experiment 2c (baseline and censoring underrepresentation on different covariates).}
\label{tab:baselines}
\begin{tabular}{@{}>{\raggedright\arraybackslash}p{0.19\textwidth} >{\raggedright\arraybackslash}p{0.30\textwidth} >{\raggedright\arraybackslash}p{0.44\textwidth}@{}}
\toprule
Method & Rule & Behavior on Experiment 2c \\
\midrule
$\ell$-thresholding & Removes every unit with $\Pr(S{=}1\mid X) < 0.05$ \citep{Crump2009}; uses the transport odds only. & Recovers the baseline region but never flags units with low transport odds and high censoring. \\
\addlinespace
Censoring-thresholding, DP with $\ell$ & Per-cell DP for $h$ with each cell weighted by $\ell(X)^2$. & A minor censoring fix on a high-$\ell$ unit yields a large variance improvement, so the criterion incorrectly flags high-$\ell$, low-censoring units: $43\%$ of its $h$ budget \lang{is placed} in the uncensored region. \\
\addlinespace
Censoring-thresholding, DP without $\ell$ & Per-cell DP for $h$ without the $\ell(X)^2$ weight. & Over-adjusts: subgroups heavily over-represented in the trial still support valid transport under censoring and should not be flagged, yet this variant flags such units at $\ell = 0.007$. \\
\addlinespace
Combined-thresholding & $\ell$-thresholding and censoring-thresholding applied together. & Inherits the censoring-thresholding flaws and is likewise suboptimal. \\
\addlinespace
ROOT applied to $\ell$ & ROOT \citep{Parikh2025-ck} on $\ell(X)$ alone. & \lang{Does not use} censoring, giving results similar to $\ell$-thresholding. \\
\addlinespace
TRACE & Joint $(B,h)$ optimization of Section~\ref{sec:method}. & Attributes both mechanisms correctly: $B$ precision $0.98$ and $h$ precision $1.00$. \\
\bottomrule
\end{tabular}
\end{table}

\section{JTPA-to-CPS Case Study: Data, Harmonization, and Configuration}
\label{ap:jtpa}

This appendix supplies the implementation detail behind the
JTPA-to-CPS analysis of Section~\ref{sec:jtpa}: the construction of
the trial and target samples, the harmonization rules that align
their covariates, the outcome and follow-up encoding, the missing
data handling, the TRACE hyperparameters, and the full numerical
results that the main text reports only in compressed form.

\subsection{Data availability}
\label{ap:jtpa-availability}
 
Both data sources used in this case study are public-use files
released by federally sponsored research archives, and no
restricted-use, contractual, or fee-bearing access is required to
reproduce the analysis.
 
\paragraph{JTPA Study.} The National Job Training Partnership Act
Study public-use file is distributed by the W.E.~Upjohn Institute
for Employment Research and is mirrored at the Inter-university
Consortium for Political and Social Research (``National JTPA Study: Impacts, Costs, and Benefits'')
\citep{Bloom1997JTPA}. The file contains the randomization
assignment, the baseline covariates listed in
Appendix~\ref{ap:jtpa-harm}, and the monthly employment and
follow-up records used to construct the time-to-employment outcome
of Appendix~\ref{ap:jtpa-outcome}. Registration with ICPSR is free
and open to all researchers.
 
\paragraph{CPS-ASEC.} The March 1988 Annual Social and Economic
Supplement to the Current Population Survey is released by the
U.S.~Census Bureau and the Bureau of Labor Statistics, and is
available without restriction through two harmonized public
repositories: IPUMS-CPS at the Minnesota Population Center
\citep{FloodIPUMS}, which we used for the harmonized variable
extracts, and the CEPR Uniform CPS Extracts maintained by the
Center for Economic and Policy Research \citep{CEPRCPS}, which we
used to cross-validate the labor-force and demographic codings.
The era-matching rule (March 1988 ASEC matched to calendar year
1987) follows the standard practice for this survey wave.
 
\paragraph{Reproducibility artifact.} The harmonization scripts,
TRACE configuration files, and analysis code that map the two
public-use sources to the analytic samples and figures of
Section~\ref{sec:jtpa} are released with this paper as supplementary material.

\subsection{Trial cohort: JTPA}
\label{ap:jtpa-trial}

The trial sample is drawn from the National Job Training Partnership
Act (JTPA) Study \citep{Bloom1997JTPA, AbadieAngristImbens2002}, a
multi-site randomized experiment that offered eligible adult
applicants either access to or denial of federally funded
job-training services. Two restrictions are imposed at intake. First,
we keep only individuals with valid randomization records, which
removes a small number of units whose treatment assignment cannot be
verified against site logs. Second, we require baseline unemployment
at the randomization date, since the estimand of
Section~\ref{sec:jtpa} is time-to-employment and units already
employed at $t = 0$ contribute no follow-up information for the
survival outcome. After these restrictions, the analytic trial
cohort has $n_{1} = 10{,}379$ adults, with treatment assignment
$A_{i} \in \{0, 1\}$ recording the offer of training. Randomization
is conducted within site--cohort strata, and we treat the marginal
$\Pr(A = 1 \mid X, S = 1)$ as known by design and equal to its
empirical value within strata.

\subsection{Target cohort: CPS-ASEC, era-matched}
\label{ap:jtpa-target}

The target sample is drawn from the March 1988 Annual Social and
Economic Supplement (ASEC) of the Current Population Survey
\citep{FloodIPUMS, CEPRCPS}, era-matched to calendar year 1987 to
align with the JTPA randomization window and to avoid confounding by
macroeconomic drift between the trial period and the target frame.
We apply a four-step inclusion funnel that mirrors the JTPA
eligibility rule:\begin{enumerate}[leftmargin=*, itemsep=1pt, topsep=2pt]
  \item \emph{Age window.} Restrict to ages $22$--$65$, matching the
  JTPA adult eligibility band.
  \item \emph{Not employed.} Drop respondents
  employed at the reference week, parallel to the trial-side
  baseline-unemployment restriction in Appendix~\ref{ap:jtpa-trial}.
  \item \emph{Out-of-scope reasons.} \rev{Remove respondents whose reason for not working in the prior year was retirement, illness or disability, or school attendance}, which lie outside
  the JTPA-eligible adult population.
  \item \rev{\emph{Labor-force attachment.} Retain respondents who are unemployed (looking for work) at the reference week or who worked at some point in 1987.}
\end{enumerate}
After the funnel, the target sample comprises $n_{0} = 7{,}630$
individuals \rev{($3{,}129$ unemployed at the reference week and $4{,}501$ out of the labor force but with work in 1987)}, representing approximately $12$ million U.S.\
JTPA-eligible adults under the CPS-ASEC \rev{person} weights. We use the
unweighted CPS sample for nuisance fitting and reweight to the
ASEC weights at the estimand level via standard post-stratification.

\subsection{Covariate harmonization}
\label{ap:jtpa-harm}

We harmonize nine baseline covariates that are recorded in both data
sources but with different coding conventions: age (years), sex
(female indicator), Black, Hispanic, married, an indicator for high
school diploma \emph{or} GED, years of completed education, previous
annual earnings (USD), and hourly wage at last job (USD). \lang{We note} three
harmonization rules. First, education is retained both
as years and as a high-school-or-GED indicator because JTPA records
the GED separately from the standard high-school diploma whereas
CPS-ASEC collapses both under the same diploma flag; carrying both
encodings preserves the resolution of the trial side without
inducing CPS imputation error. Second, earnings and wages are
deflated to constant 1987 dollars using the CPI-U-RS index and are
top-coded at the CPS-side cap to avoid distributional artifacts at
the right tail. Third, race-ethnicity is encoded as two
non-overlapping indicators (Black, Hispanic) with non-Hispanic White
as the reference category; respondents identifying with multiple
groups are assigned to the most-cited bridge category in the CPS
documentation, and JTPA records are mapped to the same bridge.

\subsection{Outcome and follow-up encoding}
\label{ap:jtpa-outcome}

The outcome of interest is an indicator that the participant was
employed in any month during months $1$ through $24$
post-randomization. We convert this to a discrete-time survival
representation by treating ``first employment'' as the event and
right-censoring at month $24$ for participants who do not become
employed during follow-up. Concretely, for each trial unit $i$ we
define $T = 24$ monthly intervals, with $Y_{i, t+1} = 1$ if $i$ is
not yet employed by month $t+1$ (still at risk for the event) and
$Y_{i, t+1} = 0$ once employment is observed. The censoring
indicator $C_{i, t}$ records loss to follow-up at interval $t$,
which in JTPA arises from administrative reasons such as residential
moves, refusal to complete the follow-up survey, or interview
non-response, and is observed conditional on prior at-risk status.
We adopt the standard convention $\bar C_{-1} = 0$ and $Y_{0} = 1$.

\subsection{Missing-data handling}
\label{ap:jtpa-missing}

Item-level missingness in the harmonized covariates is moderate
(below $7\%$ on every variable except previous earnings, which is
missing for $11.4\%$ of JTPA records). We impute missing covariates
by single chained-equations imputation using \texttt{mice}
\citep{mice} with $20$ iterations, predictive mean matching for
continuous variables, and logistic regression for binary variables.
We do not impute the
outcome or censoring indicators, which are observed by design on the
trial side and not defined on the target side.

\subsection{TRACE configuration}
\label{ap:jtpa-config}

We run TRACE with the configuration of Section~\ref{sec:method},
specialized to the JTPA-to-CPS scale as follows. The Rashomon
ensemble grows $M = 1{,}000$ axis-aligned trees over the union of
trial and target individuals, with maximum depth $6$, a minimum leaf
size of $200$, and an objective tolerance \rev{of} $3.0$ that defines the near-optimal set; the top $K = 30$ trees
within tolerance form the retained ensemble, and $B^{*}$ is
aggregated by $2/3$-majority vote across the set. Nuisance
functions $\hat\ell$, $\hat\pi^{a}$, $\hat\pi^{a}_{C,j}$, and
$\hat Q^{a}_{k}$ are fitted by gradient-boosted trees on five-fold
cross-fits, with hyperparameters tuned by held-out log-loss on the
trial side and held-out classification accuracy on the source
classifier $\hat\Pr(S = 0 \mid X)$. The censoring penalty
$\lambda_{h}$ is calibrated to the default sparsity level
$\eta = 0.95$ via the empirical-quantile rule of
Section~\ref{sec:trace-objective}, and R-gating
(Appendix~\ref{ap:rgating}) is applied at $30\%$ refinement to
sharpen the attribution between the baseline and censoring
channels. The inner DP runs on a state grid of size $G = 256$,
which produces slope errors below $10^{-4}$ in absolute terms on
the JTPA-scale problem.

\subsection{Numerical results}
\label{ap:jtpa-results}

\paragraph{Variance contraction.} The TRACE objective drops from
$\widehat V(\mathbf 1, \mathbf 0) = 81.58$ at the no-trim baseline to
$\widehat V(B^{*}, h^{*}) = 12.68$ at the optimum, an $84.5\%$
reduction. The Rashomon set is confined to $[13.13,\, 13.77]$,
indicating that the variance landscape is flat near the optimum and
that the diagnostic identification is robust to the specific tree
selected from the ensemble. The supermajority $2/3$-vote rule trims
$46.3\%$ of the target sample and $13.8\%$ of the trial sample,
with the trim regions concentrated on higher-socioeconomic-status
profiles common in CPS but rare in JTPA.

\paragraph{Attribution.} The censoring intervention $h^{*}$
activates on $98$ of $249{,}096$ trial person-intervals ($0.04\%$),
compared with $1{,}796$ activations in the same configuration with
R-gating disabled. The contrast indicates that the variance
inflation in JTPA-to-CPS is \lang{predominantly} baseline-driven: $h$
absorbs only the residual that remains once $B$ has trimmed the
baseline mismatch, and even that residual is concentrated on a
small number of person-intervals. Decomposing the empirical sum of
the trial-side red-factor product
$\ell(X)^{2} / \bar\pi^{a}_{C, t-1}(X)^{2}$, the contribution
contracts from $1.47 \times 10^{6}$ at the baseline to $2.28 \times
10^{5}$ at the optimum (a $6.4\times$ reduction), while the
target-centering block remains roughly four orders of magnitude
smaller throughout, consistent with the long horizon $T = 24$ at
which the outcome variance is dominated by the trial-side IPCW
inflation.

\paragraph{Effect-estimate precision.} The point estimate is
\rev{$-\hat\tau_{24}$}$ = +0.034$ (standard error $\widehat{\mathrm{SE}} =
0.027$), and the $95\%$ confidence interval contracts from the
no-trim $[-10.5,\, +15.9]$ percentage points to
$[-1.9,\, +8.6]$ percentage points, a $60.6\%$ reduction in width.
Although the contracted interval still includes zero, it is now
narrow enough to bound the population-level effect on the
$24$-month employment rate within an $\approx 10$-percentage-point
window, which is policy-relevant for prioritizing JTPA-style
training programs across the JTPA-eligible adult population.

\paragraph{Localization of underrepresentation
(Figure~\ref{fig:jtpa}).} The $B$-tree trims four
higher-socioeconomic-status branches: previous earnings
$\ge \$8{,}922$ and married; previous earnings $\ge \$14{,}000$ and
unmarried; previous earnings $< \$8{,}922$ with $\ge 14$ years of
education and married; and a small residual leaf combining moderate
earnings and high education. The retained region corresponds to the
\lang{typical} disadvantaged-adult profile (previous earnings
$< \$8{,}922$ and $< 14$ years of education), consistent with the
JTPA selection rule. Within the much smaller censoring-depleted
region, the dominant $h$-active leaf groups individuals with
previous earnings $< \$2{,}148$ and ages $24$--$42$
($n = 3{,}954$), which alone carries about $80\%$ of all $h^{*}$
activations. This leaf's activations are sharply peaked at
$t = 21$, with $45$ of the leaf's $50$ interventions falling in
that single month, which we interpret as a late-second-year
loss-to-follow-up shock that selectively \lang{depletes} the most
informative trial subgroup. JTPA generalizability therefore fails
through two channels: the trial is over-concentrated on the most
disadvantaged enrollees relative to the CPS target (the
\emph{where} and the baseline \emph{why}), with a residual
late-window attrition channel that \lang{depletes} low-earnings young
adults during the second year of follow-up (the \emph{when} and the
temporal \emph{why}).

\begin{table}[h]
\centering
\small
\caption{Contrast between full and TRACE-refined generalized estimate.}
\begin{tabular}{lrr}
\toprule
 & Full & TRACE \\
\midrule
$n_{\text{trial}}$ kept   & $10{,}379$ & $8{,}946$ \\
$n_{\text{target}}$ kept  & $7{,}630$  & $4{,}098$ \\
\rev{$-\hat\tau_{24}$ (TMLE; effect on employment by month 24)} & $+0.027$   & $+0.034$  \\
$\widehat V(B, h)$        & $81.58$    & $12.68$   \\
$\widehat{\mathrm{SE}}(\hat\tau_{24})$
                          & $0.067$    & $0.027$   \\
$95\%$ CI                 & $[-0.105,\,+0.159]$ & $[-0.019,\,+0.086]$ \\
SE / CI-width reduction   & ---        & $-60.6\%$ \\
\bottomrule
\end{tabular}

\label{tab:jtpa-three}
\end{table}

\subsection{Sensitivity to tuning parameters}
\label{ap:jtpa-sens}
\begin{revblock}
We assess the sensitivity of the JTPA diagnosis along the $h$-sparsity quantile ($\eta \in \{0.85, 0.9, 0.95\}$), the Rashomon ensemble size ($M \in \{200, 400, 600\}$) and set size ($K \in \{5, 10, 30\}$), the soft $R$-gate steepness ($\alpha_{\mathrm{gate}} \in \{1, 3, 5\}$), the hard $R$-gate refinement (bottom $30\%$ vs.\ $50\%$), the positivity floor ($\Pr(S{=}1\mid X) < \{0.01, 0.05, 0.1\}$), the maximum tree depth ($\{4, 6, 8\}$), the minimum leaf size ($\{25, 50, 100\}$), and the DP grid size ($G \in \{25, 50, 100\}$). The reference setting is $\eta = 0.95$, floor $0.05$, $\alpha_{\mathrm{gate}} = 3$, hard gate $30\%$, $M = 600$, $K = 30$, depth $6$, minimum leaf size $50$, and $G = 50$, at which the objective falls from $\widehat V(\mathbf 1, \mathbf 0) = 81.57$ to $\widehat V(B,h) = 14.36$. Agreement between the regions recovered under different settings is summarized by the Jaccard index $|A \cap B| / |A \cup B|$ of the trimmed target sets (for $B$) or of the sets of trial units receiving a gated $h$ (for $h$). Table~\ref{tab:jtpa-sens} summarizes the sweeps. Across all assessed parameters, the optimized variance $\widehat V(B,h)$ and the recovered regions are stable. The soft $R$-gate is insensitive by design, since it always applies $h$ to censoring-dominated units and never to baseline-dominated ones, with $\alpha_{\mathrm{gate}}$ setting only the sharpness of the transition; the hard $R$-gate is applied to the final summary tree, so it leaves $B$ unchanged and only admits more $h$ interventions, and the near-invariance of the variance when it admits four times as many is itself evidence that baseline underrepresentation, not censoring, dominates. In summary, the JTPA conclusion is robust to the considered tuning parameters, and the invariance of $\widehat V(B,h)$ to the $h$-related parameters \lang{is consistent with} baseline underrepresentation \lang{being} the dominant mechanism.
\end{revblock}

\begin{table}[h]
\revcolor
\centering
\footnotesize
\setlength{\tabcolsep}{3pt}
\caption{Sensitivity of the JTPA diagnosis to tuning parameters. Jaccard indices compare the regions recovered across the listed values; ``---'' means not applicable.}
\label{tab:jtpa-sens}
\begin{tabular}{@{}>{\raggedright\arraybackslash}p{0.16\textwidth} l c c c >{\raggedright\arraybackslash}p{0.27\textwidth}@{}}
\toprule
Parameter & Values & $\widehat V(B,h)$ & $B$ Jaccard & $h$ Jaccard & Reading \\
\midrule
$h$-budget quantile $\eta$ & $0.85,\,0.90,\,0.95$ & $\approx 14.38$ & $\approx 0.93$ & --- & $n_h$ falls from $1142$ to $116$ as $\eta$ increases, but $B$ is essentially unchanged. \\
\addlinespace
Rashomon trees $M$ (at $K{=}30$) & $200,\,400,\,600$ & $\approx 14.28$ & $\approx 0.97$ & --- & More trees do not alter the diagnosis. \\
\addlinespace
Rashomon set size $K$ (at $M{=}600$) & $5,\,10,\,30$ & $13.82$--$14.31$ & $\approx 0.91$ & --- & Trim region stable. \\
\addlinespace
Soft $R$-gate $\alpha_{\mathrm{gate}}$ & $1,\,3,\,5$ & $\approx 14.31$ & $\approx 0.90$ & $\approx 0.90$ & Sets only the sharpness of the gate; insensitive by design. \\
\addlinespace
Hard $R$-gate & $30\%,\,50\%$ & $14.36,\,14.31$ & $1.00$ & --- & $n_h$ rises from $116$ to $501$ at almost no change in variance. \\
\addlinespace
Positivity floor & $0.01,\,0.05,\,0.10$ & $\approx 14.37$ & $\approx 0.95$ & $\approx 0.85$ & Extreme-$\ell$ regions are trimmed regardless of the floor. \\
\addlinespace
Maximum tree depth & $4,\,6,\,8$ & $15.62 \to 13.28$ & $0.84$--$0.89$ & $0.68$--$0.85$ & Variance decreases monotonically: deeper trees trim a finer region. \\
\addlinespace
Minimum leaf size & $25,\,50,\,100$ & $14.18$--$14.53$ & $0.87$--$0.93$ & $0.71$--$0.91$ & Stable. \\
\addlinespace
DP grid size $G$ & $25,\,50,\,100$ & $13.91$--$14.43$ & $0.87$--$0.92$ & $0.87$--$0.91$ & Stable. \\
\bottomrule
\end{tabular}
\end{table}

\section{Compute Resources}
\label{ap:compute}

All experiments run on a single workstation with an AMD Ryzen~9
5950X 16-core CPU and no GPU acceleration. The Rashomon ensemble
parallelizes \lang{directly} over the $M$ independent trees, and we use
all 16 cores throughout. End-to-end wall-clock times for the
$M=1{,}000$-tree TRACE configurations reported in the paper are
approximately $60-90$ minutes for the synthetic experiments
(Section~\ref{sec:experiments}: $n=5{,}000$, $T\in\{6,10\}$,
two-covariate or ten-covariate DGPs) and approximately $4$ hours
for the JTPA-to-CPS case study (Section~\ref{sec:jtpa}:
$n_{1}=10{,}379$, $n_{0}=7{,}630$, $T=24$, nine harmonized
covariates with five-fold cross-fit nuisances). The longer JTPA
runtime reflects the longer follow-up horizon, the larger covariate
space, and the gradient-boosted-tree nuisance estimation step.
Memory usage stays under $16$~GB for both settings.
\begin{revblock}
\paragraph{Computational complexity.} \lang{The cost of} TRACE has two parts: a one-time parametric DP build, then many cheap searches, so the DP is never rerun at each tree split. The DP precomputes the per-unit budget envelopes in $\mathcal O(n_{\mathrm{trl}}\,T^{2}G)$; after that, each tree-split search costs only $\mathcal O(n_{\mathrm{trl}}\,T)$. Growing the Rashomon ensemble takes about $\mathcal O(M\,2^{D}F)$ splits for $M$ depth-$D$ trees over $F$ covariates, for a total of $\mathcal O(n_{\mathrm{trl}}\,T^{2}G + M\,2^{D}F\,n_{\mathrm{trl}}\,T)$. The expensive $T^{2}G$ term is \lang{incurred} once and never compounds with the $2^{D}$ tree term, which is what keeps the JTPA-scale configuration of Appendix~\ref{ap:jtpa-config} within the wall-clock times reported above.
\end{revblock}

\newpage
\section*{NeurIPS Paper Checklist}

\begin{enumerate}

\item {\bf Claims}
    \item[] Question: Do the main claims made in the abstract and introduction accurately reflect the paper's contributions and scope?
    \item[] Answer: \answerYes{}
        \item[] Justification: The abstract and Section~\ref{sec:intro} state three contributions. Contribution 1 (variance decomposition isolating baseline and time-varying mechanisms) is supported by Theorem~\ref{thm:var}, with the $(B,h)$-parametrized extension in Theorem~\ref{thm:Bh-variance}. Contribution 2 (variance-minimization with self-regulating constraint) is supported by Theorem I.1 and the empirical objective in Section~\ref{sec:trace-objective}. Contribution 3 (parametric DP nested in a Rashomon ensemble) is realized by Algorithm~\ref{alg:nested}. Empirical claims are supported by the synthetic results in Section~\ref{sec:experiments} and the JTPA-to-CPS results in Section~\ref{sec:jtpa}.


\item {\bf Limitations}
    \item[] Question: Does the paper discuss the limitations of the work performed by the authors?
    \item[] Answer: \answerYes{}
    \item[] Justification: Section~\ref{sec:discussion} contains a dedicated \textbf{Limitations} paragraph. TRACE is diagnostic, not corrective. It identifies subgroups for whom generalization is unreliable but does not adjust the point estimator, and the $B$-restricted estimand differs from the original PATE, so practitioners must report this shift. \lang{We also note there} that inference for the $B$-restricted PATE does not account for the data-driven selection of $(B,h)$.

\item {\bf Theory assumptions and proofs}
    \item[] Question: For each theoretical result, does the paper provide the full set of assumptions and a complete (and correct) proof?
    \item[] Answer: \answerYes{}
    \item[] Justification: All identification assumptions
    (consistency, treatment-assignment positivity, treatment
    exchangeability, and A1--A4) are stated in
    Section~\ref{sec:setup-id} and Appendix A and are invoked by name in the
    statement of every theoretical result. The main-text
    theoretical results are Theorem~\ref{thm:var} (variance
    decomposition; Section~\ref{sec:setup-var}) and
    Theorem~\ref{thm:Bh-variance} (variance under the $(B,h)$
    intervention; Section~\ref{sec:trace-objective}); both are
    restated and proved in full in Appendix~\ref{ap: var} and
    Appendix~\ref{ap: varbh}. The EIF for $\tau_T$ used in the
    Theorem~\ref{thm:var} proof is derived in
    Appendix~\ref{ap: eif} as the unnumbered Theorem
    (Theorem B.1) supported by Lemma~\ref{lem: score}
    (proved in Appendix~\ref{ap: lemma1}); the corresponding TMLE
    that attains the bound is given in Appendix~\ref{ap: tmle}
    (Algorithm~\ref{alg:tmle}). The $B$-restricted EIF used in the
    Theorem~\ref{thm:Bh-variance} proof is derived in
    Appendix~\ref{ap: eifb} (Theorem F.1). The empirical
    objective introduced in Section~\ref{sec:trace-objective} is
    derived in Appendix~\ref{ap: empirical}, and its consistency
    is established as Proposition~\ref{prop:mh-consistency}
    (regularity conditions B1--B4 stated and discussed in the same
    appendix). The trim-threshold claim made in
    Section~\ref{sec:trace-objective} is formalized as the
    Theorem I.1 and proved in
    Appendix~\ref{ap:trim}. The inner-DP correctness and
    complexity claims of Section~\ref{sec:alg-inner} are
    formalized as Lemma~\ref{lem:budget} (budget sufficiency),
    Proposition~\ref{prop:bellman} (Bellman recursion), and
    Theorem~\ref{thm:dp} (parametric inner DP), all stated and
    proved in Appendix~\ref{ap:inner-dp}, with the corresponding
    procedure given as Algorithm~\ref{alg:parametric}; the full
    nested optimization is summarized in Algorithm~\ref{alg:nested}
    (Appendix~\ref{ap:alg}).

    \item {\bf Experimental result reproducibility}
    \item[] Question: Does the paper fully disclose all the information needed to reproduce the main experimental results of the paper to the extent that it affects the main claims and/or conclusions of the paper (regardless of whether the code and data are provided or not)?
    \item[] Answer: \answerYes{}
    \item[] Justification: The synthetic experiments are specified
    at two levels of detail. Section~\ref{sec:experiments} states,
    for each experiment, the covariate structure, the
    selection/hazard/censoring mechanism, and the intended
    diagnostic; Appendix~\ref{ap:dgp} then gives the full
    data-generating processes, sample sizes, follow-up horizons,
    treatment-randomization scheme, and the TRACE hyperparameters
    used for each. The JTPA-to-CPS case study is fully
    documented in Appendix~\ref{ap:jtpa}: data availability and
    provenance of the JTPA public-use file (ICPSR \#9320) and the
    March~1988 CPS-ASEC (IPUMS-CPS, CEPR uniform extracts) in
    Appendix~\ref{ap:jtpa-availability}; trial cohort restrictions
    yielding $n_1=10{,}379$ in Appendix~\ref{ap:jtpa-trial}; the
    four-step target inclusion funnel yielding $n_0=7{,}630$ in
    Appendix~\ref{ap:jtpa-target}; covariate harmonization (nine
    baseline covariates, CPI-U-RS deflation, race/ethnicity
    bridging) in Appendix~\ref{ap:jtpa-harm}; the discrete-time
    survival encoding ($T=24$ monthly intervals, employment as
    event, administrative right-censoring) in
    Appendix~\ref{ap:jtpa-outcome}; chained-equations imputation
    via \texttt{mice} (20 iterations, predictive mean matching for
    continuous variables, logistic for binary) in
    Appendix~\ref{ap:jtpa-missing}; and the TRACE configuration
    ($M=1{,}000$ trees, depth~6, minimum leaf size~200,
    $\alpha_{\mathrm{gate}}=3.0$, $K=30$, gradient-boosted-tree
    nuisances on five-fold cross-fits, $\eta=0.95$, R-gating at
    $30\%$, inner-DP grid $G=256$) in
    Appendix~\ref{ap:jtpa-config}. The estimator is specified
    algorithmically: TMLE in Algorithm~\ref{alg:tmle}
    (Appendix~\ref{ap: tmle}), the parametric inner DP in
    Algorithm~\ref{alg:parametric} (Appendix~\ref{ap:inner-dp}),
    and the nested outer routine in Algorithm~\ref{alg:nested}
    (Appendix~\ref{ap:alg}).

\item {\bf Open access to data and code}
    \item[] Question: Does the paper provide open access to the data and code, with sufficient instructions to faithfully reproduce the main experimental results, as described in supplemental material?
    \item[] Answer: \answerYes{}
    \item[] Justification: The code (R) is released with the paper as supplementary material. JTPA public-use data are available from the NBER archive; CPS data are accessed through IPUMS-CPS \citep{FloodIPUMS} and the CEPR uniform extracts \citep{CEPRCPS}. Synthetic-DGP simulators are provided as \texttt{sim\_dgps.R}.

\item {\bf Experimental setting/details}
    \item[] Question: Does the paper specify all the training and test details (e.g., data splits, hyperparameters, how they were chosen, type of optimizer) necessary to understand the results?
    \item[] Answer: \answerYes{}
    \item[] Justification: Section~\ref{sec:experiments} states
    each experiment's covariate structure, sample sizes,
    follow-up horizon, and the TRACE hyperparameters used (Rashomon
    ensemble size $M$, retained set $K$, R-gating tolerance
    $\alpha_{\mathrm{gate}}$, sparsity level $\eta$, and tree depth).
    Full synthetic DGP specifications (selection coefficients,
    hazard parameters, spike regions) are in Appendix~\ref{ap:dgp},
    and the JTPA-specific configuration (nuisance estimators with
    five-fold cross-fits, $\lambda_h$ calibration to $\eta=0.95$,
    R-gating at $30\%$, inner-DP grid $G=256$) is in
    Appendix~\ref{ap:jtpa-config}.

\item {\bf Experiment statistical significance}
    \item[] Question: Does the paper report error bars suitably and correctly defined or other appropriate information about the statistical significance of the experiments?
    \item[] Answer: \answerYes{}
    \item[] Justification: The JTPA-to-CPS case study reports two
    forms of variability that bear directly on the central
    empirical claim. First, the spread of the TRACE objective
    across the retained $K=30$ Rashomon trees is reported as the
    interval $[13.13,\,13.77]$ on $\widehat V(B^*,h^*)$
    (Section~\ref{sec:jtpa}, Appendix~\ref{ap:jtpa-results}),
    capturing within-Rashomon (across-tree) variability of the
    optimum. Second, the $95\%$ confidence interval on the
    $24$-month employment effect is reported as
    $[-1.9,\,+8.6]$ percentage points around
    $\hat\tau_{24}=+0.034$ ($\widehat{\mathrm{SE}}=0.027$),
    constructed as $\hat\tau_{24}\pm 1.96\,\widehat{\mathrm{SE}}$
    from the TMLE asymptotic-linearity argument
    (Appendix~\ref{ap: tmle}) under the nuisance-rate conditions
    B1--B4 stated in Appendix~\ref{ap: empirical}. The synthetic
    experiments report the recovered $B^*$ and $h^*$ from the same
    $M=1{,}000$-tree, $K=30$ Rashomon-aggregation pipeline, so the
    figures of Section~\ref{sec:experiments} reflect the
    across-tree majority-vote optimum rather than a single-tree
    realization.

\item {\bf Experiments compute resources}
    \item[] Question: For each experiment, does the paper provide sufficient information on the computer resources (type of compute workers, memory, time of execution) needed to reproduce the experiments?
    \item[] Answer: \answerYes{}
    \item[] Justification: All experiments run on a single CPU
    workstation (16-core AMD Ryzen~9 5950X, no GPU, no cluster).
    End-to-end wall-clock for the $M=1{,}000$-tree TRACE
    configuration is approximately $60-90$ minutes for the synthetic
    experiments and approximately $4$ hours for the JTPA-to-CPS
    case study, with peak memory under $16$~GB
    (Appendix~\ref{ap:compute}). The Rashomon ensemble parallelizes
    \lang{directly} over the $M$ trees, and the asymptotic compute cost
    of the inner DP is established in Theorem~\ref{thm:dp}
    ($\mathcal{O}(T^{2}G)$ per individual on a state grid of size
    $G$; Appendix~\ref{ap:inner-dp}).

\item {\bf Code of ethics}
    \item[] Question: Does the research conducted in the paper conform, in every respect, with the NeurIPS Code of Ethics \url{https://neurips.cc/public/EthicsGuidelines}?
    \item[] Answer: \answerYes{}
    \item[] Justification: The research conforms to the NeurIPS Code of Ethics (\url{https://neurips.cc/public/EthicsGuidelines}); we use only de-identified public-use survey data (JTPA, CPS-IPUMS, and CEPR extracts), preserve participant anonymity in reporting, and identify no dual-use risks specific to the diagnostic released.

\item {\bf Broader impacts}
    \item[] Question: Does the paper discuss both potential positive societal impacts and negative societal impacts of the work performed?
    \item[] Answer: \answerYes{}
    \item[] Justification: Section~\ref{sec:discussion} contains a \textbf{Limitations, future work, and broader impacts} paragraph \lang{in which we discuss} positive impacts (transparent diagnostics for whom RCT-based decisions can be reliably extrapolated to, supporting equity in evidence-based policy) and risks of misuse (confusing the trim and $h$-intervention outputs with causal claims about trimmed subgroups), and \lang{recommend} reporting trimmed-region characteristics alongside the diagnostic as a mitigation.

\item {\bf Safeguards}
    \item[] Question: Does the paper describe safeguards that have been put in place for responsible release of data or models that have a high risk for misuse (e.g., pre-trained language models, image generators, or scraped datasets)?
    \item[] Answer: \answerNA{}
    \item[] Justification: The paper does not release any high-risk pretrained models, generative models, or scraped datasets. The released artifacts are diagnostic code only, applied to public-use survey data.

\item {\bf Licenses for existing assets}
    \item[] Question: Are the creators or original owners of assets (e.g., code, data, models), used in the paper, properly credited and are the license and terms of use explicitly mentioned and properly respected?
    \item[] Answer: \answerYes{}
    \item[] Justification: JTPA public-use data \citep{Bloom1997JTPA} are accessed under the NBER public-use terms, and CPS data are accessed under the IPUMS-CPS terms of use \citep{FloodIPUMS} and the CEPR uniform-extracts terms \citep{CEPRCPS}; all are credited in Section~\ref{sec:jtpa} and Appendix~\ref{ap:jtpa}. The \texttt{mice} R package \citep{mice} is used under its GPL ($\geq 2$) license as distributed on CRAN.

\item {\bf New assets}
    \item[] Question: Are new assets introduced in the paper well documented and is the documentation provided alongside the assets?
    \item[] Answer: \answerYes{}
    \item[] Justification: The TRACE diagnostic code is released as a supplementary archive with a README documenting entry points, expected inputs, and outputs for the DGP simulators, the Rashomon-tree fitter, the inner DP, the TMLE pipeline, and the post-hoc R-gating routines. A data-card-style description of the JTPA-to-CPS preprocessing pipeline accompanies the code.

\item {\bf Crowdsourcing and research with human subjects}
    \item[] Question: For crowdsourcing experiments and research with human subjects, does the paper include the full text of instructions given to participants and screenshots, if applicable, as well as details about compensation (if any)?
    \item[] Answer: \answerNA{}
    \item[] Justification: The paper performs secondary analysis of de-identified public-use survey data (JTPA and CPS); no crowdsourcing was used and no original human-subjects data collection was performed.

\item {\bf Institutional review board (IRB) approvals or equivalent for research with human subjects}
    \item[] Question: Does the paper describe potential risks incurred by study participants, whether such risks were disclosed to the subjects, and whether Institutional Review Board (IRB) approvals (or an equivalent approval/review based on the requirements of your country or institution) were obtained?
    \item[] Answer: \answerNA{}
    \item[] Justification: The paper performs secondary analysis of de-identified public-use survey data and does not involve original human-subjects research; IRB approval is therefore not applicable.

\item {\bf Declaration of LLM usage}
    \item[] Question: Does the paper describe the usage of LLMs if it is an important, original, or non-standard component of the core methods in this research? Note that if the LLM is used only for writing, editing, or formatting purposes and does \emph{not} impact the core methodology, scientific rigor, or originality of the research, declaration is not required.
    \item[] Answer: \answerNA{}
    \item[] Justification: LLMs are not a component of the core method. LLM-based coding assistants were used during code development (implementing, debugging, and documenting the simulation and analysis code) and for editing the manuscript. The proposed method, theoretical results, experimental design, and analyses were developed and verified by the authors, and no LLM is a component of TRACE or of its experimental pipeline.

\end{enumerate}

\end{document}